\documentclass[preprint,superscriptaddress,floatfix,amssymb,amsmath,aps,prc,preprintnumbers,letterpaper,showpacs]{revtex4}
\usepackage[dvips]{graphicx}
\usepackage{latexsym,epsfig,bm,psfrag,subfigure}
\usepackage{color}         
\usepackage{dcolumn}                 
\usepackage[bookmarksnumbered,bookmarksopen,colorlinks,citecolor=blue,linkcolor=blue]{hyperref} %
\usepackage{graphicx}
\usepackage{latexsym}

\begin{document}
\renewcommand{\vec}{\boldsymbol}
\newcommand{\mc}{M_{\mathrm{c}}}
\newcommand{\mn}{M_{\mathrm{n}}}
\newcommand{\mnc}{M_{\mathrm{nc}}} 
\newcommand{\mr}{M_{\mathrm{R}}}
\newcommand{\gma}{\gamma}
\newcommand{\gmat}{\tilde{\gamma}}
\newcommand{\pc}{\vec{p}_{c}}
\newcommand{\pn}{\vec{p}_{n}}
\newcommand{\bes}{\ensuremath{{}^{7}\mathrm{Be}}}
\newcommand{\be}{\ensuremath{{}^{8}\mathrm{B}}}
\newcommand{\besst}{\ensuremath{{}^{7}\mathrm{Be}^{\ast}}}
\renewcommand{\S}[2]{{}^{#1}S_{#2}}
\renewcommand{\P}[2]{{}^{#1}P_{#2}}
\newcommand{\gone}{h_{(\S{3}{1})}}
\newcommand{\gtwo}{h_{(\S{5}{2})}}
\newcommand{\gthree}{h_{(\S{3}{1}^{*})}}
\newcommand{\aone}{a_{(\S{3}{1})}}
\newcommand{\atwo}{\ensuremath{a_{(\S{5}{2})}}}
\newcommand{\rone}{r_{(\S{3}{1})}}
\newcommand{\rtwo}{r_{(\S{5}{2})}}
\newcommand{\hone}{h_{(\P{3}{2})}}
\newcommand{\htwo}{h_{(\P{5}{2})}}
\newcommand{\hpt}{h_{Pt}}
\newcommand{\hthree}{h_{(\P{3}{2}^{*})}}
\newcommand{\honet}{\tilde{h}_{(\P{3}{1})}}
\newcommand{\htwot}{\tilde{h}_{(\P{5}{1})}}
\newcommand{\Xone}{{X}_{(\S{3}{1})}}
\newcommand{\Xtwo}{{X}_{(\S{5}{2})}}
\newcommand{\Xonet}{\tilde{X}_{(\S{3}{1})}}
\newcommand{\Xtwot}{\tilde{X}_{(\S{5}{2})}}
\newcommand{\V}[1]{\vec{V}_{#1}}
\newcommand{\fdu}[2]{{#1}^{\dagger #2}}
\newcommand{\fdd}[2]{{#1}^{\dagger}_{#2}}
\newcommand{\fu}[2]{{#1}^{#2}}
\newcommand{\fd}[2]{{#1}_{#2}}
\newcommand{\T}[2]{T_{#1}^{\, #2}}
\newcommand{\Td}[2]{T_{#1}^{\dagger\;\;\; #2}}
\newcommand{\e}{\vec{\epsilon}}
\newcommand{\es}{\e^{*}}
\newcommand{\cw}[2]{\chi^{(#2)}_{#1}}
\newcommand{\cwc}[2]{\chi^{(#2)*}_{#1}}
\newcommand{\cwf}[1]{F_{#1}}
\newcommand{\cwg}[1]{G_{#1}}
\newcommand{\ke}{k}
\newcommand{\kest}{k_{\ast}}
\newcommand{\kc}{k_{C}}
\newcommand{\upartial}[1]{\partial^{#1}}
\newcommand{\dpartial}[1]{\partial_{#1}}
\newcommand{\etae}{\eta}
\newcommand{\etab}{\eta_{B}}
\newcommand{\etaest}{\eta_{\ast}}
\newcommand{\etabst}{\eta_{B\ast}}
\newcommand{\vecpt}[1]{\hat{\vec{#1}}}
\newcommand{\uY}[2]{Y_{#1}^{#2}}
\newcommand{\dY}[2]{Y_{#1 #2}}
\newcommand{\llangle}{\langle\langle}
\newcommand{\rrangle}{\rangle\rangle}


\title{Models, measurements, and effective field theory: proton capture on Beryllium-7 at next-to-leading order }


%

\author{Xilin Zhang} \email{xilinz@uw.edu}
\affiliation{Physics Department, University of Washington, 
Seattle, WA 98195, USA} 
\affiliation{Institute of Nuclear and Particle Physics and Department of
Physics and Astronomy, Ohio University, Athens, OH\ \ 45701, USA}
\author{Kenneth M.~Nollett} \email{kenneth.nollett@mail.sdsu.edu}
\affiliation{Department of Physics and Astronomy, 
University of South Carolina,
712 Main Street, Columbia, South Carolina 29208, USA} 
\affiliation{Department of Physics, San Diego State University,
5500 Campanile Drive, San Diego, California 92182-1233, USA} 
\affiliation{Institute of Nuclear and Particle Physics and Department of
Physics and Astronomy, Ohio University, Athens, OH\ \ 45701, USA}

\author{D.~R.~Phillips} \email{phillid1@ohio.edu}
\affiliation{Institute of Nuclear and Particle Physics and Department of
Physics and Astronomy, Ohio University, Athens, OH\ \ 45701, USA}

\date{\today\\[20pt]}

\begin{abstract}
We employ an effective field theory (EFT) that exploits the separation of scales in the $p$-wave halo nucleus $^8\mathrm{B}$ to describe the process
$^7\mathrm{Be}(p,\gamma)^8\mathrm{B}$ up to a center-of-mass energy of 500 keV.
 The calculation, for which we develop the lagrangian and power counting
in terms of velocity scaling, is carried out up to next-to-leading order (NLO) in the EFT expansion. The Coulomb force between $^7$Be and proton plays a major role in both scattering and radiative capture at these energies. The power counting we adopt implies that Coulomb interactions must be included to all orders in $\alpha_{\rm em}$. We do this via EFT Feynman diagrams computed in time-ordered perturbation theory, and so recover existing quantum-mechanical technology such as the Lippmann-Schwinger equation and the two-potential formalism for the treatment of the Coulomb-nuclear interference. Meanwhile the strong interactions and the E1 operator are dealt with via EFT expansions in powers of momenta, with a breakdown scale set by the size of the ${}^7$Be core, $\Lambda \approx 70$ MeV. Up to NLO the relevant physics in the different channels that enter the radiative capture reaction is encoded in ten different EFT couplings. 
The result is a model-independent parametrization for the reaction amplitude in the energy regime of interest. To show the connection to previous results we fix the EFT couplings using results from a number of  potential model and microscopic calculations in the literature. Each of these models corresponds to a particular point in the space of EFTs. The EFT structure therefore provides a very general way to quantify the model uncertainty in calculations of $^7\mathrm{Be}(p,\gamma)^8\mathrm{B}$. We provide details of this projection of models into the EFT space and show that the resulting EFT parameters have natural size. We also demonstrate that the only N$^2$LO corrections in $^7\mathrm{Be}(p,\gamma)^8\mathrm{B}$ come from an inelasticity that is practically of N$^3$LO size in the energy range of interest, and so the truncation error in our calculation is effectively N$^3$LO. The key LO and NLO results  have been presented in our earlier papers. The current paper provides further details on these studies.  We also discuss the relation of our extrapolated $S(0)$ to the previous standard evaluation.

\end{abstract}


\maketitle

\section{Introduction}

The nuclear reaction $^7\mathrm{Be}(p,\gamma)^8\mathrm{B}$ creates $^8\mathrm{B}$ nuclei inside the Sun, where they quickly decay to produce neutrinos. These $^8\mathrm{B}$ neutrinos constitute most of the solar neutrino spectrum above 2 MeV \cite{Adelberger:2010qa, Robertson:2012ib} and thus nearly the entire signal in chlorine- and water-based detectors \cite{Robertson:2012ib}.   Constraints on neutrino properties and solar interior composition based on this signal depend on comparisons of  detected and theoretical neutrino production rates, which require the $^7\mathrm{Be}(p,\gamma)^8\mathrm{B}$ cross section  \cite{Robertson:2012ib}.  However, this cross section must be extrapolated from experimental data above 100 keV  down to solar energies around 20 keV using a theoretical model, and the error associated with model selection dominates recent evaluations \cite{Davids:2003aw,Descouvemont:2004hh,Adelberger:2010qa}.  

The so-called Halo effective field theory (halo EFT) developed in recent years~\cite{vanKolck:1998bw,Kaplan:1998tg,Kaplan:1998we,Bertulani:2002sz,Bedaque:2003wa,Hammer:2011ye,Rupak:2011nk,Canham:2008jd,Higa:2008dn,Ryberg:2013iga,Hammer:2017tjm} is well suited to the study of proton capture on $^7$Be, because there is a nice separation of scales in the $^8$B system near the proton threshold.  The effective size of the $\bes$ nucleus can be estimated by looking at its lowest break-up channel, $\bes\rightarrow\, ^3\mathrm{He}+\,^4\mathrm{He}$, which has a $1.5874$ MeV~\cite{AME2012II} threshold.
 This translates to an effective binding momentum of $\sqrt{2 \mr' 1.5874} \approx 70 $ MeV ($\mr^\prime$ is the ${}^3$He-${}^4$He reduced mass \mbox{$\approx 12/7 M_N$}).  In developing our EFT below, we take this as the high momentum scale $\Lambda$, where the effective theory breaks down, corresponding to a short distance scale $\Lambda^{-1} \sim 3 $ fm.  The other scales relevant to low-energy direct capture are small compared with $\Lambda$.  The binding energy of $\be$ in the $\bes+p$ breakup channel is $B= 0.1364$ MeV \cite{AME2012II,AME2016II}, which translates to the binding momentum $\gamma = \sqrt{ 2 \mr B} =14.96 $ MeV. Here the reduced mass $\mr = \mn \mc /(\mn + \mc )$ while $\mn= 938.272$, and $\mc = 6534.18 $ MeV.  (In our notation, $n$ denotes a valence
``nucleon'' and $c$ denotes a ``core'' from which a larger nucleus is constructed.)  The static Coulomb interaction between $\bes$ and the proton is important near and below threshold, so the Coulomb momentum $\kc \equiv Q_c Q_n \alpha_{\rm em} \mr = 23.9487$ MeV (with $Q_c$ and $Q_n$ the particle charges) is also a key parameter.  This
may be written in terms of the channel momentum $k$ and the usual Sommerfeld parameter as $k_C = k\eta$, and it also is small compared to $\Lambda$.  Besides  ``elastic'' channels containing the \bes~ground state, there is a low energy excited state of $\bes$, with the excitation energy $E_\ast = 0.4291$ MeV; the corresponding momentum $\gamma_\Delta \equiv \sqrt{ 2\mr E_\ast} = 26.5352$ MeV is again small compared to $\Lambda$. (Note that these numbers are slightly different from those in our previous work~\cite{Zhang:2014zsa}, because here we update the $^8$B and \bes\ masses to the latest mass evaluation~\cite{AME2012II,AME2016II}.) All these momenta correspond to large distance scales $\sim 10 $ fm.  As a result, we can consider $\bes$ as a ``core'' particle with one low-energy excitation, and $\be$ as a shallow bound state with both $\bes+p$ and $\besst+p$ channels. Meanwhile the $s$-wave interaction between $\bes$ and $p$ in the initial state of our reaction has scattering lengths that are markedly larger than the short-distance scale
 in both
spin channels, $\approx 25$ and $\approx -7$ fm \cite{Angulo2003a}.  In that sense, the $s$-wave interactions between core and proton are abnormally strong.  Dynamics associated with all of these low-momentum scales---$\gamma$, $k_C$, $\gamma_\Delta$, as well as $1/a_{({}^5S_2)}$ and $1/a_{({}^3S_1)}$---can be accounted for in the EFT framework.  Based on this, an order-by-order expansion in a parameter that is $\sim \gamma/\Lambda \approx 0.2$ exists for the scattering and reaction amplitudes.

The dominance of large-length-scale contributions near threshold, which is a prerequisite for rapid convergence of an EFT, has always been prominent in models of the $^7\mathrm{Be}(p,\gamma)\be$ reaction.  Existing models based on ordinary quantum mechanics include potential models in which \be~is made of structureless protons and \bes~nuclei interacting through an effective potential~\cite{christyduck,barker80,davidstypel03,esbensen04,huang10,   navratil06}, phenomenological $R$-matrix models that avoid an explicit potential by reducing its effects to a few fitted parameters \cite{barker95,barker00}, and  ``microscopic'' calculations in which \be~is a collection of eight interacting nucleons.  Until recently, computational limits restricted microscopic models severely, but they nonetheless incorporated important effects like core excitation and wave function antisymmetry \cite{johnson92,descouvemont94,Descouvemont:2004hh}.   Computational limits are less severe now, and true {\it ab initio} calculations with bare nucleon-nucleon interactions and much more complete computational bases have become possible  \cite{Navratil:2011sa}.  As \textit{ab initio} models increase in completeness and complexity, they hold the promise of reducing  cross section uncertainties by producing reliable constraints complementary to those provided by experiment.

In all models, the dominance of regions with large cluster separation and zero strong interaction is the most important feature of low-energy direct capture.  The main difficulty lies in quantifying the influence of the short-range interaction.  It is our hope that an EFT formalism can provide a convenient language for understanding and comparing features of all models, and for fitting experimental data with a minimum of tacit assumptions.  EFT might also provide a check on the computational consistencies of more complex models and a convenient parameterization for disseminating their results.  It should also be useful for consistently stitching together multiple types of information from different experiments and calculations.

We have studied this reaction and its isobaric analog in EFT and presented our results in a series of short reports \cite{Zhang:2013kja, Zhang:2014zsa, Zhang:2015ajn, Zhang:2015vew} (see the discussion at the end of this section).  This paper serves to present all the technical details that have not been shown in those short papers and to discuss the relation to other models. Note that the same reaction has been studied in EFT in Ref.~\cite{Ryberg:2014exa} when our earlier reports were finished. Some content of the present paper is parallel to that in Ref.~\cite{Ryberg:2014exa}, as will be mentioned in the main text.  
In Section~\ref{sec:toymodel},  a simple model is used to illustrate the EFT Lagrangian and power counting in a Feynman diagrammatic approach. Similar theories have been developed before in systems without  Coulomb effects \cite{Bedaque:2003wa,Hammer:2011ye, Rupak:2011nk}.  A system with $s$-wave nuclear scattering in the presence of strong Coulomb interaction was also studied in EFT \cite{Kong:1999sf, Higa:2008dn}, and the related capture has been studied in \cite{Ryberg:2013iga, Ryberg:2015lea}. Here our power counting is based on so-called velocity scaling \cite{Luke:1996hj}, in order to handle the effective mass scale in the nonrelativistic dynamics.
The power counting for the EM interaction is also made transparent.  
 Then we identify the relevant leading order (LO) and next-to-leading order (NLO) diagrams for $s$- and $p$-wave scattering, as well as particular diagrams that contribute to the capture reaction.  Our $s$- and $p$-wave scattering calculations are based on a series of time-ordered-perturbation-theory diagrams, which yields the Lippmann-Schwinger expansion (LSE). The LSE is briefly developed in Appendix~\ref{app:Hformalism}, where we demonstrate that the Feynman diagrams and the corresponding diagrams in the LSE are the same. However, the LSE calculation is more suitable for the non-relativistic case at hand and also exposes the connection between EFT and conventional quantum-mechanical calculations, especially for the capture reaction.  In addition,  Coulomb effects and Coulomb wave functions (discussed in Appendix~\ref{app:coulombwf}) are well developed in the space coordinate, and the LSE enables straightforward transformation into coordinate space.
The developed Lagrangian, power counting, and calculational techniques are applied directly in Section~\ref{sec:realsticsystem} to study the $\bes+p$ system. The major challenge there is to handle the spin degrees of freedom and the low-energy excitation of the core, but the overall structure of the Lagrangian and the power counting is the same as in the previous section. 

Section~\ref{sec:capture} is devoted to the capture reaction. The relevant LO and NLO diagrams are identified and calculated one-by-one using time-ordered perturbation theory.  Section \ref{sec:modelcomparison} begins with a discussion of the EFT in the language of existing capture models and vice versa. It concludes with the results of fitting our EFT to a selection of models \cite{Davids:2003aw,esbensen96,navratil06,Descouvemont:2004hh}  from the literature.  This fitting lacks the ambiguity of fitting experimental data, because a model can be computed exactly and provides more information than just the total $S$-factor.  The fitted parameters locate published models unambiguously in the space of EFT parameter values, and show that our power counting works for those models.   

In our previous LO calculation \cite{Zhang:2014zsa}, we used the measured binding energy and scattering lengths along with \textit{ab initio} asymptotic normalization coefficients (ANCs) of the $\be$  bound state to fix couplings and find a $S(E)$ curve in  good agreement with available data (within the uncertainty of the EFT).  The results showed significant dependence of $S(E)$  on the $s$-wave scattering lengths when  all other parameters were kept fixed. 
A mistake there is corrected in the current paper, but the conclusion remains intact. The same LO calculation was applied successfully to the isospin mirror of the present reaction, i.e. $^7\mathrm{Li}+ n\rightarrow {}^8\mathrm{Li} + \gamma$ in Ref.~\cite{Zhang:2013kja}. By comparing these two calculations \cite{Zhang:2013kja, Zhang:2014zsa}, we found that isospin breaking occurs at a momentum scale at or above the breakdown scale $\Lambda$, so that the EFT parameters in the two systems are not the same when the EM interaction is switched on and off.  For the NLO amplitude developed in this paper, we took a different strategy to fix parameters.  We applied Bayesian methods to analyze the modern direct capture data, constrain EFT parameters, and obtain stringent constraints on the low energy $S(E)$ even without tight constraints on individual parameters. The major results were reported in Ref.~\cite{Zhang:2015ajn} with some details of the Bayesian analysis in Ref.~\cite{Zhang:2015vew}; we summarize these in Sec.~\ref{sec:realistic}.   We conclude with a short summary of major results.  

\section{A simple model} \label{sec:toymodel}
In this section we use a simplified model to explain the power-counting rules for the EFT Lagrangian and Feynman diagrams. We then identify the LO and NLO diagrams for proton-core scattering in the $s$- and $p$-waves, and show that this reproduces the Coulomb-modified effective range expansions (ERE) for the two scattering phase shifts. The energy variable $E$ in the resulting $T$-matrix operator $T(E)$ is then continued from the positive- (scattering) to the negative-(bound state) $E$ region in order to locate and study the shallow bound-state pole. This pole is the analog, in this simple model, of the $\be$ pole in the $\bes$-proton scattering amplitude. This section adds details to our previous brief reports~\cite{Zhang:2013kja,Zhang:2014zsa} and also lays the ground work for the realistic study of the $\bes$-proton system in the next section. Although the power-counting discussion relies heavily on Ref.~\cite{Luke:1996hj}, we reproduce it in full here, in order to make this paper self-contained. In the process we tailor the arguments of Ref.~\cite{Luke:1996hj} to the context of light-nuclear reactions. 

\subsection{Lagrangian}

The EFT Lagrangian is:
\begin{eqnarray}
\mathcal{L}&=& c^{\dagger}  \left[i \partial_0- e Q_c A_0 +\frac{\left(\overset{\rightarrow}{\nabla} - i e Q_c \vec{A}\right)^2}{2 M_c}\right]  c 
+n^{\dagger}  \left[i \partial_0- e Q_n A_0 +\frac{\left(\overset{\rightarrow}{\nabla} - i e Q_n \vec{A}\right)^2}{2 M_n}\right]  n  \notag \\
&& -\phi^{\dagger}  \left[i \partial_0- e Q_{nc} A_0 +\frac{\left(\overset{\rightarrow}{\nabla} - i e Q_{nc} \vec{A}\right)^2}{2 M_{nc}} + \Delta_\phi \right]  \phi 
 +  h_s \phi^{\dagger} n c   \notag + {\rm c.c.}\\&&
 + \pi^{\dagger\, i}  \left[i \partial_0- e Q_{nc} A_0 +\frac{\left(\overset{\rightarrow}{\nabla} - i e Q_{nc} \vec{A}\right)^2}{2 M_{nc}} +\Delta_\pi \right]  \pi_i + h_p \pi^{\dagger\, i}\, n \,\widetilde{\vec{V}}_{Ri}\,  c + {\rm c.c.} \ .  \label{eqn:toylag}
\end{eqnarray}
Here $c$, $n$, $\phi$, and $\pi^{\pm 1,0}$ are the core, proton (``nucleon''), $s$-wave dimer, and $p$-wave dimer fields and ${\rm c.c.}$ stands for complex conjugation. Repeated indices are implicitly summed, as they are throughout the paper. In the simple model of this section the $c$ and $n$ are spin-zero particles. Their masses are $\mc$ and $\mn$ and their charges are $Q_c$ and $Q_n$. Meanwhile the $\phi$ and $\pi$ spins correspond to the $s$-wave scattering state (zero) and the $p$-wave bound state (one). These dimer fields are introduced to simplify the EFT calculation~\cite{Kaplan:1996nv, Hammer:2011ye, Higa:2008dn, Ryberg:2014exa,Ryberg:2015lea}. Both have mass $\mnc\equiv \mn+\mc$, and charge $Q_{nc}\equiv Q_c + Q_n$.  $\Delta_\phi$ and $\Delta_\pi$ denote the dimers' unrenormalized binding energies. Note that the extra ``$-$'' for the free-$\phi$-field piece of the Lagrangian is introduced to reproduce a positive $s$-wave effective range, as will become clear in later discussion---see Eq.~(\ref{eq:DphiandERE2}). The interaction associated with $h_s$ is a contact $nc\phi$ coupling which leads to $nc$ $s$-wave scattering.  A similar term generates the $p$-wave interaction, but it is proportional to the relative velocity 
\begin{eqnarray}
\widetilde{\vec{V}}_R \equiv \left(\vec{V}_n -\vec{V}_c-e \frac{Z_{eff}}{\mr} \vec{A}\right) \ ,
\end{eqnarray}
where operator $\vec{V}_n$ ($\vec{V}_c$) picks up the velocity of the $n$ ($c$) particle, and  $Z_{eff}/\mr\equiv Q_{n}/\mn-Q_{c}/\mc$, with $\mr\equiv \mn \mc/(\mn+\mc)$  the reduced mass for the $cn$ system. In this coupling, the relative velocity dependence is required by Galilean invariance;  the photon coupling results from minimal substitution on particle momenta. 

It should be pointed out that if we take the usual convention that, under time reversal, a field with spin $J$ and spin projection $m$ transforms as  $\psi^{m}(\vec{x}, t) \stackrel{T}{\rightarrow}  (-1)^{J-m} \psi^{-m}(\vec{x}, -t)$, then both the $s$- and $p$-wave interactions in Eq.~(\ref{eqn:toylag}) are even under time reversal. Thus, under this standard convention for a spin-$J$ field, the factor of $i$ in the $\pi$-$n$-$c$ coupling that was present in our previous publications~\cite{Zhang:2013kja,Zhang:2014zsa} should, in fact, be absent. However, this change makes no difference to any physical amplitude that was calculated in those papers.

Adding the free Lagrangian for the photon, which is just the canonical one, $\mathcal{L}_{\gamma}=-\frac{1}{4} F^{\mu\nu} F_{\mu\nu}$ with $F^{\mu\nu} \equiv \partial^{\mu} A^{\nu}-\partial^{\nu} A^{\mu}$, to the matter Lagrangian of Eq.~(\ref{eqn:toylag}), then specifies the dynamics---apart from some higher-order $cn\gamma$ contact interactions which will be discussed below.  

\subsection{Velocity scaling}

Even without considering Coulomb interactions, the Lagrangian (\ref{eqn:toylag}) naively exhibits three distinct energy/momentum scales: the high ($\Lambda$) and low ($k_{\mathrm{low}}$) momentum scales associated with the short- and long-distance dynamics, and the reduced mass $\mr$. The appearance of particle masses obscures the power-counting discussion~\cite{Hammer:2011ye}. The so-called velocity scaling proposed in Ref.~\cite{Luke:1996hj} solves this problem by guaranteeing the correct scaling of momenta and energies for a non-relativistic theory. The low-momentum and low-energy scales are rewritten as, respectively, $\mr V \equiv k_{\mathrm{low}}$  (relative momentum) and $\mr V^2 $ (relative energy). 

Velocity scaling proceeds by defining the scaling factors for space and time as $\lambda_x\equiv 1/\left(\mr V \right)$, and $\lambda_t\equiv 1/\left(\mr V^2\right)$, since these are the typical space and time scales of interest in our EFT.  We then scale space and time with these factors, defining new, dimensionless, co-ordinates, via $\vec{x}\rightarrow \lambda_x \vec{X} $ and $t \rightarrow \lambda_t T$. The corresponding momentum and energy variables are $P^0$ and $\vec{P}$, defined by $\vec{p} \rightarrow \vec{P}/\lambda_x $ and $ p^0 \rightarrow P^0/\lambda_t $; again $\vec{P}$ and $P^0$ are then of order 1 in the EFT power counting. Meanwhile, matter fields are scaled by $\lambda_x^{-3/2}$ while $A^{\mu}$ is scaled by $\left(\mr \lambda_x^3\right)^{-1/2}$, so that the normalization of the free Lagrangian is the same after rescaling. Defining a rescaled Lagrange density, $\tilde{\mathcal{L}}$, from the original action $S$, via $S \equiv \int d^3 \vec{X} dT \tilde{\mathcal{L}}$ we find that the matter and minimal-substitution part of this rescaled Lagrange density is:
\begin{eqnarray}
\tilde{\mathcal{L}}&=& c^{\dagger}  \left[i \partial_0- \frac{e Q_c}{\sqrt{V}}  A_0 +\frac{\left(\overset{\rightarrow}{\nabla} - i e Q_c \sqrt{V} \vec{A}\right)^2}{2} (1-f) \right]  c 
+n^{\dagger}  \left[i \partial_0- \frac{e Q_n}{\sqrt{V}} A_0 +\frac{\left(\overset{\rightarrow}{\nabla} - i e Q_n \sqrt{V} \vec{A}\right)^2}{2} f \right]  n  \notag \\
&& -\phi^{\dagger}  \left[i \partial_0- \frac{e Q_b}{\sqrt{V}} A_0 +\frac{\left(\overset{\rightarrow}{\nabla} - i e Q_b \sqrt{V} \vec{A}\right)^2}{2} f(1-f) + \Delta_\phi \lambda_t\right]  \phi  \notag \\ 
&& + \pi^{\dagger\, i}  \left[i \partial_0- \frac{e Q_b}{\sqrt{V}} A_0 +\frac{\left(\overset{\rightarrow}{\nabla} - i e Q_b \sqrt{V} \vec{A}\right)^2}{2} f(1-f) +\Delta_\pi \lambda_t \right]  \pi_i   \notag \\
&& + \frac{\tilde{h}_s}{\sqrt{V}} \phi^{\dagger} n c + \tilde{h}_p \sqrt{V}  \pi^{\dagger\, i}\, n\,  (\vec{V}_n -\vec{V}_c -e Z_{eff} \sqrt{V}\vec{A})_i\,  c + {\rm c.c.} \label{eqn:Ltoy1}
\end{eqnarray}
Here $f\equiv \mc/\mnc$, and $\tilde{h}_l \equiv h_l \sqrt{M_R}$ for $l=s,p$. For simplicity, the symbols for the scaled fields, space-time derivatives $\partial_0$ and $\nabla$, and $\vec{V}_{n,c}$ are kept the same as before, but now the natural expectation for all free-particle terms is that they are of order $1$, since, e.g., $\vec{V}_n -\vec{V}_c$, is the relative velocity operator in units of the low velocity scale $V$. For a matter field with four-momentum $(P^0,\vec{P})$ the propagator is now $1/\left(P^0-1/2 \vec{P}^2\right)$. 

Essentially by construction then, the only dependence on mass in the Lagrangian (\ref{eqn:Ltoy1}) comes through the fraction $f$. 
The velocity scaling proposed in Ref.~\cite{Luke:1996hj} indeed makes it explicit that the velocity $V$ is what determines the suppression or enhancement of different terms in the EFT. As with the more standard Lagrangian written in terms of momenta, explicit factors of $V$ in strong-interaction vertices will be compensated by factors of $V_\Lambda$ buried in couplings, e.g., the appearance of a $\sqrt{V}$ in the denominator (numerator) of the $s$-wave ($p$-wave) interaction term means that, once the natural scaling of $\tilde{h}_s$ ($\tilde{h}_p$) is taken into account, that term will be enhanced by a factor of $\sqrt{V_\Lambda/V}$ (suppressed by a factor of $\sqrt{V/V_\Lambda}$).

Photon-matter interactions reveal the full benefit of velocity scaling, and in Eq.~(\ref{eqn:Ltoy1}) we have also included the interactions with photon fields that minimal substitution produces. The factors of $\sqrt{V}$ in the minimal couplings of $A_0$ and $\vec{A}$ photons are simply a consequence of the (different) roles of time and space derivatives in non-relativistic dynamics: the $A^0$ photon coupling is proportional to $\frac{1}{\sqrt{V}}$, while the transverse ($\vec{A}$) one is $\propto \sqrt{V}$, so transverse photons are suppressed by a factor of $V$ relative to $A^0$ photons. Here, in contrast to the strong interactions, the suppression is by $V/c$, i.e. the velocity $V$ is to be measured in units of the speed of light---since these minimal-substitution vertices are not sensitive to the breakdown velocity $V_\Lambda$ it is not that ratio that controls the suppression, but the (significantly smaller) $V/c$. 
Meanwhile, free-photon propagation is also clearer in terms of velocity scaling: for an on-shell transverse photon (e.g., the photon radiated in the reaction of interest) its momentum and energy both scale as $\mr V^2$, but for an off-shell photon (e.g., a photon exchanged between the charged core and the charged proton) the two scale as $\mr V$ and $\mr V^2 $. This is made explicit by defining a rescaled free-photon Lagrangian density $\tilde{\mathcal{L}_\gamma}$, through $S_\gamma \equiv \int d^3 \vec{X} dT \tilde{\mathcal{L}}_\gamma$, which is:
\begin{eqnarray}
\tilde{\mathcal{L}}_\gamma &=& \frac{1}{2} A_i \left[(\nabla^2 -V^2 \partial_0^2 )\delta_{ij} -\partial_i \partial_j \right] A_j
-\frac{1}{2} A^0 \nabla^2 A^0 - V \partial_i A^0 \partial_0 A_i \ , \notag
\end{eqnarray}
 For the $A^0$ piece of the photon field that generates the Coulomb potential the rescaled propagator is then $1/\vec{K}^2$, while transverse ($\vec{A}$) photons have a propagator $\frac{(-)}{\vec{K}^2-(V {K^0})^2} (\delta_{ij}-\frac{K_i K_j}{\vec{K}^2}) $. In the last propagator, the factor $(V K^0)^2$ in principle should be expanded in geometric series. However, if the transverse photon goes on shell (becomes a ``radiation photon") that series needs to be resummed. A detailed discussion of this distinction can be found in  Ref.~\cite{Luke:1996hj}. Since we will, for the most part, consider only internal photon lines that obey the kinematics $\vec{K} \sim \mr V$ and $K^0 \sim \mr V^2 $ we do not reproduce that discussion here. 
 
 \subsection{Power counting for $nc$ strong-interaction parameters}

 \label{sec:toypowercounting}
 
 \begin{table}
\begin{ruledtabular} 
   \begin{tabular}{cccc}
	  $\Delta_\phi \lambda_t$  & $\tilde{h}_s/\sqrt{V} $ & $\Delta_\pi \lambda_t $ &  $ \tilde{h}_p \sqrt{V} $  \\ \hline
     $1$  & $ \sqrt{\frac{V_\Lambda}{V}} $ &  $ 1$  &  $\sqrt{\frac{V}{V_\Lambda}} $ \\ 
     $ \frac{V_\Lambda}{V} $ & $ \sqrt{\frac{V_\Lambda}{V}} $ & $ 1 $ & $ \sqrt{\frac{V}{V_\Lambda}} $ \\ 
\end{tabular} \caption{The first row shows NDA assignments of scales for strong interaction couplings, while in the second row are those used to reproduce the dynamics in our problem. } \label{tab:NDAstrong}
	\end{ruledtabular}
	\end{table}
	
The power-counting for the Lagrangian will then be complete if we can determine how many powers of the high scale (now $V_\Lambda$) the strong-interaction parameters $\tilde{h}_s$, $\Delta_\phi$, $\tilde{h}_p$, and $\Delta_\pi$ carry. Naive dimensional analysis (NDA) applied to the rescaled Lagrangian yields  the scalings shown in the first row of Table~\ref{tab:NDAstrong}. 

\begin{figure}
\centering
\includegraphics[width=12cm, angle=0]{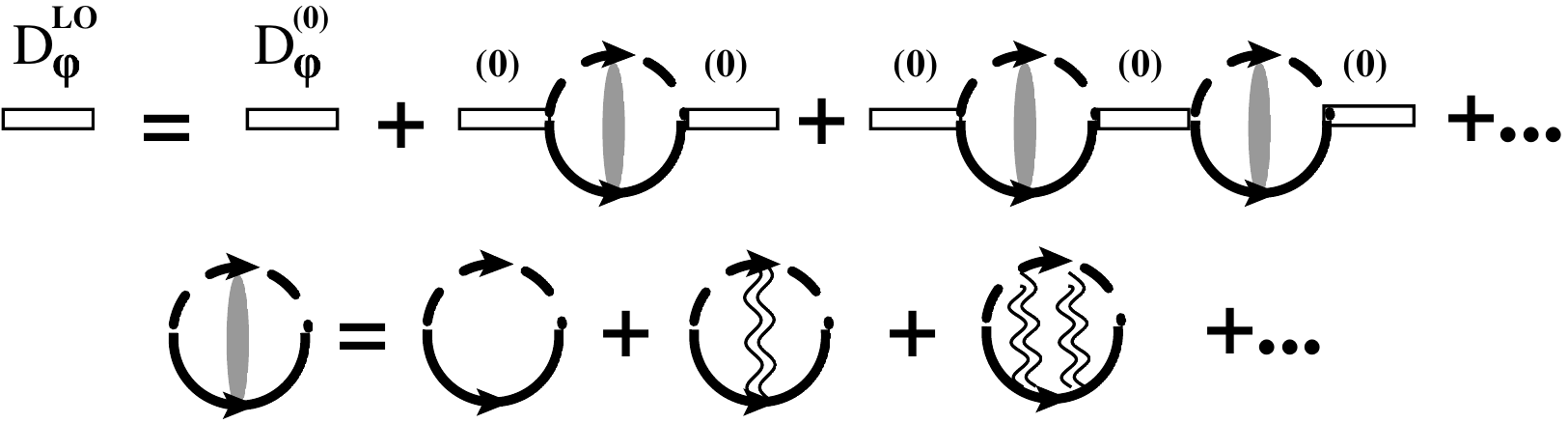}
\caption{The top equation shows the leading-order fully dressed propagator for $\phi$, $D_{\phi}^\mathrm{LO}$ (LHS), that results from the resummation of the bubble diagrams (RHS). The propagators in these diagrams are the free ones $D_{\phi}^{(0)}$, as discussed in the main text. The lower equation expands the filled oval as a series of $0$, $1$, $2$, \ldots Coulomb photon exchanges (double wavy lines) between the charged proton (solid line) and core (long dashed line). This line labeling will be used throughout the entire paper.} \label{fig:Dphi_LO}
\end{figure}

We now consider the Dyson series for the  $\phi$ propagator, $D_{\phi}^\mathrm{LO}$, shown on the first line of Fig.~\ref{fig:Dphi_LO}. Based on the rescaled Lagrangian in expression~(\ref{eqn:Ltoy1}), we can estimate the size of these diagrams by counting factors of $V$ in vertices and propagators.  If we adopt NDA scaling for $\Delta_\phi$ then it is order one, and, since the rescaled particle momentum are also $O(1)$, the free propagator of the $s$-wave dimer $\phi$ field, $D_\phi^{(0)}$ is $O(1)$. The diagrams that constitute the leading part of the $\phi$ self energy are defined on the second line of Fig.~\ref{fig:Dphi_LO}. In this paragraph we consider only the self-energy bubble without any Coulomb interaction: the first diagram on the right-hand side (RHS) of the lower line of Fig.~\ref{fig:Dphi_LO}. In contrast to standard EFT power counting there is no need to keep track of factors from loops, since the scaled-momentum integration in the loop calculation is always of order $1$. However, the one-loop self energy is enhanced, due to the presence of a factor $\tilde{h}_s^2 \sim \frac{V_\Lambda}{V}$. Then, for a natural $\Delta_\phi$, each term in the Dyson series for the $\phi$ propagator is larger than the last, thus vitiating a diagrammatic expansion for the dressed propagator $D_\phi$.  Following Refs.~\cite{Kaplan:1998tg,Kaplan:1998we} we ensure that each term in the Dyson series is of the same EFT order by enhancing the unrenormalized $\phi$ mass, to  $\Delta_\phi \lambda_t \sim \frac{V_\Lambda}{V}$ as shown in the second row of table~\ref{tab:NDAstrong}. With this counting each term in the first line of Fig.~\ref{fig:Dphi_LO} is of the same size ($\sim V_\Lambda/V$). Since we have kept the $\tilde{h}_s$ scaling unchanged this constitutes a fine tuning between the NDA estimate of the self-energy bubble and the size of the dimer's bare mass. 
The fact that the rescaled $\Delta_\phi$ $\sim V_\Lambda/V$ 
while the kinetic, $P^0 - f(1-f) {\vec P}^2/2$, piece of the inverse propagator is still $\sim 1$ then also justifies dropping the kinetic piece of the $\phi$ propagator at leading-order in the EFT expansion~\footnote{The factor $f(1-f) \approx 0.1$ in $\bes$-p system, which may suppress the kinetic term further in this context. In general, $f(1-f) \leq 1/4$.}. In other words, under the scaling in the second line of Table~\ref{tab:NDAstrong}, the  scaled $s$-wave dimer propagator can be taken to be static at LO: $D_\phi^{(0)} = \frac{(-)}{ \Delta_\phi \lambda_t } \sim  \frac{V}{V_{\Lambda}}$.

For the $p$-wave dynamics, we follow the NDA assignments.  The $p$-wave bubble is then suppressed by a factor of $V/V_\Lambda$---in contradistinction to the $s$-wave bubble. We will see that this is indeed the correct power-counting conclusion, except in certain special kinematic regions. Such a power counting, in which the self-energy of the $p$-wave dimer is suppressed relative to its kinetic part, has been used in earlier EFT studies~\cite{Bedaque:2003wa, Hammer:2011ye} of systems sharing the same feature of  a low-energy $p$-wave resonance. Those studies, were, however, for neutron-core scattering, and so did not consider the role of the Coulomb interaction. 

\subsection{Power counting with Coulomb}

Turning our attention, then, to the power counting of such Coulomb interactions, we first point out that 
$s$-wave scattering with strong Coulomb effects was previously studied in this EFT in Refs.~\cite{Kong:1999sf,Higa:2008dn,Ryberg:2015lea}. We will now reiterate these arguments, albeit in the context of velocity-scaling, for the $\phi$ propagator, and discuss their extension to the $\pi$ propagator. As already discussed,  the self-energy bubble without Coulomb-photon exchange, i.e., the first diagram on the RHS of the lower line of Fig.~\ref{fig:Dphi_LO}, $\sim \sqrt{\frac{V_{\Lambda}}{V}} \times 1 \times \sqrt{\frac{V_{\Lambda}}{V}} \sim \frac{V_\Lambda}{V}$. As we move from left-to-right in that figure each diagram has an extra exchange of a Coulomb photon. That results in the diagram acquiring an additional factor associated with the product of the two $A_0$-photon vertices $\sim \frac{\alpha_{\rm em} Q_c Q_n \mr}{k} =\frac{\kc}{k}  \equiv \eta$. (Recall that the $A_0$-photon propagator is $O(1)$ in terms of rescaled momenta.) This factor, $\eta$, is known as the Sommerfeld parameter.  In the energy region of interest here it is $\sim 1$. The loop integrations also generate factors $\sim 1$, and so, as long as $\eta \sim 1$, resummation of the ladder of Coulomb photon exchange diagrams is mandatory. This then defines the LO $s$-wave dimer self-energy: $\Sigma_\phi \sim \frac{V_\Lambda}{V} $. Such a self-energy, which includes the sum of the exchange of zero, one, two, \ldots Coulomb-photon exchanges will henceforth be denoted by a shaded bubble. It follows that, in the kinematic regime $k \sim \kc$, the addition of Coulomb photons  does not change the order of 
the self energy from the order computed with the free $nc$ Green's function. It is just that now the self energy must be computed using a core-proton Green's function that includes one-Coulomb-photon exchange to all orders in $\alpha_{\rm em}$.
That self energy is still---as in the $\alpha_{\rm em}=0$ case---resummed in a geometric series, as per the upper line of Fig.~\ref{fig:Dphi_LO}, and this procedure generates the LO $\phi$ propagator, $D_{\phi}^\mathrm{LO}$, in the $nc$ system.

\begin{figure}
\centering
\includegraphics[width=12cm, angle=0]{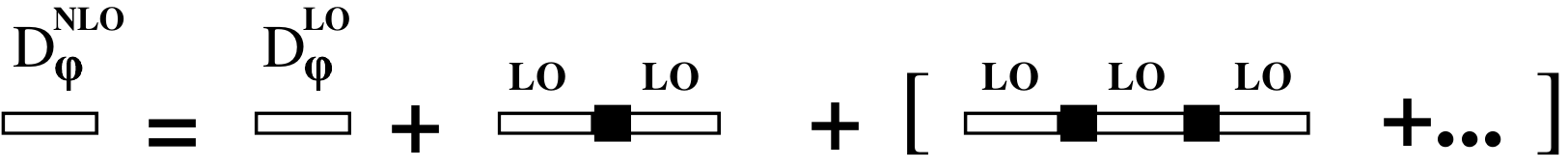}
\caption{The  NLO $\phi$ propagator $D_\phi^\mathrm{NLO}$ as a sum of zero and one insertions of the dimer kinetic energy term that encodes the effective range. The bracketed diagrams containing two, three, \ldots insertions are strictly higher order, but resumming them allows us to exactly match the ERE. } \label{fig:Dphi_NLO}
\end{figure}

To compute the dressed propagator at NLO, $D_{\phi}^\mathrm{NLO}$, the second diagram shown in Fig.~\ref{fig:Dphi_NLO} should be included. The vertex depicted there as a small filled box is the $\phi$-field kinetic term that got demoted to NLO when we chose to enhance $\Delta_\phi$ over its NDA estimate. We see that the  diagram  with the single insertion of this vertex $\sim  \frac{V}{V_{\Lambda}} \times 1 \times  \frac{V}{V_{\Lambda}}$, which makes it NLO compared to the LO $\phi$ propagator, $D_\phi^\mathrm{LO}$ (we established that is $\sim \frac{V}{V_\Lambda}$). Note that if we wish to recover the effective-range expansion exactly, not just order by order in the effective range, then we must resum a geometric series involving this kinetic-energy operator. This is the content of the terms in square brackets in Fig.~\ref{fig:Dphi_NLO}. However, strictly speaking, only the second diagram on the RHS of Fig.~\ref{fig:Dphi_NLO} is NLO. 

\begin{figure}
\centering
\includegraphics[width=12cm, angle=0]{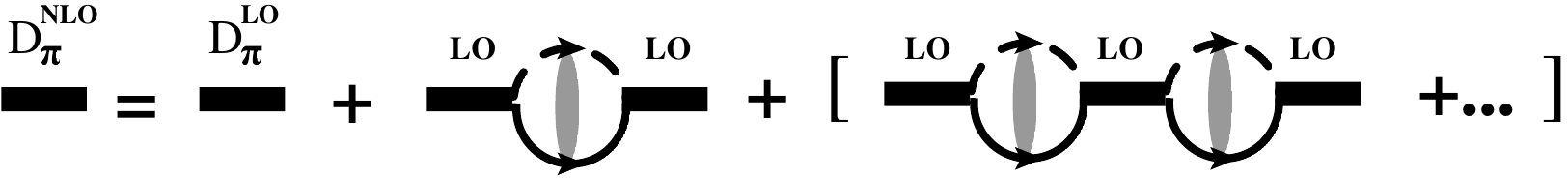}
\caption{The NLO $\pi$ propagator, $D_\pi^\mathrm{NLO}$, is a sum of shaded SE ($\Sigma_\pi$) insertions, by using the free propagator as the LO one, $D_\pi^\mathrm{LO}$. Again shading the bubble diagram means a sum of  diagrams with $0$ to $\infty $ number of Coulomb photon exchanges. In order to differentiate the $\pi$ propagator from the $\phi$ one, here, and in the following discussion, we use a filled rectangle to label $D_\pi$. } \label{fig:Dpi_LONLO}
\end{figure}

The situation for the $p$-wave dimer propagator, $D_{\pi}$, is different. In Fig.~\ref{fig:Dpi_LONLO}, we show the Dyson series for this case.  Since $\Delta_\pi \lambda_t \sim 1$, the rescaled free propagator $D_\pi^{(0)} = \frac{1}{P^0 - 2/1 \vec{P}^2 f (1-f) + \Delta_\pi \lambda_t}$ is also of order $1$. Meanwhile, the $p$-wave self-energy $\Sigma_\pi \sim \frac{V}{V_\Lambda}$, since the loops generate factors of order one, for an arbitrary number of Coulomb-photon exchanges, as long as $k \sim \kc$, and the coupling $\tilde{h}_p$ gives suppression by a factor of $V$. 
Therefore the second diagram on Fig.~\ref{fig:Dpi_LONLO}'s RHS is $\sim \frac{V}{V_\Lambda}$ and hence is NLO. However, as argued in Refs.~\cite{Pascalutsa:2002pi,Bedaque:2003wa,Hammer:2011ye}, when the center-of-mass energy (i.e. $P^0-\vec{P}^2/2\mnc$ in terms of the unscaled momentum) is close to $\Delta_\pi$, the leading-order propagator becomes larger than 
the NDA expectation. In this regime the entire series shown on the RHS of Fig.~\ref{fig:Dpi_LONLO} must be resummed. This is the regime that is pertinent to $p$-wave bound states in the proton-core system, and since we are interested in $\be$ as a $p$-wave proton-$\bes$ bound state we must use that resummation here. 

\begin{figure}
\centering
\includegraphics[width=6cm, angle=0]{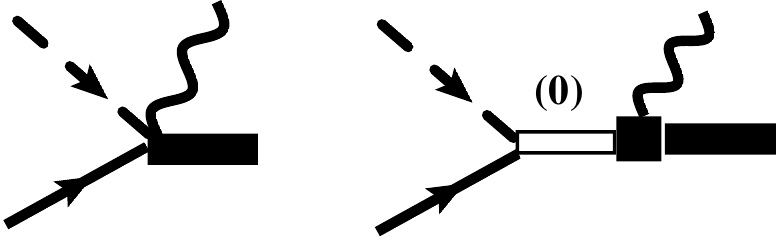}
\caption{A LO diagram (1st) and a NLO diagram (2nd) for $s$-wave to $p$-wave bound state radiative capture. The LO is due to the gauged $n-c-\pi$ coupling. In the 2nd diagram, the filled box is for the contact coupling term discussed in Eq.~(\ref{eqn:contacttoy}). Comparing the two tells the size of the coupling $L_{E1}$. } \label{fig:stop}
\end{figure}

\subsection{Power counting for radiative capture} \label{sec:toypowercounting2}

Turning our attention now to electromagnetic reactions: for a general diagram and interaction vertex the power counting is similar to that we have done so far for scattering. The size 
of a diagram can be established by the number of factors of $V$ that it carries. Consider the diagrams $A$ and $B$ that both describe the same capture reaction. If diagram $A$ 
carries a factor $V^{N_A}$ and diagram $B$ carries a factor of $V^{N_B}$ then NDA states that their ratio is:
\begin{equation}
\frac{V^{N_A-N_B}}{V_{\Lambda}^{N_A-N_B}}=\left(\frac{k_{\mathrm{low}}}{\Lambda}\right)^{N_A-N_B}\ ,
\end{equation} 
since the explicit factors of $V$ from the Lagrangian must be compensated by factors of $V_\Lambda$ in couplings. (Note that this assumes that the same types of photon, longitudinal or transverse, appear in both diagrams $A$ and $B$. Otherwise factors of $V$, not $V/V_\Lambda$, will appear.)

First we compute the order of a graph  that results from the gauged $h_p$ coupling in the Lagrangian, which is just one of a set of LO diagrams for the capture reaction. (The full set can be found in Fig.~\ref{fig:stopcaptureLO}.) This first diagram in Fig.~\ref{fig:stop} is order $V$, since it includes a factor of $\sqrt{V}$ from the $\pi n c$ coupling, and an additional factor of $\sqrt{V}$ from the coupling to the radiated ($\vec{A}$) photon.

Since EFT includes all interactions consistent with underlying symmetries the Lagrangian (\ref{eqn:Ltoy1}) must be supplemented by terms that are gauge-invariant by themsleves. The leading such term that describes the E1 transition between the $s$-wave and the $p$-wave dimer,
\begin{eqnarray}
-i e Z_{eff} L_{E1} {\pi^{\dagger}}^i \vec{E}_i \phi  \ ,  \label{eqn:contacttoy1}
\end{eqnarray}
contributes to the capture reaction via the second diagram in Fig.~\ref{fig:stop}, where the filled box means the $L_{E1}$ coupling. This is a contact term and it renormalizes loop graphs that appear at the same order in the EFT exapnsion. 
The factor of $i$ ensures that this coupling is invariant under time reversal. The factor was omitted in the Lagrangians for short-distance electromagnetic operators that were given in Ref.~\cite{Hammer:2011ye}, but this does not affect any of the results presented there.

After the velocity scaling discussed at the beginning of this section is applied, the contact term changes to 
\begin{eqnarray}
-i e Z_{eff} \tilde{L}_{E1} V^{\frac{3}{2}} {\pi^{\dagger}}^i \vec{E}_i \phi  \ . \label{eqn:contacttoy}
\end{eqnarray}
Here the overall factor of the reduced mass has been eliminated by defining $\tilde{L}_{E1}=L_{E1} M_R$. 
After this scaling the second diagram in Fig.~\ref{fig:stop} is $\sim \tilde{h}_s \frac{1}{\sqrt{V}} D_\phi^{(0)} e Z_{eff} \tilde{L}_{E1} V^{\frac{3}{2}}$. Assuming NDA for $\tilde{h}_s$ and $\tilde{L}_{E1}$, and inserting the power counting for the $s$-wave propagator identified above, $D_\phi$ $\sim V/V_\Lambda$, produces an overall scaling of  $V^2$ for the graph involving the E1 contact operator that is proportional to the LEC $L_{E1}$. Thus, according to NDA, the ratio of the second and first diagram should be of order $V/V_{\Lambda}$, which shows that the E1 contact term contributes to the capture into a $p$-wave bound state at NLO. (The counting is different if the reaction proceeds from a $p$-wave scattering state, into an $s$-wave bound state~\cite{Ryberg:2015lea}.) This also means that $\tilde{L}_{E1} \sim 1/V_{\Lambda}$ and hence $L_{E1} \sim 1/\Lambda $.

\subsection{Toy amplitude for $s$-wave scattering up to NLO} \label{sec:swavetoymod}

Having established the power counting through the use of velocity scaling we now return to expressions in terms of momenta. The discussion can be continued in terms of velocities, and this has the benefit of yielding dimensionless integrals. But the connection with previous work in halo EFT is  more straightforward if amplitudes are written in terms of momenta. 

The power-counting discussion of the previous subsection is based on an expansion in Feynman diagrams. However, in practice, time-ordered perturbation theory is more suited for our  calculations. In particular, the use of time-independent quantum-mechanical perturbation theory allows us to employ the Lippmann-Schwinger equation (LSE) for resummations, such as the one that takes place for Coulomb interactions between the proton and the core. This, in turn, allows us to identify Coulomb wave functions---with all their well-known properties---in our calculation. 

\begin{figure}
\centering
\includegraphics[width=12cm, angle=0]{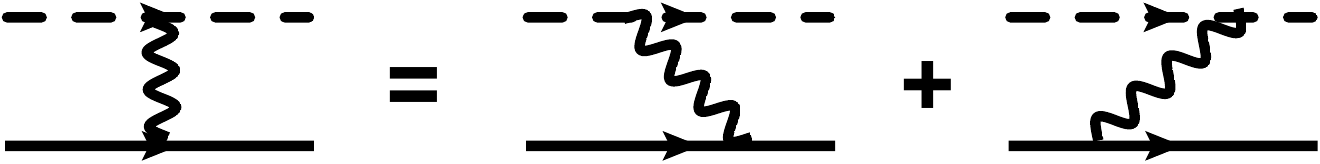}
\caption{Diagrams for $nc$ scattering due to transverse photon exchange. The LHS is the Feynman diagram. The corresponding time-ordered perturbation theory diagrams are on the RHS.} \label{fig:transversephotonexchange}
\end{figure}

In fact, since particle-antiparticle pair production does not exist in this EFT, the intermediate states that occur in a given Feynman diagram all have fixed particle content. For proton-core scattering 
this diagram is the same as a particular contribution to the LSE time-ordered perturbation theory series. The only exception is transverse photon exchange between charged particles, for which an example is shown in Fig.~\ref{fig:transversephotonexchange}. Its LHS is the one-transverse-photon exchange Feynman diagram, which in fact equals the sum of two time-ordered perturbation theory graphs on the RHS. However in our problem, radiative corrections turn out not to affect the result at the accuracy we seek. 
Therefore, in the following calculation, we generate Feynman diagrams but then use the corresponding time-ordered perturbation theory expression to do the matrix element computation. This in no way affects the power counting, since the time-ordered and Feynman graphs are equivalent.  The LSE in the context of our EFT is developed in Appendix~\ref{app:Hformalism}, which includes a brief discussion of quantization, Fock-state definition, and calculations of various matrix elements corresponding to vertices and propagators in a Feynman diagram. Our notation is also defined there. 
(While Ref.~\cite{Kong:1999sf} used the LSE in their EFT calculation, the connection to the original field theory is not fully explained there.).

First we compute the $\phi$ propagator, defined in Eq.~(\ref{eqn:DphidefHformalism}) as a matrix element between $|\phi\rangle$ plane wave states. The discussion of power counting above implies  the LO free propagator is static: $D_{\phi}^{(0)} $  should be defined as  $\langle \vec{P}_\phi' \vert \frac{1}{\Delta_{\phi}}  \vert \vec{P}_\phi \rangle = (2\pi)^3 \delta(\vec{P}_\phi-\vec{P}_\phi') \frac{(-)}{\Delta_{\phi} } \equiv (2\pi)^3\delta(\vec{P}_\phi-\vec{P}_\phi')  D_{\phi}^{(0)} \left(\vec{P}_\phi', \vec{P}_\phi, E \right) $; the ``$-$'' is due to the negative norm of $|\phi\rangle$. Notice that a general matrix element always has the $(2 \pi)^3 \delta()$ factor due to total three-momentum conservation in time-ordered perturbation theory. 

The $\phi$ self energy, $\Sigma_\phi$, is given by the sum of the series of zero to infinitely many Coulomb-photon exchanges depicted in the second line of Fig.~\ref{fig:Dphi_LO}. The matrix 
element is written as:
\begin{align}
 &(2\pi)^3 \delta(\vec{P}_\phi-\vec{P}_\phi') \Sigma_{\phi} \left(\vec{P}_\phi;\, E \right)=\\
 & \qquad \langle \vec{P}_\phi' \vert  W_s \left[ \frac{1}{E-H_0+ i 0^{+}} + \frac{1}{E-H_0+ i 0^{+}} W_C \frac{1}{E-H_0+ i 0^{+}} + \cdots \right] W_s   \vert \vec{P}_{\phi} \rangle  \notag \\ 
& \qquad \qquad =  \langle \vec{P}_\phi'\vert W_s \frac{1}{E-H_C+ i 0^{+}} W_s   \vert \vec{P}_{\phi} \rangle \label{eqn:phiselfEcal2}
\end{align}
where we have summed the series algebraically in the third line by defining the Coulomb Hamiltonian, $H_C=H_0+W_C$. Meanwhile, as discussed in 
 Appendix~\ref{app:Hformalism}, $W_s$ is the strong potential in this system, and it produces an $nc \rightarrow$ dimer transition. The first $W_s$ (reading from left to right) annihilates $n$ and $c$ particles from the intermediate state and creates a $\phi$ in the final state, while the second $W_s$ annihilates the $\phi$ in the initial state and creates a $nc$ pair in the intermediate state. 

To compute $\Sigma_\phi$, we make use of the Coulomb-distorted two-body states $|\vec{P}, \cw{\vec{p}}{\pm}\rangle$ that are the eigenstates of $H_C$ (see Eq.~(\ref{eqn:CoulombWFdef})). Here we omit the $nc$ subscripts on momenta in the $nc$ Fock-space state; dimer momenta are still indicated as such. 
We also introduce the intrinsic states, defined in the $nc$ center-of-mass frame, for the Coulomb Hamiltonian. With the intrinsic Hamiltonian, $\overline{H}_C$ defined as
\begin{equation}
\overline{H}_C =-\frac{ \nabla^2 }{2 \mr} + \frac{e^2 Q_cQ_n}{4 \pi r }
\end{equation}
the intrinsic states satisfy $\overline{H}_C |\cw{\vec{p}}{\pm} \rrangle = \overline{E} |\cw{\vec{p}}{\pm} \rrangle $ with the relative energy $\overline{E}\equiv E-\frac{\vec{P}_\phi^2}{2 \mnc }$ equal to the energy of the $nc$ pair in its center-of-mass frame.  The matrix element of $W_s$ can then be re-expressed as:
\begin{align}
&\langle\vec{P}_{\phi} | W_s |\vec{P}, \cw{\vec{p}}{\pm}\rangle = (2 \pi)^3 \delta(\vec{P}-\vec{P}_\phi) \llangle \phi |\overline{W}_s| \cw{\vec{p}}{\pm}  \rrangle;
&\qquad \llangle \phi |\overline{W}_s| \cw{\vec{p}}{\pm}  \rrangle \equiv h_s \cw{\vec{p}}{\pm}(0).
\end{align}
Note that since the $nc \phi$ Lagrangian is a point coupling the probability amplitude for the conversion of an $nc$ pair into a dimer is the product of the coupling $h_s$ and the size of the $nc$ wave function at $\vec{r}=0$. 

By introducing these intrinsic wave functions, which are the solutions of quantum-mechanical one-body problems, we can use standard quantum-mechanical results to do our calculation.
Inserting a complete set of eigenstates of $H_C$, and using these definitions, the matrix element becomes
\begin{align}
&\int \frac{d \vec{P}}{(2 \pi)^3}\frac{d\vec{p}}{(2 \pi)^3} \langle \vec{P}_\phi'\vert W_s \frac{1}{E-H_C+ i 0^{+}}  |\vec{P},  \cw{\vec{p}}{+} \rangle   \langle\vec{P},  \cw{\vec{p}}{+} | W_s   \vert \vec{P}_{\phi} \rangle  \notag \\
& \qquad = (2\pi)^3 \delta(\vec{P}_\phi-\vec{P}_\phi') \int \frac{d\vec{p}}{(2 \pi)^3} \llangle \phi \vert \overline{W}_s \frac{1}{\overline{E}-\overline{H}_C+ i 0^{+}}  |\cw{\vec{p}}{+} \rrangle   \llangle \cw{\vec{p}}{+} | \overline{W}_s   \vert \phi \rrangle  \label{eqn:phiselfEcal}
\end{align}
It follows that the self energy is only a function of the Gallilean invariant combination $E - \frac{\vec{P}_\phi^2}{2 \mnc } \equiv \overline{E}$, i.e.
\begin{eqnarray}
\Sigma_\phi(\vec{P}_\phi;\, E)&=&h_s^2 J_0(\overline{E});\\
J_{0}(\overline{E})&\equiv& \int \frac{d^{3}\vec{q}}{(2\pi)^{3}} \frac{\cwc{\vec{q}}{+}(0)\cw{\vec{q}}{+}(0)}{\overline{E}-\frac{q^{2}}{2\mr}} =2 M_R \int \frac{d^{3}\vec{q}}{(2\pi)^{3}} C_{\eta_{q},0}^{2} \left[\frac{\ke^{2}}{q^{2}(\ke^{2}-q^{2}+i\epsilon)}-\frac{1}{q^{2}}\right]  \ , \label{eq:J0expression}\nonumber\\
\end{eqnarray}
with $k\equiv \sqrt{2 \mr (\overline{E}+ i 0^{+})}$. Similar results have been derived in, e.g., Refs.~\cite{Higa:2008dn,Ryberg:2013iga}. Details about $\cw{\vec{q}}{\pm}(\vec{r})$ and the definition of $C_{\eta_q,0}$ ($\eta_q\equiv \kc/\vert\vec{q}\vert$ is the Sommerfeld parameter) can be found in Appendix~\ref{app:coulombwf}. The integration diverges, however the integrand has been split into a finite and a divergent piece, as in the second step of Eq.~(\ref{eq:J0expression}). The first part yields $\kc H(\eta)/\pi$ with  
\begin{eqnarray}
H(\eta)&\equiv& \frac{(-)}{\pi}\int_{0}^{+\infty} \frac{ C_{\eta_q,0}^{2} d\eta_q}{\left(\eta_q^{2}-\eta^{2}+i\epsilon\right)} =\psi(i\eta)+\frac{1}{2i\eta}-\ln(i\eta)  \ ,
\end{eqnarray}
$\eta\equiv \kc/k$, and $\psi(z)$ the digamma function  \cite{MathHandBook1}.  The divergent term can be analytically continued in terms of the space-dimension variable  $d$. The following integration is involved \cite{Kong:1999sf}, 
\begin{eqnarray}
\int_{0}^{+\infty} \frac{\eta_q^{2-d} d\eta_q}{e^{2\pi\eta_q}-1} &=& \frac{\zeta(3-d)\Gamma(3-d)}{(2\pi)^{3-d}} \ , 
\end{eqnarray}
with $\zeta(x)$ the Riemann zeta function. When $\epsilon\rightarrow 0$,  $\zeta(\epsilon)=-\frac{1}{2}-\frac{\epsilon}{2} \ln(2\pi) $, $ \zeta(1+\epsilon)=\frac{1}{\epsilon} \left(1+C_{E} \epsilon + \cdots \right) $, and $\Gamma(\epsilon)=\frac{1}{\epsilon} \left(1-C_{E} \epsilon + \cdots \right)$ with $C_E$ the Euler constant; at other integers $n$, $\zeta(n)$ is finite.
We use the power-divergence subtraction (PDS) scheme \cite{Kaplan:1998we, Higa:2008dn} and subtract the pole at $d=2$. We then use the MS scheme to remove the pole in $d=3$ that is associated with the Coulomb interaction and obtain 
\begin{eqnarray}
J_{0}(\overline{E})&=& -\frac{M_R}{\pi}\left[\kc \left(H(\eta)-\ln\left(\frac{\mu \sqrt{\pi}}{\kc}\right) +\frac{3}{2} C_E -1 \right) +\mu\right] \ . 
\end{eqnarray}%
where $\mu$ is the dimensionful scale introduced to ensure the correct overall dimensions of $J_0$. 

As mentioned in the power-counting discussion of the previous section, summing up all the bubble insertions shown in Fig.~\ref{fig:Dphi_LO} gives $D_{\phi}^\mathrm{LO} = -\left(\Delta_{\phi} + \Sigma_{\phi}\right)^{-1}$. The NLO contribution to $D_\phi$, as shown in Fig.~\ref{fig:Dphi_NLO}, comes from the insertion of the operator $-\phi^{\dagger} \left(i \partial_0 -\frac{\nabla^2}{2 \mnc}\right) \phi$, i.e. the second diagram on Fig.~\ref{fig:Dphi_NLO}'s RHS. However in order to match the conventional ERE, we can sum all of the bracketed diagrams in Fig.~\ref{fig:Dphi_NLO}. This leads to $D_{\phi}^\mathrm{NLO}=-\left(\overline{E}+\Delta_\phi + \Sigma_\phi\right)^{-1}$. If we then impose the renormalization conditions: 
\begin{subequations}
\begin{alignat}{2}
\frac{(-)}{a_{0}} &\equiv \frac{ 2\pi \Delta_\phi}{h_s^{2} \mr} - 2 \mu + 2 \kc \left[\ln\left(\frac{\mu \sqrt{\pi}}{\kc} \right)- \frac{3}{2} C_E + 1\right]  \ , \label{eq:DphiandERE1} \\
\frac{r_{0}}{2} &\equiv \frac{\pi}{h_s^{2} \mr^2} \ , \label{eq:DphiandERE2}
\end{alignat}
\end{subequations}
we get 
\begin{eqnarray}
- \frac{2 \pi}{h_s^2 \mr} D_{\phi}^{-1}= -\frac{1}{a_{0}}+\frac{r_{0}}{2} \ke^{2}- 2\kc H(\etae)  \ . \label{eqn:swavepropaandr}
\end{eqnarray} 
Even though here we work only at the level of the dimer propagator we have chosen to already write things in terms of $a_0$ and $r_0$, which are the scattering length and effective range in the $s$-wave ERE for scattering. We make three points before moving on to discuss the
scattering $T$-matrix in the next paragraph. First, setting $\kc=0$, Eqs.~(\ref{eq:DphiandERE1}) and~(\ref{eq:DphiandERE2}) recovers the corresponding relationships for a system without Coulomb effects \cite{Hammer:2011ye,Zhang:2013kja}. Second, the overall ``$-$'' sign in the $\phi$'s free lagrangian in expression~(\ref{eqn:toylag}) is responsible for generating a positive $r_0$ in Eq.~(\ref{eq:DphiandERE2}). Third, when PDS and MS are employed, $\Delta_\phi$ is renormalized by the self-energy loop diagram, but $h_s$ is not. 

\begin{figure}
\centering
\includegraphics[width=15cm, angle=0]{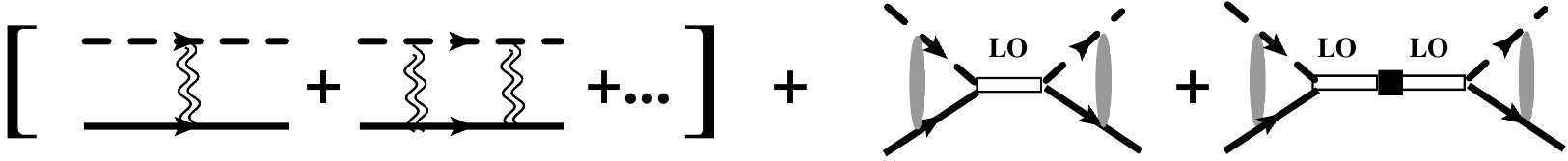}
\caption{The general scattering $T$-matrix caused by the Coulomb potential and strong interaction in $s$-wave channel. The bracketed diagrams are the pure Coulomb  scattering. The last two diagrams are the strong interaction LO and NLO $T$-matrix. } \label{fig:Tmatrix}
\end{figure}

The scattering $T$-matrix diagrams are shown in Fig.~\ref{fig:Tmatrix}. The bracketed diagrams are due to pure Coulomb scattering. 
They can be analytically computed (see, e.g.,~\cite{Goldberger1964}), and won't be dealt with here. The so-called strong-interaction $T$-matrix, i.e., the total $T$-matrix with pure Coulomb scattering subtracted is
\begin{equation}
\langle\vec{P}', \vec{p}'| T - W_C - \left(W_{C} \frac{1}{E-H_C+i0^{+}}  W_C\right)  |\vec{P},\vec{p}\rangle.
\end{equation} 
Up to NLO it is due to the last two diagrams in Fig.~\ref{fig:Tmatrix}. In terms of matrix elements, it is
\begin{eqnarray}
\langle\vec{P}', \vec{p}'|\left( 1+ W_C \frac{1}{E-H_C+i 0^{+}} \right)   W_s   \left(\frac{1}{E-H_C+i 0^{+}}\right)_{\mathrm{NLO}}  W_s  \left( 1+ \frac{1}{E-H_C+i 0^{+}}  W_C\right) |\vec{P},  \vec{p}\rangle  \label{eqn:strongTs}\nonumber\\
\end{eqnarray}
If the Green's function $\left(\frac{1}{E-H_C+i 0^{+}}\right)_{\mathrm{NLO}}$ were sandwiched between $|\phi\rangle$ states it would be $D_{\phi}^\mathrm{NLO}$. Thus, in the following we define 
\begin{eqnarray}
T_s \equiv W_s   \left(E-H_C+i 0^{+}\right)_{\mathrm{NLO}}^{-1}  W_s \ , \label{eqn:Tsdef}
\end{eqnarray} 
 with the first (second) $W_s$---reading from left to right---annihilates (creates) the $\phi$ particle in the intermediate state and creates (annihilates) an $nc$ pair in the final (initial) state.  We then use the dimer completeness relation:
 \begin{equation}
 (-) \int \frac{d \vec{P}}{(2 \pi)^3} |\vec{P}_\phi \rangle   \langle\vec{P}_\phi|
 \end{equation}
 (note the minus sign) and the previously computed dimer field propagator in Eq.~(\ref{eqn:swavepropaandr}), as well as Eq.~(\ref{eqn:CoulombWFdef}), to express the matrix element in Eq.~(\ref{eqn:strongTs}) as 
\begin{eqnarray}
\langle\vec{P}', \cw{\vec{p}'}{-}|T_s|\vec{P}, \cw{\vec{p}}{+}\rangle &=&-(2\pi)^3\delta(\vec{P}-\vec{P}') \frac{2\pi}{\mr} \frac{\cwc{\vec{p}'}{-}(0) \cw{\vec{p}}{+}(0) }{-\frac{1}{a_0}+\frac{1}{2} r_0 \ke^2 - 2 k_C H(\etae) } \ .  \label{eqn:TsinFockspace}
\end{eqnarray}
This matrix element can again be simplified to a delta function times the matrix element between the intrinsic states:
\begin{equation}
\langle \vec{P}', \cw{\vec{p}'}{-}|T_s|\vec{P}, \cw{\vec{p}}{+} \rangle=(2\pi)^3\delta(\vec{P}-\vec{P}') \llangle \cw{\vec{p}'}{-}|\overline{T}_s|\cw{\vec{p}}{+}\rrangle \ .
\end{equation}
Since the states $|\cw{\vec{p}}{+}\rrangle$ are already defined, the intrinsic operator $\overline{T}_s$ is then defined by its matrix elements on this basis. 
On shell the strong interaction $\overline{T}_s$ operator, evaluated on this, the ``intrinsic Coulomb basis", can be expressed in terms of a phase shift $\delta_0(\overline{E})$~\cite{Higa:2008dn,Ryberg:2015lea}
\begin{eqnarray}
\llangle \cw{\vec{p}'}{-}|\overline{T}_s(\overline{E})|\cw{\vec{p}}{+} \rrangle &\equiv&(-)\frac{2\pi}{\mr} \frac{e^{2i\sigma_{0}}}{\ke\left(\cot\delta_{0}-i\right)}  \ . 
\label{eq:intrinsicphaseshift}
\end{eqnarray}
Comparing Eqs.~(\ref{eqn:TsinFockspace}) and (\ref{eq:intrinsicphaseshift}), and using the relation $\mathrm{Im} [2\kc H(\eta)]=C_{\etae,0}^{2} \ke$, gives the Coulomb-modified ERE up to $O(k^2)$: 
\begin{eqnarray}
C_{\etae,0}^{2} \ke (\cot\delta_{0}-i)=-\frac{1}{a_{0}}+\frac{r_{0}}{2} \ke^{2}- 2\kc H(\eta)  \ . \label{eqn:EREswave}
\end{eqnarray}

This derivation clarifies how the Coulomb modified wave function $\cw{\vec{p}}{\pm} (\vec{r})$ appears in halo EFT: it is the co-ordinate-space representation of the intrinsic Coulomb basis $|\cw{\vec{p}}{\pm} \rrangle$. The separation of transverse photons from Coulomb photons through velocity scaling also delineates the order at which corrections due to those photons must be considered. (See Sec.~\ref{sec:higherorder}.)
As far as strong interactions are concerned, the next correction to the ERE, which must be $ \propto k^4$, occurs in the EFT via an operator that appears in the Lagrangian only at $\mathrm{N}^3$LO in the $\frac{V}{V_\Lambda}=\frac{k_\mathrm{low}}{\Lambda}$ EFT expansion~\cite{Mehen:1998tp}.

Eq.~(\ref{eqn:EREswave}) justifies the use of the notation $a_0$ and $r_0$ in the renormalization conditions in Eqs.~(\ref{eq:DphiandERE1}) and~(\ref{eq:DphiandERE2}).  
Since $a_0$ and $r_0$ are observables they are $\mu$-independent and this, in turn, determines the $\mu$-dependence of $\Delta_\phi$. 
$\Delta_\phi$ must absorb both the $d=2$ divergence that gives the $\mu$ in Eq.~(\ref{eq:DphiandERE1}) and the $\ln \mu$ Coulomb ($d=3$) divergence proportional to $\kc$. Thus the short-distance physics is affected by Coulomb photons, and so the separation of physics between dimer and $cn$ parts of the Fock space becomes dependent on the treatment of that short-distance physics, i.e. only model-dependent statements can be made about it. This, in turn, means that one cannot define a scheme- and scale-independent strong proton-core scattering length~\cite{Kong:1999sf,Gegelia:2003ta}.

\subsection{Toy amplitude for $p$-wave scattering up to NLO, and computation of shallow bound-state properties}

Fig.~\ref{fig:Dpi_LONLO} shows the diagrams for calculating the $\pi$ propagator $D_{\pi,j}^{i} $. As discussed in Sec.~\ref{sec:toypowercounting}, if we are in the kinematic regime where
the free dimer propagator has a singularity then we must resum the LO self energy to all orders, i.e.~compute the entire series shown in Fig.~\ref{fig:Dpi_LONLO}~\cite{Pascalutsa:2002pi,Bedaque:2003wa}. As will become clear by the end of this section, this is equivalent to requiring resummation in the vicinity of a value of momentum $k$ that is both within the domain of the EFT (i.e. $k < \Lambda$) and satisfies $\frac{1}{a_1}=\frac{1}{2} r_1 k^2$, where $r_1$ is the $p$-wave effective range and $a_1$ the $p$-wave scattering volume.
 In our case this condition is satisfied at the $\be$ bound-state pole. It is not satisfied in $\bes$-proton $p$-wave scattering, and so if our calculation were concerned with that process we could terminate the series at NLO, i.e. include the self-energy only perturbatively (cf. Ref.~\cite{Hammer:2011ye}) and still have NLO accuracy. However, here the properties of the $\be$ bound state are crucial to the calculation of $\bes(p,\gamma)\be$, so we resum the $p$-wave Dyson series to all orders to obtain the $\pi$ propagator.

The self-energy bubble in Fig.~\ref{fig:Dpi_LONLO} corresponds to $\langle \vec{P}_\pi',\pi^j \vert W_p \left(E-H_C+ i 0^{+}\right)^{-1} W_p   \vert \vec{P}_\pi, \pi^i  \rangle$. Due to rotational symmetry, we can always define $D_{\pi,j}^{i} \equiv D_{\pi} \delta_{j}^{i}$  and $\textstyle \Sigma_{\pi,j}^{i} = \delta_{j}^{i} \Sigma_\pi$. The latter can be evaluated through 
\begin{align}
\Sigma_{\pi,j}^i=&\int \frac{d \vec{P}}{(2 \pi)^3}\frac{d\vec{p}}{(2 \pi)^3} \langle \vec{P}_\pi',\pi^j \vert W_p \frac{1}{E-H_C+ i 0^{+}}  |\vec{P},  \cw{\vec{p}}{+} \rangle   \langle\vec{P},  \cw{\vec{p}}{+} | W_p   \vert \vec{P}_\pi,\pi^i \rangle  \notag \\
=& (2\pi)^3 \delta(\vec{P}_\pi-\vec{P}_\pi') \int \frac{d\vec{p}}{(2 \pi)^3} \llangle \pi^j \vert \overline{W}_p \frac{1}{\overline{E}-\overline{H}_C+ i 0^{+}}  |\cw{\vec{p}}{+} \rrangle   \llangle \cw{\vec{p}}{+} | \overline{W}_p   \vert \pi^i \rrangle.  \label{eqn:piselfEcal}
\end{align}
Here we again introduced the solutions of a one-body quantum mechanics problem, as we did in the previous section, i.e.~we write $\langle\vec{P}_\pi, \pi^j | W_p |\vec{P}_{nc}, \cw{\vec{p}_{nc}}{\pm}\rangle = (2 \pi)^3 \delta(\vec{P}_{nc}-\vec{P}_\pi) \llangle \pi^j |\overline{W}_p| \cw{\vec{p}}{\pm}  \rrangle$. This time though, the overlap of the $p$-wave dimer and the $nc$ state is given by:
\begin{equation}
\llangle \pi^{j}|\overline{W}_{p}| \cw{\vec{p}}{\pm}  \rrangle = i  \frac{h_p}{\mr }  \partial_j  \cw{\vec{p}}{\pm}(0). 
\end{equation}
The cm-frame energy is, once again, $\overline{E}\equiv E-\vec{P}_\pi^2/\left(2\mr\right)$, so we write
$\textstyle \Sigma_\pi (\overline{E}) \equiv \frac{h_p^2}{\mr^{2}} J_{1}(\overline{E})$,  with
\begin{eqnarray}
J_{1}(\overline{E})&\equiv& \int \frac{d^{d} \vec{q}}{(2\pi)^{d}} \frac{1}{d} \frac{\left(\upartial{j}\cwc{\vec{q}}{+}(0)\right)\left(\dpartial{j}\cw{\vec{q}}{+}(0)\right)}{\overline{E}-\frac{\vec{q}^{2}}{2\mr}+i\epsilon} \ , \notag \\
&=& (-)\frac{\mr}{6\pi}\ke^{2}(1+\etae^{2}) 2\kc H(\etae) +\frac{\mr \kc^{2}}{6\pi} \left[2\kc \left( \ln\left(\frac{\mu \sqrt{\pi}}{\kc}\right) - \frac{3}{2} C_E + \frac{4}{3}\right)-3\mu (1+\frac{\pi^{2}}{3})\right] \notag \\
&+& \frac{\mr \ke^{2}}{6\pi} \left[2\kc \left( \ln\left(\frac{\mu \sqrt{\pi}}{\kc}\right) - \frac{3}{2} C_E + \frac{4}{3}\right)-3\mu\right]  \ .
\end{eqnarray}
Eq.~(\ref{eqn:dcw0}) has been used in this derivation, and the divergent parts are again removed using PDS and MS. Summing up the bubble diagrams gives  $D_{\pi}=\left(\overline{E}+\Delta_\pi-\Sigma_{\pi} \right)^{-1}$. The parameters are renormalized to give scattering volume $a_1$ and effective range $r_1$ (i.e. the first two parameters in the ERE of the $p$-wave scattering phase shift):   
\begin{subequations}
\begin{alignat}{2}
\frac{(-)}{a_{1}} &\equiv (-)6\pi \frac{\mr \Delta_\pi}{h_p^{2}} +2\kc^{3} \left( \ln\left(\frac{\mu \sqrt{\pi}}{\kc}\right) - \frac{3}{2} C_E + \frac{4}{3}\right) -3\mu \kc^{2}\left(1+\frac{\pi^{2}}{3}\right) \ , \label{eq:DandERE1} \\
\frac{r_{1}}{2} &\equiv (-)\frac{3\pi}{h_p^{2}} +2\kc \left( \ln\left(\frac{\mu \sqrt{\pi}}{\kc}\right) - \frac{3}{2} C_E + \frac{4}{3}\right) -3 \mu \  \ . \label{eq:DandERE2}
\end{alignat}
\end{subequations}
We notice in the above expressions that---in contrast to the $s$-wave case---both $\Delta_\pi$ and $h_p$ are renormalized by the self-energy loop diagram in  PDS; both also absorb $\kc$-dependent logarithmic divergences for which we have used MS. 
As for $\Delta_\phi$ in the $s$-wave case, the presence of $\kc$-dependent pieces in the LECs of the ``strong" Lagrangian means that isospin symmetry is broken---and here the breaking has more physical consequences than in the $s$-wave case, since it affects both $r_1$ and $a_1$. Notice again that setting $\kc=0$, the two equations reproduce the corresponding relationships for the system without Coulomb effects~\cite{Hammer:2011ye,Zhang:2013kja}.

The propagator can then be expressed as
\begin{eqnarray}
(-) \frac{6\pi \mr}{h_p^2} \left(D_\pi^\mathrm{NLO}\right)^{-1} =-\frac{1}{a_{1}}+\frac{r_{1}}{2} \ke^{2}-\ke^{2}(1+\etae^{2}) 2\kc H(\etae) \ .
\end{eqnarray}
The corresponding strong-interaction $T$ matrix, evaluated on the Coulomb basis, is:
\begin{eqnarray}
\langle\vec{P}',\cw{\vec{p}'}{-}|T_p(E)|\vec{P},\cw{\vec{p}}{+}\rangle &=&- (2\pi)^3 \delta \left(\vec{P}-\vec{P}'\right) \frac{6\pi}{\mr} \frac{\vec{\partial}\cwc{\vec{p}'}{-}(0) \cdot \vec{\partial} \cw{\vec{p}}{+}(0) }{-\frac{1}{a_1}+\frac{1}{2} r_1 \ke^2 - k^2 (1+\eta^2) 2  k_C H(\etae) }  \notag \\ 
&\equiv & (2\pi)^3 \delta \left(\vec{P}-\vec{P}'\right) \llangle \cw{\vec{p}'}{-}|\overline{T}_p(\overline{E})|\cw{\vec{p}}{+}\rrangle . \label{eqn:Tpmatrix}
\end{eqnarray}
Here $T_p \equiv W_p   \left(E-H_C+i 0^{+}\right)^{-1}  W_p$, with, as before, the $W_p$ operators producing the transition from the dimer state to the $nc$ state.  The on-shell strong interaction $T$-matrix is related to the strong phase shift $\delta_1(\overline{E})$, 
\begin{eqnarray}
\llangle \cw{\vec{p}'}{-}|\overline{T}_p(\overline{E})|\cw{\vec{p}}{+}\rrangle= - \frac{6\pi}{\mr}  \frac{e^{2i\sigma_{1}}}{\ke\left(\cot\delta_{1}-i\right)} \ . 
\end{eqnarray}
As a result, the phase-shift ERE can be expressed in the convention used in Ref.~\cite{Hamilton:1973xd}:  
\begin{eqnarray}
\left( 3 C_{\etae,1} \right)^{2} \ke^{3}(\cot\delta_{1}-i)=-\frac{1}{a_{1}}+\frac{r_{1}}{2} \ke^{2}-\ke^{2}(1+\etae^{2}) 2\kc H(\etae)  \ . \label{eqn:pEREconv}
\end{eqnarray}
This convention is different from the one used in Ref.~\cite{Koenig:2012bv, Ryberg:2014exa}, and has the advantage of approaching the non-Coulomb ERE in the $\kc \rightarrow 0$ limit.  In our previous work \cite{Zhang:2014zsa}, the factor $3$ in $\left( 3 C_{\etae,1} \right)^{2}$ was missing. 

The EFT can generate a bound state in the $p$-wave channel, corresponding to a $T$-matrix pole at $\overline{E}=-B < 0$ with $B \sim \mr V^2$. This requires $D_\pi^{-1}(\ke=i\gma)=0$ ($\gma \equiv \sqrt{2 \mr B}$), which, after we resum the Coulomb bubbles to all orders, as we must near the pole, means,  
\begin{align}
-\frac{1}{a_{1}}-\frac{r_{1}}{2} \gma^{2}+\gma^{2}(1-\eta_{B}^{2}) 2\kc \tilde{H}(\eta_B)=0 \ . 
\end{align}
Here $\eta_B\equiv \kc/\gma$ and $\tilde{H}(z)\equiv H(-iz)$. $D_\pi$ has other poles too, but in this, two-body, calculation we are only concerned with the shallow bound-state pole. Note that, as promised, the pole involves a cancellation between the $1/a_1$ and $r_1$ terms in the ERE, and so the nominally higher-order terms $\sim \kc \gamma^2$  must be included: their effect on the pole position is not higher-order.

Having located this pole, the remainder of this section aims to extract the so-called asymptotic normalization coefficient (ANC) of the bound-state wave function \cite{Hammer:2011ye,Nollett:2011qf,Koenig:2012bv}.  Here we will derive the relationship between this quantity and the wave-function renormalization. The latter quantity is the residue of this pole for $D_{\pi}(\overline{E})$:
\begin{align}
Z=-\frac{6\pi}{h_p^2} \bigg\{r_{1} -2\kc  \bigg[2\tilde{H}(\etab)+\etab\left(\etab^{2}-1\right)\tilde{H}^{\,'}(\etab)\bigg] \bigg\}^{-1} \ ,  \label{eqn:Zv1}
\end{align}
with $\tilde{H}^{\,'}(\etab)= \left. \frac{d \tilde{H}(\eta)}{d \eta} \right|_{\eta=\etab}$. Second, the residue of the $nc$  Green's function between states having plane-wave CM motion and a relative coordinate $\vec{r}$ can be related to $Z$ in the following manner      
\begin{align}
\mathrm{Res}&\left[\langle\vec{P}', \vec{r}' |\frac{1}{E-H}|\vec{P}, \vec{r} \rangle\right] = \mathrm{Res}\left[\langle\vec{P}',\vec{r}' |\frac{1}{E-H_c} T_p(E) \frac{1}{E-H_c}  |\vec{P},\vec{r}\rangle\right]
\notag \\ 
& = (2 \pi)^3 \delta(\vec{P}-\vec{P}') Z \frac{h_p^2}{\mr^{2}}  \delta_{j'}^{j} \, \int \frac{d^{3}\vec{q}'}{(2\pi)^{3}}    \frac{\left(\upartial{j'}\cwc{\vec{q}'}{-}(0)\right) \cw{\vec{q}'}{-}(\vec{r}')}{B+\frac{q^{'2}}{2\mr}}
\int \frac{d^{3}\vec{q}}{(2\pi)^{3}} \frac{\left(\dpartial{j}\cw{\vec{q}}{+}(0)\right) \cwc{\vec{q}}{+}(\vec{r})}{B+\frac{q^{2}}{2\mr}} \ . \label{eqn:anccalculationmethod}
\end{align}
The first step relies on the fact that the pure repulsive Coulomb potential does not produce a bound state, i.e., $\langle\vec{P}',\vec{r}' |\left(E-H_C\right)^{-1}|\vec{P},\vec{r} \rangle$ does not have a bound-state pole. To proceed further we use the partial-wave decomposition of  $\cw{\vec{q}}{\pm}(\vec{r})$ in terms of $\cwf{l}(q,r)$ [see Eq.~(\ref{eqn:cwldecomp})] in the integrations. According to Eq.~(\ref{eqn:dcw0}), $\dpartial{j}\cw{\vec{q}}{+}(0)$ aligns with $\vec{q}$, so integrating over $\vec{q}$ picks out the $\cwf{1}(q,r)$ term, which then can be related to the Whittaker functions through Eqs.~(\ref{eqn:FGvsW1}) and~(\ref{eqn:FGvsW2}). Thus we get
\begin{eqnarray}
&& \int \frac{d^{3}\vec{q}}{(2\pi)^{3}} \frac{\left(\dpartial{j}\cw{\vec{q}}{+}(0)\right) \cwc{\vec{q}}{+}(\vec{r})}{B+\frac{q^{2}}{2\mr}} \notag \\
&=& - \frac{\mr}{ 2 \pi r } \frac{\dY{1}{j}(\vecpt{r})}{\sqrt{3\pi}} \int_{-\infty}^{+\infty} \frac{q^{2}dq}{q^{2}+\gma^{2}}\left[\Gamma(2-i\eta) W_{i\eta,\frac{3}{2}}(2iqr) + \Gamma(2+i\eta) W_{-i\eta,\frac{3}{2}}(-2iqr) \right]  \ , \notag \\ 
&=&  \frac{\mr}{\sqrt{3\pi}} \gma \Gamma(2+\eta_{B})  \frac{W_{-\eta_{B},
\frac{3}{2}}(2\gma r)}{r} \dY{1}{j}(\vecpt{r}) \ . \label{eqn:master1}
\end{eqnarray}
The last step uses the fact that $q \left[\Gamma(2 \pm i\eta) W_{\mp i\eta,\frac{3}{2}}(2iqr) \right]$ is analytic in the upper (lower) half of the complex-$q$ plane including the real axis (see Appendix~\ref{app:coulombwf} for proof). Note that for any real $r$ the oscillatory behavior of $\cwc{\vec{q}}{\pm}(\vec{r})$
 guarantees convergence of the integral as long as $r \neq 0$. Equivalently, it is the Whittaker functions that ensure that when Cauchy's theorem is used to evaluate the integral over $q$ the piece on the contour at $q=\infty$ goes to zero. However, if $r=0$ this oscillatory/damping factor is absent and the integral diverges. 
As long as $r,r' \neq 0$ though, Eqs.~(\ref{eqn:master1}) and (\ref{eqn:anccalculationmethod}) show that the $nc$ Green's function has the form:
\begin{equation}
\mathrm{Res}\left[\langle\vec{P}', \vec{r}' |\left(E-H\right)^{-1}|\vec{r}, \vec{P}\rangle\right] \equiv (2 \pi)^3 \delta(\vec{P}-\vec{P}') C^2\,  \uY{1}{j}(\vecpt{r'}) \dY{1}{j}(\vecpt{r}) \frac{W_{-\eta_{B},\frac{3}{2}}(2\gma r')  W_{-\eta_{B}, \frac{3}{2}}(2\gma r)}{r' r},
\end{equation}
where we have identified $C^2$ as the squared ANC. Explicitly inserting Eq.~(\ref{eqn:master1}) into Eq.~(\ref{eqn:anccalculationmethod}) $C^2$ is then found in terms of the wave-function renormalization $Z$ as
\begin{eqnarray}
\label{eqn:anc-scaling}
C^{2}&=& \frac{h_p^{2}}{3\pi} \gma^{2} \Gamma^{2}(2+\eta_{B}) Z \ . 
\end{eqnarray}
To check our calculation against the results in Ref.~\cite{Koenig:2012bv}, we take the limit $\eta_B \gg 1$, where $H(\etae=-i\eta_{B})\approx -1/\left(12\eta_{B}^2\right)$ applies, and find $C^{2} \approx - 2\gma^{2} \Gamma^{2}\left(2+\eta_{B}\right)/\left(r_1-\frac{\kc}{3}\right)$. This is consistent with the ANC computed in Ref.~\cite{Koenig:2012bv}. However in our study the  $\eta_B \gg 1$ condition does not apply, instead we have $\eta_B \sim 1$.

\section{$\bes$--proton case: lagrangian, scatterings, and bound state} \label{sec:realsticsystem}
In this section, we apply the methods developed in Section~\ref{sec:toymodel} to study the $\bes$-proton system. The additional complexities reside in the particle spins and the presence of a low-energy excited state of the core, $\besst$. The pertinent quantum numbers $J^{P}$ are $\frac{1}{2}^{+}$ for the proton, $\frac{3}{2}^{-}$ for the $\bes$ core, $\frac{1}{2}^{-}$ for the core excitation $\besst$, and $2^{+}$ for the $\be$ ground state. We can write the gauged (but not rescaled) free Lagrangian for all the degrees of freedom in a very compact notation as:
\begin{eqnarray}
\mathcal{L}_0&=& \psi^{\dagger} \left[i \partial_0 - e\, \hat{Q}\, A_0  +\frac{\left(\overset{\rightarrow}{\nabla}   - i e \vec{A}\, \hat{Q} \right)^2}{2 \hat{M}} + \hat{\Delta} \right] \psi \ .
\end{eqnarray}
Here $\fd{\psi}{s} \equiv \left(\fd{n}{\sigma}, \fd{c}{a}, \fd{d}{\delta}, \fd{\phi}{(1)i}, \fd{\phi}{(2)\alpha}, \fd{\pi}{\alpha} \right)^{T}$, and $\fdu{\psi}{s} \equiv \left(\fdu{n}{\sigma}, \fdu{c}{a}, \fdu{d}{\delta}, -\fdu{\phi}{i}_{(1)}, -\fdu{\phi}{\alpha}_{(2)}, \fdu{\pi}{\alpha} \right)$ with  $\fd{n}{\sigma}$, $\fd{c}{a}$, $\fd{d}{\delta}$, $\fd{\pi}{\alpha}$ the fields of the proton, $\bes$, $\besst$, and $\be$; $\fu{\phi}{i}_{(1)}$ and $\fu{\phi}{\alpha}_{(2)}$ are the two $s$-wave $1^{-}$ and $2^{-}$ dimer fields respectively. All the fields' indices correspond to the spin projections with a specific convention:  $\sigma, \delta, \sigma', \delta'=\pm 1/2$, $a,a'=\pm 3/2, \, \pm 1/2$, $\alpha,\beta=\pm 2, \, \pm 1, \, 0$, and $i,j,k=\pm 1, \, 0$. The daggered fields are, e.g., $\fdu{n}{\sigma} \equiv \left(\fd{n}{\sigma}\right)^\ast$.
The mass matrix $\hat{M}=\mathrm{Diagonal}\left(M_n,M_c,M_c,M_{nc},M_{nc},M_{nc}\right)$ and the bare binding-energy matrix $\hat{\Delta}=\mathrm{Diagonal}\left(0,0,-E_{\ast},\Delta_{\phi 1},\Delta_{\phi 2},\Delta_{\pi}\right)$. ($E_{\ast}$ is the excitation energy of $\bes^{\ast}$.)  The corresponding charge matrix $\hat{Q}=\mathrm{Diagonal}\left(Q_n,Q_c,Q_c,Q_{nc},Q_{nc},Q_{nc}\right)$ with $Q_n=1$, $Q_c=4$, $Q_{nc}=Q_n+Q_c=5$.

There are three different interaction terms between the $s$-wave $\phi_{(1)}$ and $\phi_{(2)}$ dimers and the $nc$ state; these generate the strong $s$-wave $\bes$-proton and $\besst$-proton interactions,  
\begin{eqnarray}
\mathcal{L}_{S}= \gone  \fdu{\phi}{i}_{(1)}  \T{i}{ \sigma a }\fd{n}{\sigma}  \fd{c}{a}  + \gthree \fdu{\phi}{i}_{(1)}  \T{i}{\sigma \delta }  \fd{n}{\sigma}  \fd{d}{\delta} +\gtwo \fdu{\phi}{\alpha}_{(2)} \T{\alpha}{\sigma a} \fd{n}{\sigma} \fd{c}{a} + \mathrm{c.c.} \ . \label{eqn:L1} 
\end{eqnarray}
The couplings' lower indices explain the interaction channel quantum numbers; the associated $\T{...}{...}$ are the corresponding Clebsch-Gordan coefficients \cite{Zhang:2013kja}, e.g. $\T{i}{\sigma a}\equiv \llangle i | \sigma a \rrangle=(\frac{1}{2} \sigma \frac{3}{2} a|1 i)$ and $\T{\alpha}{\sigma a}=(\frac{1}{2} \sigma \frac{3}{2} a|2 \alpha)$.  $\gone$ and $\gtwo$ are related to the ``unnaturally'' large $s$-wave scattering lengths~\cite{Higa:2008dn, Zhang:2014zsa}. However, based on the fact that the $\bes$-proton inelasticity is small~\cite{Knox:1987},we assign $\gthree$, which describes the only inelastic channel at these energies, $\bes + p \leftrightarrow \besst+p$, as NLO, i.e.,  $\gthree/\gone\sim \gthree/\gtwo\sim \frac{V}{V_\Lambda} = \frac{k_{\mathrm{low}}}{\Lambda}$ \cite{Zhang:2014zsa}.

There are also two terms that generate interactions between the $p$-wave dimer $\pi$ and the $nc$ pair, as well as one coupling for the $\besst$-p-$\pi$ interaction (the spins of $\besst$ and the proton only allow such an interaction in the $s=1$ $p$-wave state): 
\begin{equation}
\mathcal{L}_{P}=\fdu{\pi}{\alpha} \left[ \hone  \T{\alpha}{ij} \T{i}{\sigma a}  +\htwo \T{\alpha}{\beta j}\T{\beta}{\sigma a}   \right] \fd{n}{\sigma} \widetilde{\vec{V}}_{Rj} \fd{c}{a} +\hthree \fdu{\pi}{\alpha} \T{\alpha}{jk} \T{k}{\delta \sigma}\fd{n}{\sigma} \widetilde{\vec{V}}_{Rj} \fd{d}{\delta} 
+\mathrm{c.c.} \ . \label{eqn:L2}
\end{equation} 
Based on NDA, we expect the three couplings to have roughly the same size, as all are natural. 

Following the same procedure used in the toy-model calculation, we compute the $\phi_{(1)}$ and $\phi_{(2)}$ propagators, which we must now write as $D_{\phi 1} \delta_j^i$ and $D_{\phi 2} \delta_\beta^\alpha$, with the sub- and super-scripts providing spin indices. We then compute the $1^{-}$ and $2^{-}$ channel strong-interaction $T$ matrix, which is defined in Eq.~(\ref{eqn:Tsdef}). Note that in the remainder of the main text of this paper, whenever we refer to the equations in the simple model discussion,  the Hamiltonian in those equations is re-derived from our realistic Lagrangian, and the Fock states include the spin degrees of freedom properly. 

The LO self-energy bubble diagrams -- see Fig.~\ref{fig:Dphi_LO} -- for the two fields come from the $\bes$-p intermediate state. The $d+n \leftrightarrow c+n$ amplitude is suppressed by $\gthree/\gone \sim \frac{ k_\mathrm{low}}{\Lambda}$ compared to elastic scattering. The
 $\besst$-p $s$-wave contribution -- which because $\besst$ is a spin-one nucleus only exists for the $\phi_{(1)}$ dimer -- is therefore suppressed by  $\sim \gthree^2/\gone^2 \sim \gthree^2/\gtwo^2 \sim k_\mathrm{low}^2/\Lambda^2 $, and is thus an N$^2$LO effect in the self-energy calculation. (Note, however, that these scalings also imply that the $\besst$ channel contributes to the total $\bes$-proton amplitude already at NLO.)
 As a result, the two $s$-wave dimer propagators have exactly the same structure as in the toy model; the renormalization conditions in Eqs.~(\ref{eq:DphiandERE1}) and~(\ref{eq:DphiandERE2}) can be used directly to connect ($\Delta_{\phi 1}$, $\gone$) to ($\aone $, $\rone$) , and ($\Delta_{\phi 2}$, $\gtwo$) to ($\atwo$, $\rtwo$); the fully dressed propagators are
\begin{eqnarray}
(-) \frac{2 \pi}{\mr \gone^2} D_{\phi 1}^{-1} &=& -\frac{1}{\aone}+\frac{\rone}{2} \ke^{2}- 2\kc H(\eta)  \equiv \left[ \mathbb{N}_{\phi 1}\left(\overline{E}\right)\right]^{-1}  \ , \notag \\
(-) \frac{2 \pi}{\mr \gtwo^2} D_{\phi 2}^{-1} &=& -\frac{1}{\atwo}+\frac{\rtwo}{2} \ke^{2}- 2\kc H(\eta)  \equiv  \left[\mathbb{N}_{\phi 2}\left(\overline{E}\right) \right]^{-1} \ .  \notag 
\end{eqnarray}
Here $k$ is the $nc$ relative momentum, equal to $\sqrt{2\mr \overline{E}+ i 0^+}$ with $\overline{E}=E-\frac{\vec{P}^2}{2\mr}$. 
The strong interaction (in)elastic $T$-matrices can then be calculated straightforwardly as the product of a $cn$ to dimer vertex, the dimer propagator, and the hermitian conjugate of the incoming vertex, as was done in Eqs.~(\ref{eqn:strongTs}) and~(\ref{eqn:TsinFockspace}). Here we only show the relevant results:
\begin{subequations}
\begin{alignat}{2}
 \llangle  \cw{\vec{p}'}{-},\,  (nc)^{i'}  |\overline{T}_s (\overline{E}) | \cw{\vec{p}}{+},\,  (nc)^i \rrangle &=  (-) \delta_{i'}^i\, \frac{2\pi}{\mr}\,  \mathbb{N}_{\phi 1}\left(\overline{E}\right)\, \cwc{\vec{p}'}{-}(0) \cw{\vec{p}}{+}(0)  \, , \label{eqn:Ts1} \\ 
 \llangle  \cw{\vec{p}'}{-},\,  (nc)^{\alpha'}  |\overline{T}_s (\overline{E}) | \cw{\vec{p}}{+},\,  (nc)^\alpha \rrangle  & =  (-) \delta_{\alpha'}^{\alpha}\, \frac{2\pi}{\mr}    \, \mathbb{N}_{\phi 2}\left(\overline{E}\right)
\, \cwc{\vec{p}'}{-}(0) \cw{\vec{p}}{+}(0)\, , \label{eqn:Ts2} \\ 
  \llangle  \cw{\vec{p}'}{-},\,  (nd)^{i'}  |\overline{T}_s (\overline{E}) | \cw{\vec{p}}{+},\,  (nc)^i \rrangle 
 & = (-) \delta_{i'}^i\, \frac{2\pi}{\mr} \frac{\gthree}{\gone} \, \mathbb{N}_{\phi 1}\left(\overline{E}\right)
\, \cwc{\vec{p}'}{-}(0) \cw{\vec{p}}{+}(0) \ . \label{eqn:Ts3}
\end{alignat}
\end{subequations}
This notation indicates that we have coupled the proton and core spins to a particular total spin $s$: $s=1$, $| (nc)^i  \rrangle \equiv \Td{\sigma a}{i}  |  n^\sigma,  c^a \rrangle $; $s=2$, $| (nc)^\alpha  \rrangle \equiv \Td{\sigma a}{\alpha}  |  n^\sigma,  c^a \rrangle $, and, for the case of the excited core, $s=1$ again: $| (nd)^i  \rrangle \equiv \Td{\sigma \delta}{i}  |  n^\sigma,  d^\delta \rrangle $. To compute the matrix elements between direct product spin states, we simply invert these relations: $|   n^\sigma,  c^a \rrangle = \T{i}{\sigma a} |(nc)^i  \rrangle +\T{\alpha}{\sigma a} |(nc)^\alpha  \rrangle $, and  $| n^\sigma,  d^\delta \rrangle = \T{i}{\sigma \delta} |(nd)^i  \rrangle +\T{\alpha}{\delta} |(nd)\,\mathrm{singlet}  \rrangle $, where the last component, involving the $S=0$ proton-$\besst$ state, does not couple to any dimers. 
For the first two matrix elements we have elastic $\bes$-proton scattering, and the initial and final asymptotic $\bes$-proton relative momenta are $\vec{p}$ and $\vec{p}'$; with $\overline{E}=\frac{\vec{p}^2}{2 \mr }= \frac{\vec{p}^{'2}}{2 \mr }$. In Eq.~(\ref{eqn:Ts3}) we consider inelastic scattering, and in that case the on-shell condition is  $\overline{E}=\frac{\vec{p}^2}{2 \mr }= \frac{\vec{p}^{'2}}{2 \mr } + E_\ast$. To get the  inelastic $T$-matrix with $\besst$ in the initial state, we simply exchange the initial and final state quantum numbers in Eq.~(\ref{eqn:Ts3}). Elastic $\besst$-proton scattering is suppressed by two orders compared to $\bes$-proton elastic scattering and hence is not discussed here. 

Summarizing, because the inelasticity $\frac{\gthree}{\gone}$ is parameterically small ($\sim  k_\mathrm{low}/\Lambda$), the ERE in the ${}^3$S$_1$ channel looks the same (up to N$^2$LO) as that in the ${}^5$S$_2$ channel, where $\besst$ cannot play a role, therefore:
\begin{eqnarray}
C_{\etae,0}^{2} \ke (\cot\delta_{\left(X\right)}-i) & =&-\frac{1}{a_{\left(X\right)}}+\frac{r_{\left(X\right)}}{2} \ke^{2}- 2\kc H(\eta) \ , X=\S{3}{1},\, \S{5}{2} \ . \label{eqn:EREswave2}
\end{eqnarray}

Turning our attention now to the $p$-waves, in the $p$-wave self-energy bubble both $\bes$-proton and $\besst$-proton contributions are at the same order: the computed ANCs (see numbers at the end of this section) suggest the $nc\pi$ and $nd\pi$ couplings are of the same order. Therefore, both loop contributions are summed to obtain the dominant piece of the one-loop self-energy, and we get the (resummed) full propagator ($\equiv D_{\pi} \delta_{\beta}^{\alpha}$)
\begin{eqnarray}
(-) \frac{6\pi \mr}{\hpt^{2}} D_{\pi}^{-1}
&=&-\frac{1}{a_{1}}+\frac{r_{1}}{2} \ke^{2}-2\kc \ke^{2}(1+\etae^{2})  H(\etae) -2\kc \frac{\hthree^{2}}{\hpt^{2}} \kest^{2}(1+\etaest^{2})  H(\etaest)\notag \\
 & \equiv & \left[\mathbb{N}_\pi\left(\overline{E}\right) \right]^{-1} \ . \label{eqn:pEREcoreexcitation}
\end{eqnarray}
Here, $\hpt^{2}\equiv \hone^2+\htwo^2$, $\gamma_\Delta \equiv\sqrt{2 \mr E^{\ast}}$, $\kest\equiv \sqrt{2\mr (\overline{E}-E^{\ast}) + i 0^{+}}=\sqrt{k^2-\gamma_\Delta^2 + i 0^{+}}$, $\etaest\equiv \kc/\kest$, and (using PDS and MS)
\begin{eqnarray}
\frac{(-)}{a_{1}} &\equiv& (-)6\pi \frac{\mr \Delta_\pi}{\hpt^{2}} +\left(1+\frac{\hthree^{2}}{\hpt^{2}}\right)\bigg[2\kc^3 \left( \ln\left(\frac{\mu \sqrt{\pi}}{\kc}\right) - \frac{3}{2} C_E + \frac{4}{3}\right)  -3\mu \kc^{2}\left(1+\frac{\pi^{2}}{3}\right)\bigg] \notag \\
&&-\frac{\hthree^{2}}{\hpt^{2}} \gma_{\Delta}^{2} \bigg[2\kc \left( \ln\left(\frac{\mu \sqrt{\pi}}{\kc}\right) - \frac{3}{2} C_E + \frac{4}{3}\right)  -3\mu \bigg] \notag \\
\frac{r_{1}}{2} &\equiv& (-)\frac{3\pi}{\hpt^{2}} +\bigg[2\kc \left( \ln\left(\frac{\mu \sqrt{\pi}}{\kc}\right) - \frac{3}{2} C_E + \frac{4}{3}\right) -3 \mu\bigg] \left(1+\frac{\hthree^{2}}{\hpt^{2}}\right) \ .
\end{eqnarray}
The (in)elastic strong interaction $T$-matrix in  the $2^+$ channel can then be calculated in the same way as Eq.~(\ref{eqn:Tpmatrix}) was, i.e.~by multiplying by the vertex factors associated with the overlap of the $nc$ and dimer states. In order to get the strong interaction $T$-matrix elements, we therefore multiply the numerical coefficients associated with a particular channel, and given in table~\ref{tab:Tp}, by the channel-independent dimer and vertex factors:
\begin{eqnarray}
(-) \frac{6\pi}{\mr }\, \mathbb{N}_{\pi}\left(\overline{E}\right)\, \partial^{j'} \cwc{\vec{p}'}{-}(0) \partial_j \cw{\vec{p}}{+}(0) \label{eqn:Tpoverallfactor}
\end{eqnarray}
Here the initial-state and final-state relative momenta are $\vec{p}$ and $\vec{p}'$ (whether the relevant channel is elastic or inelastic), but we reiterate that for on-shell inelastic scattering the $n$-$d$ relative momentum $\vec{p}$, is given by $\overline{E}=\frac{\vec{p}^2}{2\mr} + E_\ast $.
Also note that in this inelastic channel, the Coulomb potential has a negligible influence compared with the strong interaction. 
\begin{table}
\begin{ruledtabular} 
   \begin{tabular}{cccc}
$\llangle \, |\overline{T}_p| \, \rrangle $	 &  $ | \left(nc\right)^i \rrangle $ & $ | \left(nc\right)^\alpha \rrangle $  &  $ | \left(dc\right)^i \rrangle $  \\ \hline
   $ \llangle \left(nc\right)^{i'} | $    &  $\Td{i' j'}{\beta} \T{\beta}{i j} \hone \hone $ &  $ \Td{i' j'}{\beta} \T{\beta}{\alpha j}\, \hone\htwo   $  &  $ \Td{i' j'}{\beta} \T{\beta}{i j}\, {\hone \hthree} $ \\ 
     $ \llangle \left(nc\right)^{\alpha'} | $ & $ \Td{\alpha' j'}{\beta} \T{\beta}{i j} \htwo \hone $ &  $\Td{\alpha' j'}{\beta} \T{\beta}{\alpha j} \htwo\htwo  $   &  $ \Td{\alpha' j'}{\beta} \T{\beta}{i j}\,\htwo \hthree $   \\ 
		 $ \llangle \left(nd\right)^{i'} | $ &  $ \Td{i' j'}{\beta} \T{\beta}{i j}\, \hthree\hthree $   &  $ \Td{i' j'}{\beta} \T{\beta}{\alpha j}\, \hthree\htwo $ &  $ \Td{i' j'}{\beta} \T{\beta}{i j}\, \hthree\hthree $   \\
\end{tabular} \caption{The channel-dependent factors needed to obtain the matrix elements of $\overline{T}_p$ in different spin-one channels $|(nc)^i \rrangle$. These are multiplied by the overall factor of Eq.~(\ref{eqn:Tpoverallfactor}). } \label{tab:Tp}
	\end{ruledtabular}
	\end{table}

Once inelasticity is involved $a_1$ and $r_1$ are not related to the scattering phase shift as in Eq.~(\ref{eqn:pEREconv}), but since they can, in principle, be measured through the $T$-matrix, $a_1$ and $r_1$ can be used as the renormalization conditions, i.e. they do not have renormalization scale dependence.  This connection is straightforward when $k\ll \gma_{\Delta}$, i.e., far below the proton-$\besst$ threshold. There $D^{-1}$ can be expanded in Taylor series. By keeping terms up to $k^2$, we see that redefining the scattering volumes and effective range, according to
\begin{eqnarray}
\frac{(-)}{\mathcal{A}_{1}} &\equiv& \frac{(-)}{a_{1}} -2 \frac{\hthree^{2}}{\hpt^{2}} \kc \gma_{\Delta}^{2} \left(\eta_{\Delta}^{2}-1\right) \tilde{H}\left(\eta_{\Delta}\right)  \ , \\
\frac{\mathcal{R}_{1}}{2}&\equiv& \frac{r_{1}}{2} -2 \frac{\hthree^{2}}{\hpt^{2}} \kc \bigg[\tilde{H}\left(\eta_{\Delta}\right)+\frac{1}{2} \eta_{\Delta}\left(\eta_{\Delta}^{2}-1\right) \tilde{H}^{\,\mathbf{'}}(\eta_{\Delta})  \bigg] \ , 
\end{eqnarray}
recovers the elastic scattering ERE with $\mathcal{A}_1$ and $\mathcal{R}_1$ as ERE parameters, i.e. 
\begin{equation}
\left(3 C_{\etae,1}\right)^{2} \ke^{3}(\cot\delta_{1}-i) =-\frac{1}{\mathcal{A}_{1}}+\frac{\mathcal{R}_{1}}{2} \ke^{2}-2\kc \ke^{2}(1+\etae^{2})  H(\etae).
\end{equation}
That is to say, far below the threshold the core-excitation contributions are subsumed into redefined ERE parameters.
For the $\bes$-p system, $\eta_{\Delta}\equiv \gamma_\Delta/k_C =0.9$, $\tilde{H}(\eta_\Delta)\sim 0.1$, $\tilde{H}^{'}(\eta_\Delta)\sim -0.1$, so $\mathcal{A}_1 \approx a_1$, $\mathcal{R}_1 \approx r_1$. However, we emphasize that Eq.~(\ref{eqn:pEREcoreexcitation}) represents different assumptions about the analytic structure of the $\bes$-proton amplitude than those that lead to the standard ERE. 

In this situation of a core excitation we can again compute $ d D_{\pi}^{-1} / d E$ at  the $E=-B$ pole to  get the residue of $D_{\pi}$,
\begin{align}
-\frac{6\pi}{Z}=\hpt^{2} \ r_{1} -2\kc &\bigg\{\hpt^{2} \bigg[2\tilde{H}(\etab)+\etab\left(\etab^{2}-1\right)\tilde{H}^{\,\mathbf{'}}(\etab)\bigg] \notag \\
&+\hthree^{2}  \bigg[2\tilde{H}(\etabst)+\etabst\left(\etabst^{2}-1\right)\tilde{H}^{\,\mathbf{'}}(\etabst)\bigg] \bigg\}  \ , \label{eqn:2plusdimerresidue}
\end{align}
with $\etabst\equiv \frac{\kc}{\gma^{\ast}}$. Then it is easy to compute the bound-state ANCs by plugging the elastic scattering matrix elements [see Eq.~(\ref{eqn:Tpoverallfactor}) and Table~\ref{tab:Tp}] into Eq.~(\ref{eqn:anccalculationmethod}). We find 
\begin{align}
\frac{C^{2}_{(\P{3}{2})}}{\hone^{2}\gma^{2} \Gamma^{\,2}(2+\etab)}=
\frac{C^{2}_{(\P{5}{2})}}{\htwo^{2}\gma^{2} \Gamma^{\,2}(2+\etab)}=
\frac{C^{2}_{(\P{3}{2}^{\ast})}}{\hthree^{2}\gma^{\ast 2} \Gamma^{\,2}(2+\etabst)} =\frac{Z}{3\pi} \ .
\end{align}
These ANCs have been calculated using {\it ab initio} methods \cite{Nollett:2011qf}, yielding $ C_{(\P{3}{2})}^2=0.0990 (57)\, \mathrm{fm}^{-1}$, $ C_{(\P{5}{2})}^2= 0.438 (23)\, \mathrm{fm}^{-1}$ $C_{(\P{3}{2}^{\ast})}^2= 0.1215(36)\, \mathrm{fm}^{-1}$ \footnote{The ANC for the excited state is not readily accessible in transfer reaction experiments.  However, the
other ANCs quoted here have values consistent with DWBA analysis of transfer reactions, by which Ref.~\cite{Tabacaru:2005hv} found $C_{(\P{5}{2})}^2 = 0.414(43)~\mathrm{fm}^{-1}$ and inferred $C_{(\P{3}{2})}^2+C_{(\P{5}{2})}^2 = 0.466(49)~\mathrm{fm}^{-1}$.}. The first two of these come from the same wave function and Monte Carlo walk, giving them a correlation coefficient of $0.672$. 
Proper inclusion of these correlations in the Monte Carlo uncertainty was new in our latest work~\cite{Zhang:2015ajn}, as compared to the earlier Ref.~\cite{Zhang:2014zsa}.  However, we note that the errors still do not include any estimate of the uncertainty due to the Hamiltonian employed or the precision of the variational wave functions.  
\color{green}
\color{black}
We find that the ANC central values are reproduced when the parameters used in Eq.~(\ref{eqn:pEREcoreexcitation}) are    
\begin{align}
\frac{\hthree^{2}}{\hpt^{2}} = 0.2749, \ r_{1}=-0.3102 \ \mathrm{fm}^{-1}, \ a_{1}=1127.7 \ \mathrm{fm}^{3} \ .
\end{align}
Extra decimal places beyond the precision of the ANCs are provided to aid those wishing to reproduce our calculations. 
We note that these values agree with the power counting proposed above, i.e., $\hone\sim\htwo\sim\hthree$, $r_1 \sim \Lambda$, and $a_1 \sim 1/\left(\Lambda \gamma^2 \right)$. 

They are, though, a little different from those published in our previous work~\cite{Zhang:2014zsa}, because 
\begin{enumerate}
\item We corrected a mistake in that publication: there was an extra factor $\gamma^\ast/\gamma$ after $\hthree^2$ in the quantity $-\frac{6\pi}{Z}$ in that paper, as compared to the one computed in Eq.~(\ref{eqn:2plusdimerresidue}) of this paper. This changes the LO to NLO $Z$ ratio, $Z^\mathrm{LO}/Z$ from $0.87$  to $0.955$, which also increases the $C^\mathrm{LO\, 2}_{(\cdots)}/C^2_{(\cdots)}$ ratio from  $0.87$  to $0.955$ and makes our LO $S(E)$ results in better agreement with data in Ref.~\cite{Zhang:2014zsa}. Note, however, that that work's major conclusions are not changed; 

\item The binding energy is updated from $B=0.1375$ MeV in Ref.~\cite{Zhang:2014zsa} to  $0.1364$ MeV in the current work. 
\end{enumerate}
We emphasize that our latest work~\cite{Zhang:2015ajn} is not affected by these changes.

\section{Capture reaction amplitude} \label{sec:capture}

\begin{figure}
\centering
\includegraphics[width=12cm, angle=0]{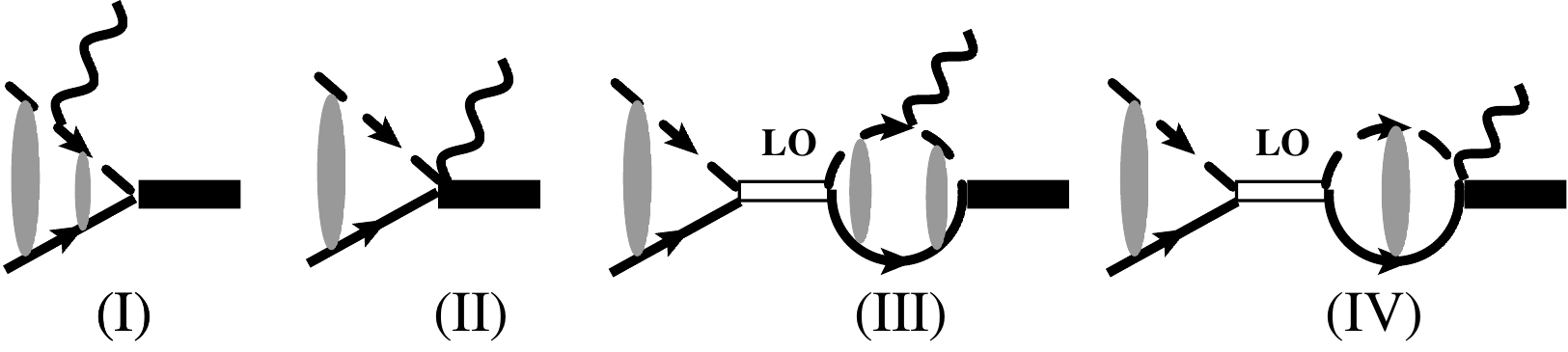} 
\caption{The LO diagrams for radiative capture to a shallow $p$-wave  bound state. The open elongated box denotes the LO propagators for the $s$-wave $\phi_{(1)}$ and $\phi_{(2)}$ dimers. The black filled box is the $\be$ state. The particle spins are not shown explicitly here, but discussed in the main text. These four diagrams $\sim V^{\frac{1}{2}} \left(\frac{V}{V_{\Lambda}}\right)^{\frac{1}{2}}$.  The corresponding diagrams with the photon coupled to the proton line are not shown here for simplicity.}\label{fig:stopcaptureLO}
\end{figure}

\begin{figure}
\centering
\includegraphics[width=10cm, angle=0]{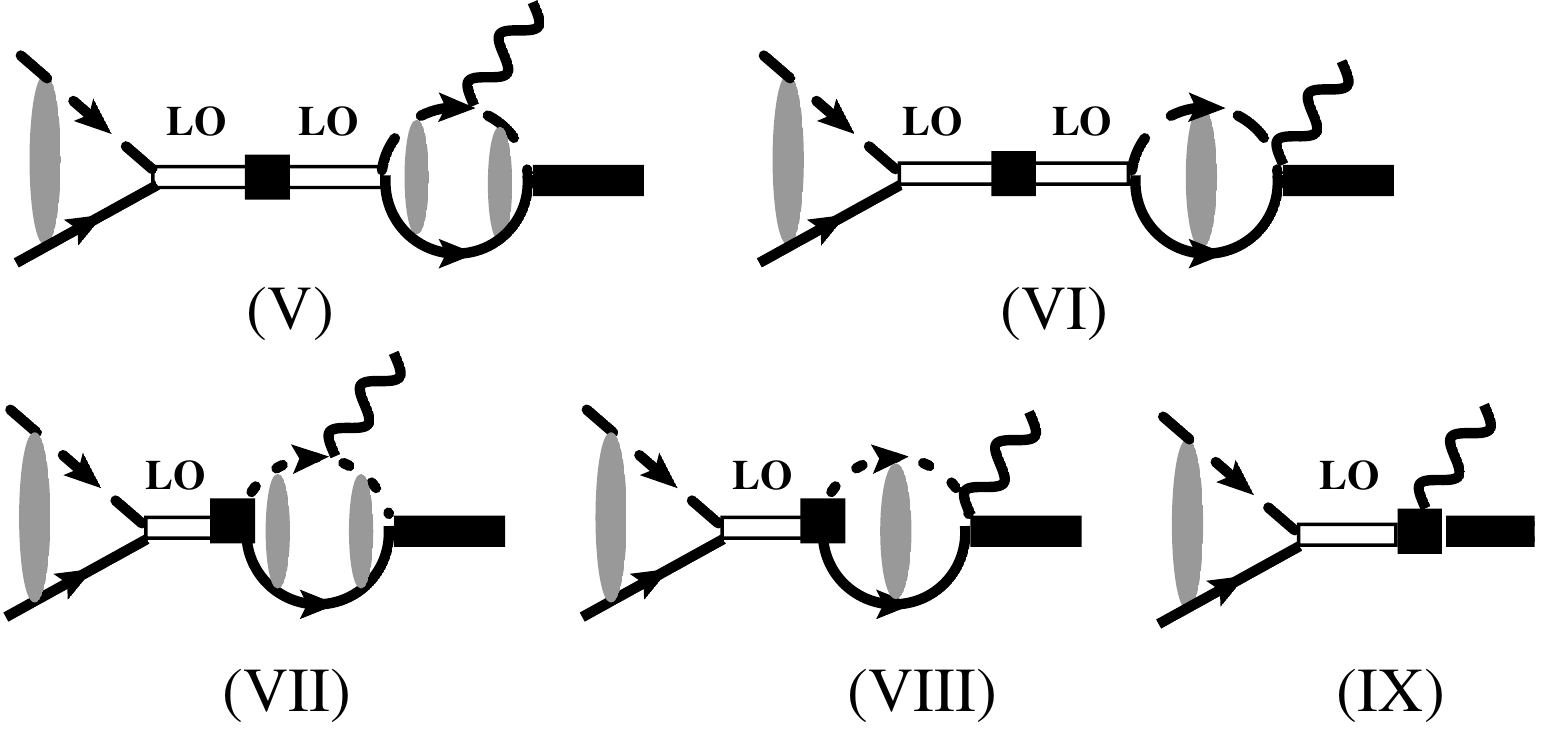} 
\caption{The NLO diagrams for radiative capture to a shallow $p$-wave bound state. They scale as $V^{\frac{1}{2}} \left(\frac{V}{V_{\Lambda}}\right)^{\frac{3}{2}}$. The filled boxes in the (V) and (VI) diagrams are the effective range correction contact coupling, as used in Fig.~\ref{fig:Dphi_NLO}; those in (VII) and (VIII) are the $\hthree$ couplings (the dotted line in the bubbles labels the $\besst$ field); the one in diagram (IX) denotes the $L_1$ and $L_2$ E1 contact couplings defined in Eq.~(\ref{eqn:L12lag}). } \label{fig:stopcaptureNLO}
\end{figure}

The capture reaction is studied in detail in this section. Figs.~\ref{fig:stopcaptureLO} and~\ref{fig:stopcaptureNLO} show the LO and NLO diagrams. According to the power counting, the LO diagrams  $\sim V^{\frac{1}{2}} \left(\frac{V}{V_{\Lambda}}\right)^{\frac{1}{2}}$, while the NLO ones $\sim V^{\frac{1}{2}} \left(\frac{V}{V_{\Lambda}}\right)^{\frac{3}{2}}$. Note that diagrams (VII) and (VIII)  differ from (III) and (IV) by having $\besst$ in the intermediate state instead of $\bes$. The last diagram, (IX), originates from the E1 contact terms from the Lagrangian, which produce NLO effects for the process of interest here:
\begin{align}
\mathcal{L}_c =-i e Z_{eff} L_{1} \fdu{\pi}{\alpha} \T{\alpha}{i j}\vec{E}_i \phi_{(1)j} -i e Z_{eff} L_{2} \fdu{\pi}{\alpha} \T{\alpha}{i \beta}\vec{E}_i \phi_{(2)\beta}+{\rm c.c.} \ .\label{eqn:L12lag}
\end{align}
These terms are built as in Ref.~\cite{Hammer:2011ye}, which itself followed Ref.~\cite{Beane:2000fi}. The structure is the same as Eq.~(\ref{eqn:contacttoy}), except that here spin degrees of freedom have been included. According to the power counting discussed at the end of Sec.~\ref{sec:toypowercounting2}, $L_{1,2}$ should scale as $1/\Lambda$.

As preparation for the full result, let's use Eq.~(\ref{eqn:EMvertexCM}) to compute the E1 matrix element of an operator, $L_{EM}$, that operates between $nc$ Fock states and produces single-photon radiation (see Fig.~\ref{fig:EMvertex}). Here we will relate the overall matrix element of $L_{EM}$ to that of the operator that acts on the $nc$ Coulomb wave function: 
\begin{align}
& \langle\vec{P}', \cw{\vec{p}'}{-},\, \vec{P}_{\gamma} , A^\lambda, n^{\sigma'}, c^{a'}   |L_{EM}|\vec{P}=\vec{0},\, \cw{\vec{p}}{+}, n^{\sigma}, c^{a}  \rangle  \notag \\
=& \left(2 \pi \right)^3\delta\left(\vec{P}'+\vec{P}_\gamma\right) \delta_{\sigma'}^{\,\sigma} \delta_{a'}^{\,a} \left[ \llangle \cw{\vec{p}_{nc}'}{-}|\frac{Q_{n}}{\mn} e^{-if \vec{P}_{\gamma} \vec{r}} \es_{\lambda} \cdot \tilde{\vec{p}}|\cw{\vec{p}}{+}\rrangle
-\llangle \cw{\vec{p'}}{-}|\frac{Q_{c}}{\mc} e^{i(1-f)\vec{P}_{\gamma} \vec{r}} \es_{\lambda}\cdot \tilde{\vec{p}}|\cw{\vec{p}}{+} \rrangle \right] \notag \\
\overset{ P_{\gamma} \rightarrow 0}{=} & \left(2 \pi \right)^3\delta\left(\vec{P}'+\vec{P}_\gamma\right) \delta_{\sigma'}^{\,\sigma} \delta_{a'}^{\,a} \llangle \cw{\vec{p'}}{-}|\frac{Z_{eff}}{\mr}  \es_{\lambda}\cdot \tilde{\vec{p}}|\cw{\vec{p}}{+} \rrangle   \ .
\end{align}
In the expression, as well as in the following discussion, $\tilde{\vec{p}}$  is the momentum operator operating on the intrinsic $| \  \rrangle $ state. The last step in the above derivation keeps only the leading-order term in an expansion in powers of $P_{\gamma} r$, i.e. it is valid in the $P_{\gamma}\rightarrow 0$ limit. The next-order terms in that expansion correspond to E2 and M1 contributions, which we will discuss in Section~\ref{sec:higherorder}.
 Note for the matrix elements between $\besst$-p states, the same results apply; they are not shown explicitly here. 

In the following the $s=1$ channel is used to illustrate the calculation details; the $s=2$ result is closely analogous and will simply be stated at the end.   We choose the frame where the total initial $nc$ momentum $\vec{P}=\vec{0}$. We consider the case that the photon is emitted, i.e. the radiative-capture reaction, so the final $nc$ system will then be recoiling with a momentum $-\vec{P}_\gamma$. We decompose the amplitude that contributes to this process into three contributions: from external capture (EC), where the photon is emitted from 
the proton or the core; from core excitation (CX), where the excited state, $\besst$, participates in the reaction, and from short-distance contributions (SD). (Contributions which serve only to ensure current conservation, e.g. diagram (II), are associated with whatever part of the amplitude they conserve the current for.) This means that the amplitude, ${\cal M}$, is written
\begin{equation}
\mathcal{M}=\mathcal{M}_\mathrm{EC}+\mathcal{M}_\mathrm{CX}+\mathcal{M}_\mathrm{SD}.
\end{equation}
At LO only external capture contributes, but the other two mechanisms enter at NLO. In fact, at NLO, and for an arbitrary renormalization scale, the distinction between EC or CX, on the one hand, and SD, on the other hand, is scheme and scale dependent. But we still find it a useful mnemonic for the computation of the different diagrams (I)--(IX), which can be classified in this way, see Table~\ref{table-diagrams}. 

\begin{table}
\begin{tabular}{|l|c|c|c|}
\hline
 & EC & CX & SD\\
\hline
LO & I-IV & &\\
\hline
NLO & V-VI & VII-VIII & IX\\
\hline
\end{tabular}
\caption{The different classes of capture mechanisms which different diagrams depicted in Figs.~\ref{fig:stopcaptureLO} and \ref{fig:stopcaptureNLO} involve, at leading order (first line, see Fig.~\ref{fig:stopcaptureLO}) and next-to-leading order (second line, see Fig.~\ref{fig:stopcaptureNLO}).}
\label{table-diagrams}
\end{table}

\subsection{External capture: diagrams I-VI}

We write the initial spin state as $|\left(nc\right)^i \rrangle $, and focus on $\langle\pi^\alpha, A^\lambda |L_{EM}|\cw{\vec{p}}{+}, \left(nc\right)^i\rangle $. Since we set the initial $\vec{P}=0$, hence $\overline{E}=E$. 
The first diagram's contribution in Fig.~\ref{fig:stopcaptureLO} is 
\begin{align}
& \langle \vec{P}_\pi, \vec{P}_\gamma, \pi^{\alpha}, A^\lambda |W_{p} \frac{1}{E-H_C+i 0^{+}} L_{EM} |\cw{\vec{p}}{+} , \left(nc\right)^i\rangle \notag \\
 &= \sum_{\lambda',\sigma',a'}  \int\frac{d \vec{p}'}{(2 \pi)^3} \frac{d \vec{P}'}{(2 \pi)^3}  \frac{d \vec{P}_{\gamma}'}{(2 \pi)^3} \langle \vec{P}_\pi, \vec{P}_\gamma, \pi^\alpha, A^\lambda | W_{p} \frac{1}{E-H_C+i 0^{+}} |\vec{P}', \cw{\vec{p}'}{+}, \vec{P}_\gamma', A^{\lambda'} , n^{\sigma'},  c^{a'} \rangle  \notag \\ 
 & \qquad \times   \langle\vec{P}', \cw{\vec{p}'}{+}, \vec{P}_\gamma', A^{\lambda'} , n^{\sigma'},  c^{a'} | L_{EM} |\cw{\vec{p}}{+} , \left(nc\right)^i\rangle \notag \\
& \stackrel{P_\gamma \rightarrow 0}{=} (2\pi)^3 \delta(\vec{P}_\pi+\vec{P}_\gamma) \int \frac{d \vec{p}' }{(2 \pi )^3} \llangle \pi^\alpha|\overline{W}_{p} \frac{1}{\overline{E}-|\vec{P}_\gamma|- \frac{|\vec{P}_\gamma|^2}{2 \mr} -\overline{H}_C+i 0^{+}}  
|  \cw{\vec{p}'}{+},  \left(nc\right)^i \rrangle \llangle  \cw{\vec{p}'}{+} |\frac{Z_{eff}}{\mr}  \es_{\lambda} \cdot \tilde{\vec{p}} |\cw{\vec{p}}{+} \rrangle  \ , \label{eqn:diagram1expression}
\end{align}
We define the matrix element without the total-momentum conserving $\delta$-function and its associated factor of $(2 \pi)^3$ as $\mathcal{M}_I$.  In the last expression, the propagator is changed to $(\overline{E}-|\vec{P}_\gamma| -\frac{|\vec{P}_\gamma|^2}{2 \mr}-\overline{H}_C)^{-1}$, because for the $nc$ system in the final state, the effective energy is the total energy $E$ minus the photon energy $|\vec{P}_\gamma|$ and the $\be$ recoil energy. However, the $\be$ recoil energy is $\frac{\vec{P}_\gamma^2}{ 2 \mnc }  \sim \frac{B^2}{2 \mnc} \sim  10^{-6} $ MeV, and so is small compared to  $B$, $|\vec{P}_\gamma|$, and the typical $\overline{E}$.

We now use the identity $\frac{Z_{eff}}{\mr}  \es_{\lambda} \cdot \tilde{\vec{p}} = i Z_{eff}\left[\overline{H}_C,\ \es_{\lambda}\cdot\vec{r}\right]$. The non-trivial part of the matrix element (\ref{eqn:diagram1expression}) can then be written 
\begin{eqnarray}
{\cal M}_I &=& - i Z_{eff}   \llangle \pi^\alpha|\overline{W}_p \es_{\lambda} \cdot \vec{r} |\cw{\vec{p}}{+} , \left(nc\right)^i \rrangle\nonumber\\
&&  -i Z_{eff} \left(B+\overline{E}\right)
\llangle \pi^\alpha |\overline{W}_p \frac{1}{-B-\overline{H}_{C}+i 0^{+}}  \es_{\lambda} \cdot\vec{r}|\cw{\vec{p}}{+} , \left(nc\right)^i\rrangle  \ .
\end{eqnarray}
The first term is calculated using Eq.~(\ref{eqn:Wpmatrixelement}):
\begin{eqnarray}
&&-i Z_{eff}  \llangle \pi^\alpha |\overline{W}_p \es_{\lambda} \cdot \vec{r} |\cw{\vec{p}}{+} , \left(nc\right)^i \rrangle \notag \\
&&=-i Z_{eff}  \T{\alpha}{ij} \sqrt{Z}  \int d^{3}\vec{r'}d^{3}\vec{r} \frac{d^{3}\vec{q}}{(2\pi)^{3}}  \frac{i \hone}{\mr} \delta(\vec{r'}) \left(\dpartial{j}' \cw{\vec{q}}{+}(\vec{r'})\right) \cwc{\vec{q}}{+} (\vec{r})  \es_{\lambda} \cdot \vec{r} \cw{\vec{p}}{+}(\vec{r}) \notag \\
&&= Z_{eff}  \T{\alpha}{ij} \sqrt{Z} \frac{\hone}{\mr} \epsilon^{\ast}_{\lambda, j} \cw{\vec{p}}{+}(0) \ . \label{eqn:CTrspacereg}
\end{eqnarray}
Here $\epsilon^{\ast}_{\lambda, j}$ is the $j$th component of the outgoing photon polarization vector (with helicity $\lambda$) in the co-ordinate system where the outgoing photon (and hence the outgoing $\be$ momentum) are aligned with the $\hat{z}$ axis. Spin projections of massive particles are also measured along this axis.  Note that since we are calculating the reaction amplitude using LSZ reduction \cite{Weinberg:1995mt}, we have multiplied by a factor of $\sqrt{Z}$, to account for the final-state wave function renormalization.

Meanwhile diagram II of Fig.~\ref{fig:stopcaptureLO}, generates an amplitude $\mathcal{M}_{II}$ that involves the gauged $p$-wave interaction term in the Lagrangian [see Eq.~(\ref{eqn:L2})]. It can be written as   
\begin{eqnarray}
{\cal M}_{II}=- \T{\alpha}{ij} \sqrt{Z} \hone  \frac{Z_{eff}}{\mr}\epsilon^{\ast}_{\lambda, j}   \cw{\vec{p}}{+}(0)  \ .  \label{eqn:tree2}
\end{eqnarray}
This cancels the first term in expression~(\ref{eqn:CTrspacereg}), leaving 
\begin{eqnarray}
\mathcal{M}_I+\mathcal{M}_{II}&=&- i  Z_{eff} \left(B+\overline{E} \right)
\llangle \pi^\alpha|\overline{W}_p \frac{1}{-B-\overline{H}_C+i 0^+ }  \es_{\lambda}\cdot\vec{r}|\cw{\vec{p}}{+} , \left(nc\right)^i \rrangle \notag \\
&=& -  \T{\alpha}{ij}  Z_{eff} C_{(\P{3}{2})} \left(B+\overline{E}\right)  \int d^{3}\vec{r} \frac{W_{-\eta_{B},\frac{3}{2}}(2\gma r)}{r} \dY{1}{j}(\vecpt{r}) \es_{\lambda} \cdot\vec{r} \cw{\vec{p}}{+}(\vec{r})  \ .  \label{eqn:LOtree}
\end{eqnarray}
We see that the EM $\vec{J}\cdot\vec{A}$ coupling used in diagram (I) and (II) is now reduced to the dipole radiation operator in Eq.~(\ref{eqn:LOtree}),  reflecting the well-known Siegert theorem \cite{PhysRev.52.787}.

Let's now turn to $\mathcal{M}_{III}$ and $\mathcal{M}_{IV}$, which represent corrections to $\mathcal{M}_I$ and $\mathcal{M}_{II}$ due to the strong initial-state interactions that generate large $s$-wave scattering lengths in both spin channels.  In order to simplify the presentation, we also compute the NLO diagrams V and VI  (corresponding amplitudes $\mathcal{M}_{V}$ and $\mathcal{M}_{VI}$) here, so that we obtain all strong-interaction corrections to external capture at once.  The sum of the corresponding diagrams without Coulomb effects has been shown to be finite in the study of the isospin mirror system, i.e. $^7\mathrm{Li}$-neutron radiative capture to $^8\mathrm{Li}$, by computing loops in terms of momentum space and PDS regularization \cite{Hammer:2011ye,Rupak:2011nk,Zhang:2013kja}. The sum of diagrams III-VI is also finite in this case, where Coulomb is included.
 After applying the identity $\frac{Z_{eff}}{\mr}  \es_{\lambda} \cdot \tilde{\vec{p}} = i Z_{eff}\left[\overline{H}_C,\ \es_{\lambda}\cdot\vec{r}\right]$, $\mathcal{M}_{III}+\mathcal{M}_{V}$ becomes 
\begin{eqnarray}
&& \llangle \pi^\alpha |\overline{W}_p \frac{1}{-B-\overline{H}_{C}+i 0^+ } \frac{Z_{eff}}{\mr} \es_{\lambda}\cdot\tilde{\vec{p}}\frac{1}{\overline{E}-\overline{H}_{C}+i\epsilon} \overline{T}_{s}|\cw{\vec{p}}{+} , \left(nc\right)^i \rrangle \ , \notag \\
&=& -iZ_{eff} (B+\overline{E})  \llangle \pi^\alpha|\overline{W}_p \frac{1}{-B-\overline{H}_{C}} \es_{\lambda}\cdot\vec{r}\frac{1}{\overline{E}-\overline{H}_{C}+i 0^+} \overline{T}_{s}|\cw{\vec{p}}{+} , \left(nc\right)^i\rrangle  \notag \\
&&-iZ_{eff} \llangle \pi^\alpha |\overline{W}_p  \es_{\lambda}\cdot\vec{r}\frac{1}{\overline{E}-\overline{H}_C+i 0^+} \overline{T}_{s}|\cw{\vec{p}}{+} , \left(nc\right)^i\rrangle \notag \\
&& +iZ_{eff} \llangle \pi^\alpha |\overline{W}_p \frac{1}{-B-\overline{H}_C}   \es_{\lambda}\cdot\vec{r}  \overline{T}_{s}|\cw{\vec{p}}{+}, \left(nc\right)^i \rrangle \ . \label{eqn:loop1}
\end{eqnarray}
Here $\overline{T}_s$ is the $s$-wave $T$-matrix operator up to NLO operating on the single-particle wave function.   The ``$-B$'' term is used in $1/\left(-B-\overline{H}_C\right)$, because the radiated photon takes away all the available energy in the reaction's final state; see the discussion below Eq.~(\ref{eqn:diagram1expression}). By evaluating the matrix element in  coordinate space, as  in Eq.~(\ref{eqn:CTrspacereg}), we find the third term in Eq.~(\ref{eqn:loop1}) is zero, and the second term cancels $\mathcal{M}_{IV}+\mathcal{M}_{VI}$---much as $\mathcal{M}_{II}$ cancels the first term in $\mathcal{M}_I$. Therefore
\begin{eqnarray}
&&\mathcal{M}_{III}+\mathcal{M}_{IV}+\mathcal{M}_{V} + \mathcal{M}_{VI}=
\nonumber\\
&& \quad -iZ_{eff} (B+\overline{E})  \llangle \pi^\alpha |\overline{W}_{p} \frac{1}{-B-\overline{H}_C} \es_{\lambda} \cdot\vec{r}\frac{1}{\overline{E}-\overline{H}_C+i 0^+} \overline{T}_{s}|\cw{\vec{p}}{+} , \left(nc\right)^i\rrangle \notag \\
&& \qquad =-\T{\alpha}{ij} Z_{eff} C_{(\P{3}{2})} \left(B+\overline{E}\right)  \times \notag \\
&& \qquad \quad  \int d^{3}\vec{r} \frac{W_{-\eta_{B},\frac{3}{2}}(2\gma r)}{r} \dY{1}{j}(\vecpt{r}) \es_{\lambda} \cdot\vec{r} e^{i\left(\sigma_{0}+\delta_{(\S{3}{1})}\right)} \frac{\sin\delta_{(\S{3}{1})}}{\ke r} (\cwg{0}(\ke, r)+i\cwf{0}(\ke,r)). \label{eqn:LOloop}
\end{eqnarray}
To obtain this coordinate space matrix element, the momentum integration technique used in Eq.~(\ref{eqn:master1}) has been applied twice. Note that the final result again reflects the Siegert theorem. Since we have included diagrams up to NLO, the phase shift $\delta_{\left(\S{3}{1}\right)} (k)$ is defined in terms of the ERE parameters $\aone$ and $\rone$, see  Eq.~(\ref{eqn:EREswave2}).

We note that the above derivation produces a $\mathcal{M}_{III}+\mathcal{M}_{IV}+\mathcal{M}_{V} + \mathcal{M}_{VI}$ that only has an $s$-wave initial state contribution. In contrast, for $\mathcal{M}_I+\mathcal{M}_{II}$,  the angular integration in Eq.~(\ref{eqn:LOtree})  picks up $s$- and $d$-wave initial state contributions. These can be separated by applying a partial-wave decomposition to $\cw{\vec{p}}{+}(\vec{r})$ in the expression. From now on we define ${\mathcal M}_\mathrm{EC}=\mathcal{M}_I + \mathcal{M}_{II} + \mathcal{M}_{III} + \mathcal{M}_{IV} + \mathcal{M}_{V} + \mathcal{M}_{VI}$ and reintroduce the (nc) spin index, dimer angular momentum, and photon helicity labels that we previously suppressed. However, angular-momentum 
conservation then guarantees
\begin{equation}
{\cal M}_{{\rm EC}, \alpha i \lambda} \equiv \T{\alpha}{ij} \mathcal{M}_{{\rm EC},j \lambda}.
\end{equation} 
Summing diagrams I-VI then gives an $s$-wave contribution that is, up to NLO,
\begin{eqnarray}
&& \mathcal{M}_{\mathrm{EC},j \lambda}^{l=0}=- \sqrt{\frac{4 \pi}{3}}  Z_{eff}\, \epsilon^{\ast}_{\lambda,j}\, C_{(\P{3}{2})}  \left(B+\overline{E}\right) \frac{e^{i\left(\sigma_{0}+\delta_{(\S{3}{1})}\right)}}{\ke} \times \notag \\  
&& \quad \quad \int dr W_{-\eta_{B},\frac{3}{2}}(2\gma r)  r  \left[\sin\delta_{(\S{3}{1})} \cwg{0}(\ke,r)+\cos\delta_{(\S{3}{1})}\cwf{0}(\ke,r)\right] \ , \\
&& \quad \quad \equiv -  e^{i\sigma_{0}}\sqrt{\frac{4\pi}{3}}   Z_{eff}\,  \epsilon^{\ast}_{\lambda,j}\, C_{\etae, 0}  C_{(\P{3}{2})} \left(B+\overline{E}\right)    \mathcal{S}_\mathrm{EC} (\S{3}{1}) \ , 
\end{eqnarray}
with $\mathcal{S}_{H}(X)$ the pertinent wave-function overlap (with the bound-state wave-function normalization/ANC absent) for an incoming scattering channel $X$ and a mechanism $H$: 
\begin{align}
\mathcal{S}_\mathrm{EC}(X) \equiv  \int dr W_{-\eta_{B},\frac{3}{2}}(2\gma r)  r  \left[\frac{ C_{\etae,0}\cwg{0}(\ke,r)}{-a_{(X)}^{-1} +\frac{1}{2}r_{(X)} k^2 -2\kc H(\eta)} \right. \notag\\
 \left.+\frac{\cwf{0}(\ke,r)}{C_{\etae,0}\ke} \frac{-a_{(X)}^{-1} +\frac{1}{2} r_{(X)} k^2 -2\kc {\rm Re} \left[ H(\eta)\right]}{-a_{(X)}^{-1} +\frac{1}{2}r_{(X)} k^2 -2\kc H(\eta)}\right] \ . \label{eqn:Sintegrand}
\end{align}
 This notation is the same as in our previous work, Refs.~\cite{Zhang:2014zsa,Zhang:2015ajn}.
Note that the $r \rightarrow 0$ part of this integral gives a finite result since~\cite{JoachainQCT1975}:
\begin{eqnarray*}
&&\cwf{l}(k,r)\overset{kr\rightarrow 0}{\rightarrow} C_{\eta,l} (kr)^{l+1},\\
&&\cwg{0}(k,r)\overset{kr\rightarrow 0}{\rightarrow} C_{\eta,0}^{-1},\\ 
&&\cwg{l\neq0}(k,r)\overset{kr\rightarrow 0}{\rightarrow} \left[(2l+1) C_{\eta,l}\right]^{-1} (kr)^{-l},\\  
&&W_{-\eta_{B},l+\frac{1}{2}}(2\gma r) \overset{kr\rightarrow 0}{\rightarrow} (2\gamma r)^{-l} \Gamma(2 l +1) /\Gamma(l+1+\eta_B)\, (\mathrm{when}\, l\neq 0 ).
\end{eqnarray*} The integral is also finite at large $r$ because $W_{-\eta_{B},l+\frac{1}{2}}(2\gma r) \propto e^{-\gamma r} (\gamma r)^{-\eta_B}$ there.  

The only $d$-wave contribution, which is also an EC contribution, is from expression~(\ref{eqn:LOtree}):
\begin{eqnarray}
&&\mathcal{M}_{{\rm EC}, j\lambda}^{l=2}= - Z_{eff} C_{(\P{3}{2})} \left(B+\overline{E}\right) \times \notag \\
&& \quad \int d^{3}\vec{r} \frac{W_{-\eta_{B},\frac{3}{2}}(2\gma r)}{r} \dY{1}{j}(\vecpt{r}) \es_{\lambda} \cdot\vec{r} \left(4\pi i^{2} e^{i\sigma_{2}}\right)\frac{\cwf{2}(k,r)}{kr} \uY{2}{\alpha}(\vecpt{r})\dY{2}{\alpha}(\vecpt{p}) \notag \\
&& \equiv - \sqrt{\frac{2}{3}} 4\pi  e^{i\sigma_{2}}  Z_{eff}  C_{(\P{3}{2})}\,   \epsilon^{\ast}_{\lambda,i}\,   \T{j}{i\alpha} \dY{2}{\alpha}(\vecpt{p})   C_{\etae,0}   \left(B+\overline{E}\right)    \mathcal{D}_\mathrm{EC}  \ ,
\end{eqnarray}
with
\begin{eqnarray}
\mathcal{D}_\mathrm{EC} &\equiv&   \int dr  {W_{-\eta_{B},\frac{3}{2}}(2\gma r)}   r  \frac{\cwf{2}(\ke,r)}{C_{\etae,0} \ke} \ . \label{eqn:Dintegrand}
\end{eqnarray}
Here the formula $\int d\Omega_{\vecpt{r}} \dY{1}{j}(\vecpt{r}) \uY{1}{i}(\vecpt{r}) \uY{2}{\alpha}(\vecpt{r}) =(-)\sqrt{\frac{1}{2\pi}} \T{j}{i\alpha} $ has been used.  Since $\cwf{2}(\ke,r)\overset{\ke r\rightarrow 0}{\rightarrow} C_{\etae,2} (\ke r)^{3}$,  the integration is also finite. 
In summary, the EC contribution to radiative capture is, up to NLO, represented by diagrams I-VI, which yield:
\begin{eqnarray}
&& \langle \vec{P}_\pi, \vec{P}_\gamma, \pi^{\alpha}, A^\lambda | L_{EM} |\cw{\vec{p}}{+} , \left(nc\right)^i\rangle = - \left(2 \pi \right)^3\delta\left(\vec{P}_\pi + \vec{P}_\gamma\right) \frac{4\pi}{\sqrt{3}} \T{\alpha}{ij} Z_{eff} C_{\etae,0} C_{(\P{3}{2})} \left(B+\overline{E}\right)    \times \notag \\ 
&& \qquad \qquad \left[e^{i\sigma_{0}} \epsilon^{\ast}_{\lambda,j} \dY{0}{0}(\vecpt{p})\mathcal{S}_\mathrm{EC}(\S{3}{1})+e^{i\sigma_{2}} \epsilon^{\ast}_{\lambda,k}  \sqrt{2} \T{j}{k\beta} \dY{2}{\beta}(\vecpt{p}) \mathcal{D}_\mathrm{EC} \right] \ . \label{eqn:ampreorganization}
\end{eqnarray}

\subsection{CX and SD diagrams: VII-IX}

The LO result~\cite{Zhang:2014zsa} is obtained from Eq.~(\ref{eqn:ampreorganization}) by setting both effective ranges, $r_{(X)}$ to zero. But the NLO result includes contributions from core excitation and short-distance capture, as well as the effect of finite effective range. 

After repeating the previous calculations, but with the $\besst$ field inside the bubbles, we get diagram (VII) and (VIII) contributions to the total $s$-wave amplitude $\mathcal{S}$ that are analogous to Eq.~(\ref{eqn:Sintegrand}) but involve an excited core:
\begin{eqnarray}
\mathcal{S}_\mathrm{CX}(\S{3}{1} )&=&\frac{\gthree}{\gone} \frac{C_{(\P{3}{2}^\ast)}}{C_{(\P{3}{2}})}   \int dr W_{-\etabst,\frac{3}{2}}(2\gma^{\ast} r)\,  r\,   \frac{\Gamma(1+i\etaest) W_{-i\etaest,\frac{1}{2}}(-2i \kest r)}{-\frac{1}{\aone}+\frac{1}{2} \rone \ke^2 - 2 k_C H(\etae)} \ .
\label{eqn:SCX}
\end{eqnarray}
Here $k_{\ast}$, $\etaest$, $\etabst$ have been defined below Eq.~(\ref{eqn:pEREcoreexcitation}) and below Eq.~(\ref{eqn:2plusdimerresidue}). Note that $\Gamma(1+i\eta_{\ast})  W_{-i\eta_{\ast},\frac{1}{2}}(-2i k_{\ast} r)$ equals $C_{\eta_{\ast}, 0} (\cwg{0}(k_{\ast}, r)+i\cwf{0}(k_{\ast},r))$ when $k_{\ast}^2 \geq 0$, and becomes a bound state  wave function when $k_{\ast}^2 \leq 0$. The transition between these cases is smooth. Based on this, in the limit $\gma_{\Delta}\rightarrow 0$, the core excitation terms have the same structure as the expression~(\ref{eqn:LOloop}). 

Now, the diagram (IX) contribution---due to the  $L_1$ contact term---which can be considered as a short-distance operator (SD),
\begin{eqnarray}
\mathcal{M}_{\mathrm{SD}, j \lambda}&=& -  Z_{eff}\, \gone \sqrt{Z}  L_1  \epsilon^{\ast}_{\lambda,j}   \omega\, D_{\phi_{(1)}}\, \cw{\vec{p}}{+}(0)\,, \label{eqn:Lbardef}
\end{eqnarray}
with $\omega$ the outgoing photon energy ($=|\vec{P}_\gamma|$).  This result can be simplified as 
\begin{eqnarray}
\mathcal{S}_\mathrm{SD}(\S{3}{1} ) =\frac{\sqrt{3}}{2} \frac{\overline{L}_{1}}{\gma \Gamma(2+\etab)} \frac{1 }{\frac{1}{\aone}-\frac{1}{2} \rone \ke^2 + 2 k_C H(\etae) } \ , 
\label{eqn:SSD}
\end{eqnarray}
where $\overline{L}_{1} \equiv \frac{L_{1} 2 \sqrt{3} \pi }{\gone \hone \mr} $. Eq.~(\ref{eqn:SSD}) indicates that $\overline{L}_{1}$ is independent of the renormalization scale $\mu$, otherwise its contribution to the capture reaction cross section would depend on $\mu$ [see Eq.~(\ref{eqn:sfactormaster1})], contradicting the requirement that the cross section is $\mu$-independent. In other words, the $\mu$ dependence of $L_1$ is dictated by the $\mu$-independence of $\overline{L}_{1}$. The specific factor, $2 \sqrt{3} \pi$, in the redefinition is motivated by the observation [based on Eqs.~(\ref{eq:DphiandERE2}) and~(\ref{eq:DandERE2}) from the toy model of Sec.~\ref{sec:toymodel}] that if $\mu \ll \Lambda$, then $s$- and $p$-wave couplings obey: 
\begin{align}
h_s(\mu) h_p (\mu) \mr \approx 2 \sqrt{3} \pi  \sqrt{r_0 r_1} \approx 2 \sqrt{3} \pi \ . \notag 
\end{align}

According to the discussion at the end of section~\ref{sec:toypowercounting2}, $L_{1} \sim \frac{1}{\Lambda}$ and so $\overline{L}_{1} \sim \frac{1}{\Lambda}$, as long as $r_0 r_1 \sim 1$. We write the short-distance part of the $S=2$ capture amplitude in the same way:
\begin{eqnarray}
\mathcal{S}_\mathrm{SD}(\S{5}{2} ) =\frac{\sqrt{3}}{2} \frac{\overline{L}_{2}}{\gma \Gamma(2+\etab)} \frac{1 }{\frac{1}{\atwo}-\frac{1}{2} \rtwo \ke^2 + 2 k_C H(\etae) } \ , 
\label{eqn:SSDspin2}
\end{eqnarray}

This produces EC, CX, and SD contributions in accord with the power counting. From Eq.~(\ref{eqn:Sintegrand}), we know $\mathcal{S}_\mathrm{EC} \sim \frac{1}{\gma^3}$, while Eqs.~(\ref{eqn:SCX}) and (\ref{eqn:SSD}) give $\mathcal{S}_\mathrm{CX}, \ \mathcal{S}_\mathrm{SD} \sim \frac{1}{\Lambda \gma^2}$. 

\subsection{Total amplitude, $S$-factor}

The total amplitude for radiative capture in the $s=1$ channel, up to NLO in our EFT, is then:
\begin{eqnarray}
&&{\cal M}_{\alpha i \lambda}= -  \T{\alpha}{i j} \frac{4\pi}{\sqrt{3}}  Z_{eff} C_{\etae,0} C_{(\P{3}{2})} \omega\nonumber\\ 
&& \qquad \times \left[e^{i\sigma_{0}} \epsilon^{\ast}_{\lambda,j} \dY{0}{0}(\vecpt{p})\mathcal{S}(\S{3}{1})+e^{i\sigma_{2}} \epsilon^{\ast}_{\lambda,k}  \sqrt{2} \T{j}{k\beta} \dY{2}{\beta}(\vecpt{p}) \mathcal{D}_\mathrm{EC} \right]  \label{eqn:ampreorganization2}
\end{eqnarray}
(where again  $\omega$ is the outgoing photon energy). In order to extract the overall factor of $\omega$ here we have neglected the nuclear recoil, i.e. set $B + \overline{E}=\omega$. Here 
\begin{equation}
\mathcal{S}(\S{3}{1})=\mathcal{S}_\mathrm{EC}(\S{3}{1}) + \mathcal{S}_\mathrm{CX}(\S{3}{1}) + \mathcal{S}_\mathrm{SD}(\S{3}{1}).
\end{equation}

Finally, everything works out analogously in the $s=2$ channel to yield
\begin{eqnarray}
&& {\cal M}_{\alpha \beta \lambda}= - \T{\alpha}{\beta j} \frac{4\pi}{\sqrt{3}}  Z_{eff} C_{\etae,0} C_{(\P{5}{2})} \omega  \notag \\ 
&& \qquad \times \left[e^{i\sigma_{0}} \epsilon^{\ast}_{\lambda,j} \dY{0}{0}(\vecpt{p})\mathcal{S}(\S{5}{2})+e^{i\sigma_{2}} \epsilon^{\ast}_{\lambda,k}  \sqrt{2} \T{j}{k\beta} \dY{2}{\beta}(\vecpt{p}) \mathcal{D}_\mathrm{EC} \right] \ . \label{eqn:ampreorganization3}
\end{eqnarray}
where
\begin{equation}
\mathcal{S}(\S{5}{2})=\mathcal{S}_\mathrm{EC}(\S{5}{2}) + \mathcal{S}_\mathrm{SD}(\S{5}{2}).
\end{equation}

Two points are worth noting here. First, core excitation contributes only in the spin-one channel, since the spin quantum numbers do not permit it in the $s=2$ case.  Second, since $ \mathcal{D}_\mathrm{EC}$ does not involve initial-state interactions it is the same for both $s=1$ and $s=2$ channels; however, $\mathcal{S}$ depends on the $s$-wave channel parameters through the strong interactions in the initial state. 

Finally, we use
\begin{eqnarray}
\sigma &=&  \int  d\Omega_{\vecpt{k}} \frac{\mr}{8 \pi^{2}} \frac{\omega}{\ke} \frac{1}{8}\sum_{\lambda,\sigma,a, \alpha}|\mathcal{M}|^{2} \ .
\end{eqnarray}
with $\mathcal{M}$ now expressed on the basis of particle spins $\sigma, a, \alpha$ and depending on photon polarization $\lambda$. Hence we  convert the computed matrix elements for $| (nc)^i \rrangle$ and $| (nc)^\beta \rrangle$ initial states to those for the direct product  $| n^\sigma c^a\rrangle$ initial state to get the total $S$ factor: 
\begin{eqnarray}
S(E) &=& E e^{2\pi \etae} \sigma(E) \notag \\ 
&=& \frac{e^{2\pi \etae}}{e^{2\pi\etae}-1}  \frac{5\pi}{18}  Z_{eff}^{2}  \kc \omega^3  \times \notag \\ 
&&  \left[ C_{(\P{3}{2})}^{2} \left(\mid \mathcal{S}(\S{3}{1}) \mid^{2}+2 \mid \mathcal{D}_\mathrm{EC} \mid^{2}\right) +  C_{(\P{5}{2})}^{2} \left(\mid \mathcal{S}(\S{5}{2}) \mid^{2}+2 \mid \mathcal{D}_\mathrm{EC}\mid^{2}\right)\right] \ .\label{eqn:sfactormaster1}
\end{eqnarray}

Note that the NLO halo EFT calculation of $S(E)$ ultimately depends on nine parameters once the $\be$ binding energy is fixed. Of these, four enter already at LO. They are the ANCs, $C_{\left(\P{3}{2}\right)}^2,\, C_{\left(\P{5}{2}\right)}^2$, and the $s$-wave scattering lengths, $\aone,\,\atwo$. Five more parameters are necessary to describe the NLO pieces of the result: the $s$-wave effective ranges, $\rone,\,\rtwo$, the two LECs parameterizing the short-distance pieces of the matrix element $\overline{L}_{1,2}$, and the proton-$\besst$ mixing parameter $\varepsilon_1 \equiv \frac{\gthree}{\gone} \frac{C_{(\P{3}{2}^\ast)}}{C_{(\P{3}{2}})}$~\cite{Zhang:2015ajn}.

\section{Comparison to traditional models}

\label{sec:modelcomparison}

\subsection{Conceptual relationship of halo EFT and earlier calculations}

There is close correspondence between halo EFT and several aspects of the many older models of the $\bes(p,\gamma)\be$ reaction.  In fact, our EFT has been constructed to apply to this system very generally at low momentum, so  $S$-factors and phase shifts near threshold in any model that obeys general physical principles should be reproducible with correctly-chosen EFT parameters.  This means that differences among models close to threshold should reduce to choices of EFT parameters, provided that sufficient terms of the EFT expansion have been retained.  We now discuss our EFT in terms of some types of models previously in use.  This will make contact with the extensive prior literature on the $\bes(p,\gamma)\be$ reaction, and it will provide context for matching the EFT onto literature models below.  For a complete review of prior models up to 2010, see Ref.~\cite{Adelberger:2010qa}.

\subsubsection{Correspondence of halo EFT and potential-model contributions}

Like halo EFT, a potential model treats the \bes~nucleus and proton as fundamental particles.  It models their interaction with a Woods-Saxon or similar potential, usually with both central and spin-orbit terms \cite{christyduck,aurdal70,robertson73,barker80,Jennings:1998ky,davidstypel03,esbensen04,huang10,navratil06}.  Wave functions are computed in configuration space, and the electromagnetic transition operator is written in the usual Siegert-theorem form that also appears in our Eq.~(\ref{eqn:LOloop}).  As first pointed out by Christy and Duck~\cite{christyduck}, low-energy nonresonant capture in a potential model of $^8$B is dominated by the part of the matrix element integral where the \bes~and proton are well separated, far beyond the range of strong interaction.  In this region the interaction is purely Coulombic, so the final-state wave function is proportional to a Whittaker function while the initial state consists of phase-shifted Coulomb waves.  This gives a matrix element integral very similar to our Eq.~(\ref{eqn:LOloop}), the only difference being that the potential-model integrand deviates from Whittaker and Coulomb functions at small radii.  This happens at radii $\lesssim 5$ fm, where the \bes-$p$ effective interaction differs significantly from pure Coulomb.  In this region, the initial-state wave function of the potential model has a very small amplitude due to tunneling through the Coulomb barrier.

In a previous effort to separate long- and short-range effects, Jennings \textit{et al.}~\cite{Jennings:1998ky} computed a potential model and found that the radiative-capture matrix element integrand peaks at 40 fm for $s$-wave capture at threshold.   They also presented a second calculation with the same phase shifts, but otherwise pure Coulomb interaction all the way to zero radius.  For phase shifts specified through the ERE, this is exactly our Eq.~(\ref{eqn:Sintegrand}).  
The total matrix elements for the two cases are nearly equal; even at 500 keV, Jennings \textit{et al.}~find only a 3\% difference in the $s$-wave cross section between the pure phase-shifted Coulomb initial state and the full potential model with the same phase shifts.  The overall size of this difference has to grow with energy at about the same pace as the small radius part of $\mathcal{S}_\mathrm{EC}$ in order to keep the initial wave function continuous, and this gives it about the same energy dependence as our $\mathcal{S}_\mathrm{SD}$.  Thus the short-distance part of the potential-model matrix element amounts to a cancellation between the $r \lesssim 5$ fm part of our $\mathcal{S}_\mathrm{EC}$ integral and the short-distance counterterm $\mathcal{S}_\mathrm{SD}$.  In the language of the potential model, the difference encoded in $\mathcal{S}_\mathrm{SD}$ occurs in the small-$r$ region where the effective nuclear potential dominates the shape of the wave function.  This accords with the label ``short distance'' for the $\mathcal{S}_\mathrm{SD}$ term in halo EFT, which corresponds to regions in the potential model where $r \lesssim \Lambda^{-1}$.

Potential models can in principle include excitation of \besst~by adding another channel to the wave function, with corresponding ``diagonal'' and channel-coupling terms in the potential---at the cost of more elaborate calculations and additional parameters.  This is not needed for qualitative description of the data, and to our knowledge it has only been done once in the literature \cite{Nunes1997a,tagkey1997747}.  Such a contribution corresponds to our $\mathcal{S}_\mathrm{CX}$.

\subsubsection{The Pauli principle in potential models}

An important consideration for the size of short-distance effects arises from the nature of the potential-model interaction:  it is an effective interaction that incorporates not just the strong nuclear force but also particle-exchange effects.  The projection of a nucleon-level wave function onto a product of cluster wave functions has a structure constrained by fermionic antisymmetry (first considered in the present context in Ref.~\cite{aurdal70}).  In our case, \bes~contains a practically filled $0s$ shell but open $0p$ orbitals in both $j=1/2$ and $j=3/2$ subshells.  The main consequences of antisymmetry are imposed on potential models of \be~by constructing the $s$-wave effective potential to have a nodeless deeply bound state that is regarded as belonging to the $0s$ shell and therefore forbidden by the Pauli principle.  At threshold the $l=0$ scattering state then belongs to the $1s$ shell and has a single node inside the potential well.  Since there are open $p$-shell orbitals available for the proton,  no constraint from antisymmetry guides construction of the $l=1$  effective potential.

The Pauli node in $s$-wave scattering states has two consequences for the capture reaction.  First, the $\sim 40$ MeV well depth \cite{Jennings:1998ky,davidstypel03,esbensen04,Navratil2006} needed to generate a node in potential models greatly exceeds the scattering energy near threshold, so that the short-range part of the scattering wave function has a nearly energy-independent shape below 1 MeV (as features imposed by antisymmetry would).  Short-range contributions to the capture matrix element are then largely energy independent apart from barrier penetrability;
this is in accord with the energy dependence of $\mathcal{S}_\mathrm{SD}$.  Second, the presence of the node implies some radius within the potential well where the matrix element density goes to zero, so that regions of opposite-sign density just on either side of that radius cancel.  As a result, short-range contributions in the potential model are suppressed, suggesting significant cancellation between $\mathcal{S}_\mathrm{SD}$ and the small-radius part of the $\mathcal{S}_\mathrm{EC}$ integral in the EFT.

\subsubsection{Potential-model parameters and EFT couplings}

The effective potential thus has two logically distinct roles in the capture cross section: it determines phase shifts for the external part of the initial state, and it models details of both initial and final effective wave functions within the potential well.   The \be~system lacks empirical information to constrain these features separately, so model construction requires \textit{ad hoc} assumptions that impose arbitrary correlations between them.  For example,  traditional lore has usually provided the radius and shape of the potential.  (An exception is the model of Ref.~\cite{navratil06}, in which the $p$-wave potential well was constructed to reproduce overlap functions from \textit{ab initio} calculations.)

Woods-Saxon potentials are the most common choice, and their radius and diffuseness are generally chosen to be the same in all channels. A potential model of the $\bes$-proton system then has, in principle, different well depths in the  
 $s$-wave $S=1$ and $S=2$ channels, as well as a single well depth and spin-orbit coupling in all other channels. This is a total of six parameters. These models then employ spectroscopic factors for the two $p$-wave channels, meaning that they have at least eight parameters.  After assuming values or relations between values, there are usually two or three parameters adjusted to data in actual model construction.  Work in this vein started at least as early as Ref.~\cite{aurdal70}.   
 
In such calculations the spectroscopic factors for the $\P{3}{2}$ and $\P{5}{2}$ components of the $\be$ wave function were generally taken from the shell model. A final rescaling of the overall cross section (see also discussion below) was then made, on the understanding that the overall scale of the spectroscopic factors should be adjusted to match capture data.  It is probably better to view this procedure as fixing ANCs rather than spectroscopic factors, since external capture dominates the $S$-factor at low energies.  In this regard, the role of ANCs in a potential model corresponds almost exactly to the role of ANCs in the EFTs. The same experimental or theoretical constraints can be used in both frameworks. 

As discussed above, halo EFT includes the possibility of core excitation. At NLO this is encoded in the single parameter $\varepsilon_1$ without adding significant complication to practical calculations. Since most potential models do not include core excitation this parameter has no potential-model counterpart.

A critical difference between the two approaches is that at NLO halo EFT encodes the amplitudes of short-distance contributions in contact couplings $\bar{L}_{1,2}$ with no large-distance consequences; no parameter of a potential model affects just short-distance physics.  Potential models thus have implicit correlations between phase shifts and short-range amplitudes that arise from assumed potential-well geometry but are not part of the most general parameterization of the amplitude. 

We found a clear example of implicit correlations between scattering lengths and small-radius contributions in the potential models of Davids \& Typel \cite{Davids:2003aw}.  One potential model in that work was constructed to reproduce the best-fit experimental scattering lengths, while a second model reproduces the upper-limit scattering lengths and a third reproduces the lower-limit values.  In Ref.~\cite{Davids:2003aw} the differences among the $S$-factor curves of these models were interpreted as arising from the scattering lengths.  We repeated this exercise using halo EFT, by first fitting EFT parameters to match the Davids \& Typel best-fit model and then varying $\aone$ and $\atwo$ between their experimental limits while leaving other parameters fixed (see Sec.~\ref{sec:pmspace} for more details on this procedure).  For $\aone$, we found very nearly the same dependence of $S$-factor on scattering length as in the Davids \& Typel models.  The story with $\atwo$ was very different: the dependence of the $S$-factor on $\atwo$ in the EFT is much weaker than in the potential model.  In the language of halo EFT, much of the $\atwo$ dependence of the $S$-factor found by Davids \& Typel lies in $\overline{L}_2$, not in the $\atwo$ dependence of the amplitude.  This places a significant part of the $S$-factor model dependence found in Ref.~\cite{Davids:2003aw} inside the potential well rather than in the asymptotic part of the wave function described by scattering parameters.  (A further complication here is a problem with the published $S(E)$ for one of the Ref.~\cite{Davids:2003aw} models, noted below.  The calculations discussed in this paper mainly involve a corrected version that has the stated scattering lengths.) 

While the EFT contains fewer implicit assumptions than potential models, the price of the more general parameterization of the amplitude is the need to fix nine parameters.  In a potential model this might correspond to treating the eight parameters enumerated above as free and independent, all unguided by lore beyond an expectation that most should have ``natural'' sizes.  Below, we show that it is possible to constrain enough of the EFT parameters jointly from measured $S$-factors and scattering lengths to obtain a robust extrapolation for $S(E)$.  This is presumably a simpler task for halo EFT than for a potential model, in that there is no Schr\"odinger equation to be recomputed when the potential is varied. Indeed, we have written the amplitude for radiative capture in such a form that the EFT parameters can all be varied without recomputing the integrals in Eqs.~(\ref{eqn:Sintegrand}), (\ref{eqn:Dintegrand}), and (\ref{eqn:SCX}) above.

\subsubsection{Cluster and {\it ab initio} calculations}
Microscopic models, in contrast to potential models, treat all nucleons as distinct particles and in principle require less tuning to the \be~system.  They are based on a nucleon-nucleon potential and compute wave functions of the \bes~and \be~systems by solving seven- and eight-body Schr\"odinger equations.  The simplicity of the Siegert E1 operator in configuration space makes  the capture matrix element calculation resemble a projection of \be~onto clusters, and a potential-model wave function can be viewed as an ansatz for the projection of a \be~state into a purely $\bes+p$ space.  As a result, the important features of potential models carry over to microscopic models:  the largest contributions come from the long-range asymptotic region, and short-distance features are dominated by antisymmetry. 
 However, the location of the Pauli node now arises from explicit antisymmetrization of an eight-body wave function, it is no longer imposed \textit{ad hoc}. Importantly, microscopic models include configurations not writable in terms of the \bes~ground state.  Thus, in a general sense the microscopic models' biggest advantage is that their wave functions at distances from about 1 fm to the $r \sim 1/\Lambda$ short-distance scale of halo EFT (or a potential model) is determined by the underlying nuclear forces. A disadvantage is that $S$-factors are very sensitive to some quantities like threshold energies and scattering lengths that do not typically emerge with high precision from a nucleon-nucleon potential that was not fitted to them.  Some tuning to the full eight-body system is possible, but a nucleon-level potential with few parameters can only be tuned to one or two eight-body properties at once, while one with more parameters probably requires a much more demanding refit to \be~and other observables simultaneously.  

It has only recently become possible to compute accurate energies and scattering wave functions for $A=8$ systems from nucleon-nucleon interactions that reproduce many observables of two and three-nucleon systems faithfully \cite{Navratil:2011sa}.  Such models are generally referred to as \textit{ab initio}.  For four decades prior to that work, limited computer power restricted microscopic models to greatly simplified nuclear interactions and severely truncated basis spaces.  However, a great deal of useful work along those lines was done using the resonating group method (RGM) and generator coordinate method (GCM) \cite{,descouvemont88,johnson92,descouvemont94,csoto95,Descouvemont:2004hh}; we refer to such restricted microscopic models as ``cluster models,'' and we match one of them onto an EFT below.

Cluster models work in a basis constructed from energy eigenstates of clusters within a nucleus.  For \be, this means energy levels of \bes~and $^5\mathrm{Li}$ built up from $\alpha$ particles, $^3$He nuclei, and protons.  The more excited states of the clusters are included in the basis, the more exact a calculation will be.  This approach has been extended to very large bases in \textit{ab initio} calculations \cite{Navratil:2011sa}, but in cluster models the $\alpha$ and $^3$He clusters are mostly constructed as $0\hbar\omega$ harmonic-oscillator configurations.  Core excitations like \besst~are required in the model for reasonable accuracy~\cite{Navratil:2011sa,Descouvemont:2004hh,Csoto:1997}. 

The mapping between cluster models and halo EFT is roughly the same as between potential models and halo EFT.
The differences are that cluster models have a firmer grounding in general principles, include core excitation explicitly, and should need less tuning to \be~data.  However, for poorly understood reasons, cluster models almost always predict $S$-factors  larger than the data at all energies -- often by 10\% or more.   Low-energy extrapolations using these models are typically built on the assumptions that the matrix element is entirely external capture to good approximation and that most of the uncertainty lies in the ANCs, plausibly because of the truncated model space.     One holds the computed $S(E)$ curve shape of the model fixed and multiplies it by a constant to fit capture data, just as one does to fix spectroscopic factors in a potential model.

This does not exhaust the range of published models.  However, it does cover both the ones for which we find halo EFT representations in Sec.~\ref{sec:pmspace} and those used in Ref.~\cite{Adelberger:2010qa} to produce a recommended $S(0)$.  

\subsubsection{Phenomenological $R$-matrix}
Some brief comments on  the relation of our EFT to phenomenological $R$-matrix models may also be useful.  For scattering, the connection between the latter approach and halo EFT is derived explicitly in Ref.~\cite{Hale:2013ama}. 
In the case of radiative capture the phenomenological $R$-matrix has external-capture contributions that correspond very closely to our $\mathcal{S}_\mathrm{EC}$, $\mathcal{S}_\mathrm{CX}$, and $\mathcal{D}_\mathrm{EC}$ amplitudes, but  the integrals are cut off below some radius on the order of 3 fm: apart from these lower limits on the integrals there is a nearly one-to-one mapping of our matrix-element terms onto terms of the $R$-matrix capture amplitude given in Eq.~(6) of Ref.~\cite{barker95}.  

As a matter of computation, phenomenological $R$-matrix models incorporate ANCs in exactly the same way as the EFT (apart from the small-radius cutoff).  Scattering phase shifts enter both external and internal $R$-matrix amplitudes  in much the same way they enter $\mathcal{S}_\mathrm{EC}$ and $\mathcal{S}_\mathrm{SD}$, respectively.  And the $R$-matrix parameterizes short-distance contributions to radiative capture using radiative-width parameters that are completely analogous to our $\bar{L}_{1,2}$---even down to the way they enter amplitudes.  

One difference from halo EFT is that instead of an effective-range expansion, phase shifts are encoded in a pole expansion of the $R$-matrix, each term of which has a reduced width and a level energy.  
In practice there is usually only enough information to fit one pole and all others are approximated with a single high-energy pole that provides slow energy variation at low energies. Because the ERE applies very generally, near threshold there must always be an ERE that corresponds exactly to any given pole expansion: the relationship between the $R$-matrix and ERE parameters is worked out in Ref.~\cite{Teichmann:1951}. This means, though, that phase shifts are more complicated functions of $R$-matrix parameters than they are of ERE parameters, and, as in potential models, the $R$-matrix background-pole parameterization can produce implicit correlations between ERE parameters in the fit. This may give the halo EFT formalism, with its explicit construction around the ERE, significant advantages for near-threshold data fitting. 

\subsection{Mapping potential and cluster models into the EFT parameter space}

\label{sec:pmspace}

We now examine models from the literature and demonstrate that---at least below center-of-mass energy ${E}=500$ keV---each corresponds to a specific set of EFT parameters.  Our strategy is to take model outputs as data to be fitted in the EFT and show that highly accurate fits result.  This is easier than fitting experimental data because a computed model produces more information than is available from experiment, and there are no measurement errors.  We computed phase shifts and $S$-factors for several potential models from the literature using our own code, which included separate $S$-factors for each possible  spin and orbital angular momentum channel in its outputs. For the one microscopic model considered, $S$-factors are tabulated by channel in the original publication, and its author provided a table of phase shifts that were originally published as a graph \cite{descouvemont-email}.  

We chose five different models for fitting:  three variant potential models of Davids \& Typel \cite{Davids:2003aw} that were tuned to the measured scattering lengths and their error limits; the potential model based on \textit{ab initio} inputs from Navratil \textit{et al.} \cite{Navratil2006}; and a cluster model by Descouvemont \cite{Descouvemont:2004hh}.  These provide a wide range of conditions and were important in the  $S(0)$ recommendations of Ref.~\cite{Adelberger:2010qa}.  
While performing this work, we learned that the published $S$-factors for the Davids \& Typel potential with lower-limit scattering lengths contain a programming error, so that the lower-limit curve apparently reflects $\atwo = -20.9$ fm, not $-10$ fm \cite{davids-email}.  The calculations reported here reflect a corrected version of this ``lower" Davids \& Typel potential model that reproduces the intended $\atwo=-10$ fm.  

In fitting models, we use the $\be$ proton separation energy $0.1375$ MeV from the 2003 mass evaluation (current when the models were published); the fits to data discussed in Sec.~\ref{sec:realistic} use the currently recommended $0.1364$ MeV. The $d$-wave $S$-factor has very nearly the same threshold energy dependence in every model (including all EFTs), so we first obtain ANCs for each model by fitting its $d$-wave $S$-factor separately in each spin channel. We then fit scattering lengths and effective ranges to the computed phase shifts. In the potential models there is explicitly no core excitation, so we set $\mathcal{S}_\mathrm{CX} = 0$. For the cluster model we also set $\mathcal{S}_\mathrm{CX}=0$ as a simplifying assumption for fitting, even though the model contains core excitation; it is difficult to fit uniquely from the model outputs, and we obtain a precise fit without it, perhaps because core excitation can be traded against short-distance physics at these energies. This leaves only the contact terms $\overline{L}_{1,2}$ undetermined, and we fix them from the $s$-wave $S$-factor in each spin channel.

\begin{table}[htb]
\begin{ruledtabular} 
   \begin{tabular}{ccccccccc}
$ C_{\left(\P{3}{2}\right)}^2$  & $\aone$  & $\rone $ & $\varepsilon_1$ & $\overline{L}_1 $ &    $ C_{\left(\P{5}{2}\right)}^2$ & $\atwo$ & $\rtwo$ & $\overline{L}_2$ \\ \hline
 0.201 & 16.0 & 1.18 & 0 & 1.12 & 0.534 & $-10.0$ & 3.93 & 2.69 \\
 0.201 & 25.0 & 1.36 & 0 & 1.27 & 0.533 & $-7.03$ & 5.02 & 3.10 \\
 0.201 & 34.0 & 1.45 & 0 & 1.34 & 0.533 & $-4.03$ & 8.56 & 4.19 \\
 0.109 & $-4.15$ & 6.80 & 0 & 4.80 & 0.542 & $-6.91$ & 3.57 & 3.73 \\
 0.108 & 7.19 & 0.785 & 0 & 0.725 & 0.480 & 7.19 & 0.785 & 0.725 \\
\end{tabular}  \caption{EFT parameters obtained from fits to models from the literature. The units for the ANC squared ($C^2$) are $\mathrm{fm}^{-1}$, and for the scattering length, effective range, and $\overline{L}_{1,2}$ are fm. $\varepsilon_1$ is unitless. These units are implicitly assumed and will not be shown in other places. From top to bottom the models are Davids \& Typel lower, central, and upper potential models~\cite{Davids:2003aw}, the Descouvemont cluster model~\cite{Descouvemont:2004hh}, and the 
Navratil potential model~\cite{Navratil2006}.} \label{tab:EFTparaModels}
\end{ruledtabular}
\end{table}

\begin{figure}[htb]
\centering
\includegraphics[width=10cm, angle=0]{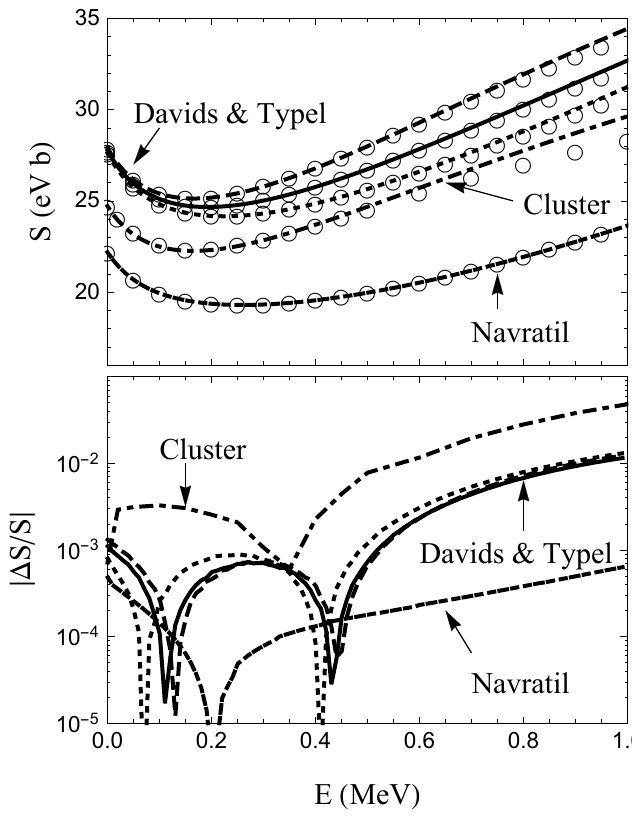}
\caption{Upper panel: $S$-factors from the EFT fits (continuous) and the original potential  \cite{Davids:2003aw,Navratil2006} and cluster \cite{Descouvemont:2004hh} models (discrete circles).  From top to bottom, the first three curves are the Davids \& Typel potential models, with scattering lengths in both channels increasing from top to bottom.  Next comes the cluster model, then at the bottom the Navratil potential model.   Lower panel:  Absolute values of $S$-factor residuals between the original models and fitted EFTs as fractions of the total.  Symbols correspond to models in the same way as in the top panel.} \label{fig:EFTvsothermods}
\end{figure}

The fitted parameters are shown in Table~\ref{tab:EFTparaModels}. In the table we quote all parameters to three significant figures. Two different implementations of our calculation agree at this level for almost all numbers in the table. But additional precision is needed to produce, e.g., the curves in Fig.~\ref{fig:EFTvsothermods}. Readers interested in higher-precision results should contact the authors. 
From top to bottom, the EFT parameters are for the Davids \& Typel lower, central, and upper potential models (corresponding to lower-limit, best-fit, and upper-limit scattering lengths), the Descouvemont cluster model with the Minnesota potential, and the Navratil potential model.  Crucially, we find that although the parameters change from row to row, almost all are consistent with NDA, i.e. $C^2 \sim \Gamma^2(2+\eta_B)\gamma^2/\Lambda$ [see Eq.~(\ref{eqn:anc-scaling})], $(\aone,\,\atwo) \sim \gamma^{-1}$, $(\rone,\,\rtwo) \sim \Lambda^{-1}$, $\overline{L}_{1,2} \sim \Lambda^{-1}$.

In the top panel of Fig.~\ref{fig:EFTvsothermods} the original $S(E)$ curves from these models are represented as circles and the fitted EFTs are shown as continuous curves. The models occur in the same top-to-bottom sequence as in Table \ref{tab:EFTparaModels}.  The fractional difference between each EFT and its original model is shown in the lower panel.  In the fitted energy range $0$ to $0.5$ MeV, these residuals are less than 0.2\% of the total $S$-factor for the potential models.  However, residuals for the Davids \& Typel models contain cancellation between $s=2$ $s$- and $d$-waves, which individually deviate from the original model by 0.4\% at 500 keV.  In the $s=1$ channels of the Davids \& Typel models, and in all channels of the Navratil model, the deviation is less than 0.1\%.  For the Descouvemont model, errors are under 1\% over the fitted range.  Extrapolating the EFT curves to 1 MeV, differences from the original models increase to about $1\%$ for potential models and $5\%$ for the cluster model.  The behavior of residuals in this 0.5 to 1.0 MeV energy range are consistent with the truncation error of our calculation EFT being N$^3$LO.

The three Davids \& Typel potentials differ in their $l\neq 1$ well depths to produce scattering lengths $\aone=16,\, 25,\, 34$ fm and $\atwo=-10,-7,-4$ fm.  
These models use spectroscopic factors from Cohen and Kurath \cite{cohenkurath67} and produce the ANCs $C_{\left(\P{3}{2}\right)}^2=0.2010$ fm$^{-1}$ and $C_{\left(\P{5}{2}\right)}^2=0.5332$ fm$^{-1}$;  ANCs fitted to the $d$-wave capture ``data" from the model in the EFT (Table \ref{tab:EFTparaModels}) are within 0.3\% of these values.
For the Navratil model the match between the fitted EFT ANCs and the original ones in the model is of the same level of precision. 

For the cluster model, the published ANCs correspond to  $C_{\left(\P{3}{2}\right)}^2=0.1116$ fm$^{-1}$ and $C_{\left(\P{5}{2}\right)}^2=0.5565$ fm$^{-1}$; however, values 2.5\% smaller were needed to fit the $d$-wave $S$-factors, apparently reflecting some unidentified difference in cluster masses or \be~separation energy between the original calculation and our EFT code.   Fitting the ERE over 0-600 keV (three tabulated energies) yielded $\atwo = -6.910$ fm although this model was tuned for $-7$ fm; this probably reflects the relatively wide energy range of our fit.   The simplification that $\varepsilon_1=0$ does not seem to have serious consequences, and our experience with experimental data (Sec.~\ref{sec:realistic}) indicates that $\varepsilon_1$ can be compensated in the fitting of $\overline{L}_1$.  

The match of the cluster model onto an EFT also differs from the other cases in that we had full information about the potential models that we computed, at several decimal places and on a dense grid in energy.  In contrast, we fitted to published information for the cluster model, which necessarily had fewer printed data and more rounding (e.g., only two digits for capture from $^3D_2$ scattering states).  It is worth noting in this context that tests of the fitting procedure with a 1 keV mismatch between the binding energies of the EFT and a potential model did not allow a fit with smaller residuals than 1\%.  The difficulties in getting EFT to fit the cluster model with the same accuracy as for the other models considered could either be due to larger higher-order effects for that model, or to these fitting issues.

We also obtained a second set of EFT parameters matched to potential models, this time using ANCs and binding energies directly from the original model and ERE parameters fitted over 0--30 keV, and still fitting $\bar{L}_{1,2}$ to 0--0.5
MeV $S$-factors as before. These EFT fits reproduce the original models at threshold even more accurately; by construction, deviations of this EFT fit from the original model grow with energy. The results for $S(E)$ 
 are not plotted, but in the 0-500 keV they match the original models about as well as the fits in Table \ref{tab:EFTparaModels} and Fig.~\ref{fig:EFTvsothermods} do. The growth with energy is again consistent with a calculation in which the leading omitted effect is N$^3$LO.

\section{Realistic analysis: recapitulation}

\label{sec:realistic}

Here, for completeness, we summarize the results of our analysis of experimental data for $\bes(p,\gamma)\be$. Further details are given in Ref.~\cite{Zhang:2015ajn}. Some details related to Bayesian priors and computational issues can also be found in Ref.~\cite{Zhang:2015vew}. 

\subsection{Data selection} We included 42 data points measuring total $S$-factors in our analysis. They come from all modern experiments with more than one data point for the direct-capture $S$-factor up to $E=500$ keV. All data lie at energies above $0.1$ MeV. We subtracted the M1 contribution of the $\be$ $1^{+}$ resonance from the data using the resonance parameters of Ref.~\cite{Filippone:1984us}. This has negligible impact for $E \leq 0.5$ MeV due to the smallness of the correction and the small uncertainty on the correction. Since we retain only points in this region, this eliminates the resonance's effects. Ref.~\cite{Adelberger:2010qa} summarizes these experiments, which are
Junghans {\it et.al., } (two experiments) \cite{Junghans:2010zz},  Filippone {\it et.al.,} \cite{Filippone:1984us}, Baby {\it et.al.,} \cite{Baby:2002hj, Baby:2002ju}, and Hammache {\it et.al.,} (two measurements published in 1998 and 2001) \cite{Hammache:1997rz, Hammache:2001tg}.
We assigned common-mode errors, listed in Ref.~\cite{Zhang:2015ajn}, according to the  published accounting of experimental systematics.  The Junghans BE1 target data were left out of the final Ref.~\cite{Adelberger:2010qa} analysis because of correlations with the BE3 data; we kept both sets because their wide energy coverage provides  valuable constraints on our model's energy dependence, most likely outweighing the disadvantage of correlations in overall normalization that are estimated to be small.  

\subsection{Analysis}

We wish to extrapolate $S(E)$ from the region of these data, $100~{\rm keV} < E < 500~{\rm keV}$, to the region of relevance for solar modeling, 30 keV and below (with peak sensitivity at 18 keV).  We used the 42 data points to constrain the nine EFT parameters, computing the posterior probability distribution function (PDF) of the parameter vector $\vec{g}$ given data, $D$, our theory, $T$, and prior information, $I$. To account for the common-mode errors in the data we introduced data-normalization corrections, $\xi_i$. Since these errors affect all data from a particular experiment in a correlated way there are only five parameters $\xi_1$--$\xi_5$: one for each experiment that has a shared normalization error of this kind.  (See Ref.~\cite{Zhang:2015ajn} for the one that does not.)

We performed a Bayesian analysis and used Markov Chain Monte Carlo to determine the posterior PDF, with details described in Refs.~\cite{Zhang:2015ajn,Zhang:2015vew}. All EFT parameters but the $s$-wave scattering lengths are assigned flat priors over ranges that correspond to, or
exceed, natural values. 
 We do, though, restrict the
parameter space by requiring that there is no $s$-wave resonance
in $\bes$-proton scattering below $0.6$ MeV.

We also constrain the EFT parameter space further by incorporating 
independent experimental information on the $s$-wave scattering lengths via 
Gaussian priors on $\left(\aone,\,
\atwo\right)$, centered at the experimental values of
Ref.~\cite{Angulo2003}, $\left(25,\, -7\right)$ fm, and with widths equal
to their reported errors, $\left(9,\, 3\right)$ fm. In fact, these numbers were extracted from an analysis of scattering data using the ERE in $s$-waves and a single-pole $R$-matrix resonance in $p$-waves, with no value of the effective range recommended.  It is possible that interesting correlation structures in the EFT parameter space would result from a full analysis of those cross-section data; an EFT analysis along these lines is deferred to a future publication. 

This ``Bayesian model averaging" samples the part of the EFT parameter space that is consistent with the scattering lengths quoted in Ref.~\cite{Angulo2003}. Regions in that space which reproduce the shape and the magnitude of the $S(E)$ data more accurately are then weighted more strongly in the final extrapolant.

\subsection{Results}

\label{sec:results}

Our tightest
parameter constraint is on the sum
$C_{\left(\P{3}{2}\right)}^2+C_{\left(\P{5}{2}\right)}^2=0.564(23)~\mathrm{fm}^{-1}$, which sets the overall scale
of $S(E)$.  Neither ANC is
strongly constrained by itself, but they are strongly anticorrelated.
The {\it ab initio} calculation of Nollett \& Wiringa~\cite{Nollett:2011qf} predicts ANCs that agree with our extraction within error bars, $C_{\left(\P{3}{2}\right)}^2+C_{\left(\P{5}{2}\right)}^2=0.537(26)~\mathrm{fm}^{-1}$, while there
is some disagreement with another {\it ab initio} prediction of 0.509 fm$^{-1}$~\cite{Navratil:2011sa} from Navratil \textit{et al.}
Comparing our results with ANCs
inferred from transfer reactions by Tabacaru
\textit{et al.}~\cite{Tabacaru:2005hv}, we found essentially the same conflict between ANCs and $S$-factors that was already recognized in Ref.~\cite{Tabacaru:2005hv}, at a level of $1.8\sigma$.

We also found that the effect
of core excitation, parameterized by $\varepsilon_1$ in the EFT, can be traded
against the short-distance part of the spin-1 E1 matrix
element, in that there is a slight non-zero signal for the 
quantity $0.33\, \bar{L}_1/\mathrm{fm} - \varepsilon_1$. The data 
do prefer a positive $\bar{L}_2$: its one-dimensional PDF yields $-0.58~{\rm fm} < \bar{L}_2 < 7.94~{\rm fm}$ at 68\% degree of belief.

We then computed the PDF of $S(E)$ at many energies and extracted each median value and 68\% interval.  At 0 keV (20 keV) we found $S=21.33^{+0.66}_{-0.69}$ eV b ($S=20.67^{+0.60}_{-0.63}$), again at 68\% degree of belief,  including all errors associated with parameter selection.
A choice of the EFT-parameter vector $\vec{g}$ that corresponds to natural coefficients, produces curves close to the median $S(E)$ curve, and has a large value of the posterior probability is given in Ref.~\cite{Zhang:2015ajn}. Ref.~\cite{Zhang:2015ajn} also supplies information on the derivatives of $S(E)$ at 0, as well as a simple parameterization
for the thermal reaction rate.

\subsection{Comparison to Solar Fusion II value}

Ref.~\cite{Adelberger:2010qa} recommends $S(0)=20.8\pm 0.7\mathrm{(expt)}\pm 1.4\mathrm{(theory)}$. In that work experimental errors were inflated by a factor of 1.65 to account for large $\chi^2$ values with respect to the models employed. Each fit in Ref.~\cite{Adelberger:2010qa} consisted of an overall rescaling of one model for $S(E)$, based on the idea that the models could accurately predict the shape of the $S$-factor but needed further adjustment of ANCs (or spectroscopic factors) to match the data. The theoretical error in Ref.~\cite{Adelberger:2010qa} was taken as half the difference between the lowest and highest extrapolated $S(0)$ obtained in this way.  The models that determined the theory error bar in Ref.~\cite{Adelberger:2010qa} were ultimately the Navratil semi-\textit{ab initio} model \cite{Navratil2006} (last line of Table \ref{tab:EFTparaModels} and lowest curve of Fig.~\ref{fig:EFTvsothermods}, upper panel) and the original Davids \& Typel low-scattering-length model that is now known to inadvertently have had $\atwo$ outside the empirical range; see Sec.~\ref{sec:modelcomparison}.  This procedure was motivated by a desire to consider a set of models that were consistent with general physical principles, and then not discriminate between them on the basis of small differences of $\chi^2$.   

Our $S(0)$ is consistent with the result of Ref.~\cite{Adelberger:2010qa}. But our total error, including model selection, and without any error inflation, is about the same size as the inflated experimental error quoted there. It is therefore markedly smaller than the combined $0.7\mathrm{(expt)}\pm 1.4\mathrm{(theory)}$ error bar given there. 
We attribute this smaller total uncertainty to two things. First, our Bayesian sampling of the EFT parameter space means we explore the full range of reasonable models of this process. We showed in Sec.~\ref{sec:pmspace} that 
differences amongst models below $E=0.5$ MeV---including the Navratil and Davids \& Typel models that defined the error bar obtained in 2011---can be encoded in nine EFT parameters.  
Our Monte Carlo sampling of that space thus includes the capture models that set the bounds in Ref.~\cite{Adelberger:2010qa}, the other models from that analysis whose results lie between those two, and models which fall elsewhere in EFT-parameter space as well. The computation of a PDF on this EFT-parameter/model space then permits discrimination---based on experimental data---regarding the shape of the $S(E)$ curve. Not all of the physics in that curve comes from the external-capture part of the matrix element, so rescaling a model curve to match the data only produces a reliable result up to a certain level of accuracy. In contrast, the Bayesian model averaging that we implemented through EFT parameterization of the capture amplitude favors regions of the model space that produce better descriptions of the $S(E)$ data.
Second, although the Navratil and Davids \& Typel models are included in this model averaging, they correspond to $\atwo=+7.2$ and $\atwo = -20.9$ fm respectively and are thus strongly disfavored by the prior on \atwo\
that we have taken from the Angulo data. In other words, neither of the models that ultimately determined the theory error bar in Ref.~\cite{Adelberger:2010qa} was consistent with the scattering lengths published in Ref.~\cite{Angulo2003}. Had those models been excluded, the range of extrapolated $S(0)$ would have been substantially narrower.  

The near-equality of our 68\% degree-of-belief interval with the experimental error estimated in Ref.~\cite{Adelberger:2010qa} appears to be entirely coincidental.  The analysis there indicated a nearly $2\sigma$ inconsistency among data sets, given the assumption that any one of nine theoretical $S(E)$ shapes was correct.   The errors in that analysis were then inflated by a factor of about 1.65 to account for the inconsistency.  Our analysis does not include this inflation factor.   
In Refs.~\cite{Zhang:2015ajn,Zhang:2015vew} we looked for, but did not find, quantitatively clear indications of inconsistency amongst the data sets in the results of our analysis.  

Our final uncertainty on the extrapolated $S(0)$ accounts for both the experimental errors and the differences among a wide class of models that are consistent with naturalness and information on the $\bes$-$p$ scattering lengths. In Bayesian model averaging there is no way to divide the degree-of-belief interval into a ``theory" and an ``experimental" error bar: the data determine the weightings of many different models, which all contribute to the final extrapolation.  We are able to take full advantage of this more quantitative accounting of model uncertainties because of the simultaneous generality and consistency with basic physics provided by the halo EFT parameterization. Bayesian model averaging over the EFT parameter space yields a more general, and more rigorous, accounting of model uncertainties than examining a range of broadly plausible models.   

\section{Effects at N$^2$LO and beyond}

\label{sec:higherorder}

The calculation we have carried out here is complete to NLO in the expansion in powers of $k_\mathrm{low}/\Lambda$.  In the numerator $k_\mathrm{low}$ any of the soft scales $k$, $\gamma$, $k_C$, $\gamma_\Delta$, or $1/\atwo$ and $1/\aone$ can appear. This suggests an expansion parameter $\approx 20$\%  for amplitudes, so errors due to higher-order effects in the $S$-factor could be as large as 10\%. However, our success in fitting the NLO halo EFT amplitude both to models and to experimental data suggests that 10\% is an overestimate of the EFT truncation error. In this section we examine various higher-order effects and attempt to assess their impact on the $S$-factor. 

\subsection{Higher orders in the proton-$\bes$ interaction}

First, it is important to recognize that although we nominally worked to NLO, we captured a large set of  higher-order corrections by resumming the range corrections in both the $s$-wave scattering ERE and the formulae for the $p$-wave ANCs. A strict NLO calculation would have re-expanded observables in powers of $\rone$ and $\rtwo$. In the single-channel $s$-wave case such a resummation improves the accuracy of the EFT amplitude from NLO to N$^2$LO, because the shape parameter (coefficient of the $k^4$ term in the ERE) affects the EFT amplitude only at N$^3$LO~\cite{Mehen:1998tp}. 

However, the presence of $\besst$ as an explicit degree of freedom in the EFT means that core excitation  enters the $s$-wave ERE at N$^2$LO (for a natural coupling). In the limit that $\gamma_\Delta$ is well below the maximum momentum of interest this could have a 5\% effect on the amplitude.  This is mitigated by the restriction of our $S$-factor study to proton-$\bes$ energies below 500 keV, which corresponds to a momentum $k_{\rm max}=28.7$ MeV: this is less than 10\% higher than $\gamma_\Delta$. The limited phase space available to the $\besst$-proton channel reduces the amount it can change the cross section in the energy domain of interest. For the natural-sized coupling assumed in this work we estimate its effect to be a few per cent at most.

Second, we recall that in this paper we consider only the Coulomb potential between the proton and the core. Transverse photons can also be exchanged between these two charged particles. However, as noted above, their effect is suppressed by $V^2/c^2 \leq 0.1$\%. This estimate is for ``potential" photons, which obey $q_0 \ll |{\bf q}|$. One might be concerned that, e.g., ``ultra-soft" photons, which have $q_0 \sim |{\bf q}| \sim E$, will produce larger effects, since they receive an infra-red enhancement from the photon propagator. However, the small amount of phase space available for these modes more than compensates for that enhancement, ultimately making ``ultra-soft" photons an $O(V^3/c^3)$ effect. 

\subsection{Higher multipoles}

Next, we turn our attention to the accuracy of the approximations that we made in evaluating the matrix element of the electromagnetic current in Section~\ref{sec:capture}. There we wrote $e^{i {\vec{P}}_\gamma \cdot {\bf r}} \approx 1$, which means we have neglected multipoles higher than E1 in the external photon field. 

In the region up to 500 keV there are no resonances in the proton-$\bes$ system, and so standard dimensional arguments give a reasonable estimate of the effect of higher multipoles. Photon radiation of M1 character will, for example, be radiated from the system with an amplitude that is $\sim V/c$ smaller than the E1 amplitude we have computed here. Since M1 and E1 photons do not interfere in the total radiative capture cross section, the M1 multipole then has an effect in the $S$-factor only at $O(V^2/c^2)$. As already mentioned above, this means neglecting M1 radiation is an approximation that is good to better than 0.1\%. We note that this approximation breaks down immediately above $E=500$ keV, due to the presence of the $1^+$ resonance in proton-$\bes$ scattering and its allowed M1 transition to the $\be$ ground state.

E2 transitions to the $\be$ ground state are also allowed, e.g., from $p$-wave proton-$\bes$ scattering states. 
Using an estimate for the amplitude for E2 radiation~\cite{Walkecka1995book}, we find that it is suppressed by $\omega \langle r \rangle/5$ compared to E1 radiation, where $\langle r \rangle$ is the size of the emitting region. (The factor in the denominator is the $(2L+1)!!$ that enters the amplitude for multipolarity $L$.) For very-low-energy capture the size of the emitting region can be tens of fermis, and $\omega$ can be as large as 500 keV.  Thus we conservatively take $\omega \langle r \rangle \approx 0.1$, suggesting an E2 amplitude that could be a couple of percent of the E1 amplitude at the upper end of our energy range. As is the case for M1 radiation, E2 and E1 amplitudes enter the $S$-factor incoherently, so even with these conservative estimates we expect that E2 transitions affect $S(E)$ by less than 0.05\%. 

These estimates of higher multipoles are supported by our potential-model calculations~\cite{Zhang:2015ajn}. Indeed, in those models they yield even smaller effects than we have estimated here, accounting for less than $0.01\%$ of the total direct-capture $S$-factor for energies up to $0.5$ MeV.

\subsection{Higher-order pieces of the E1 amplitude}

Finally, we must consider the fact that the evaluation of the direct-capture E1 amplitude in Section~\ref{sec:capture} amounts to considering only the leading term in the expansion of the pertinent spherical Bessel function $j_0(\omega r)$. The next term in the $j_0(\omega r)$ Taylor series -- sometimes called a ``retardation term'' -- is however a factor of $\omega^2 r^2/6$ smaller than the $O(\omega)$ term. It too is at most a 0.2\% effect. Moreover, the proton magnetic moment's contribution to the direct-capture E1 amplitude is suppressed by about $\omega/M_n$~\cite{Walkecka1995book}, which increases to $\approx 0.05\%$ at $0.5$ MeV, and thus its contribution to the $S$ factor are less than $0.1\%$ (the nuclear magnetic moments' contributions are even smaller). 

There also are higher-order pieces of the short-distance part of the amplitude ${\cal S}_{\rm SD}$. They are represented by higher-dimensional operators that contain derivatives acting on photon, $n$, or $c$ fields. However, parity conservation ensures that these contain at least two spatial derivatives. (The equations of motion can be used to convert all time derivatives to spatial derivatives.) They are thus suppressed by ${\vec{k}}^2 \approx 2 \mr E$ compared to the NLO short-distance effects we have included here. This makes them N$^3$LO. We have tested the impact that the inclusion of this kind of N$^3$LO term has on our Bayesian analysis \cite{Zhang:2015ajn}.  This revealed no statistical evidence for non-trivial energy dependence of either $\bar{L}_1$ or $\bar{L}_2$ in the experimental data on $\bes$(p,$\gamma)\be$.

\section{Summary}

We have studied the $^7\mathrm{Be}(p,\gamma)^8\mathrm{B}$ reaction in a low-energy effective field theory up to NLO. This yields an amplitude valid over the entire $< 500$ keV energy range directly relevant to astrophysical modeling and data extrapolation. Major results at LO and NLO were presented in previous reports. In this paper we provide details of our work.

We first discussed the EFT power counting based on velocity scaling.  The Coulomb potential is very important in our reaction, and velocity scaling---originally developed in the context of a system of two heavy quarks interacting via gluon exchange---is well-suited to its treatment. We used velocity scaling to include electromagnetic interactions in a simple EFT in which strong interactions are classified according to the power counting developed for systems with large $s$-wave scattering lengths and shallow $p$-wave bound states. In order to make a connection between EFT calculations and ordinary quantum mechanics we computed amplitudes using time-ordered perturbation theory and the Lippmann-Schwinger equation, since the intermediate states are easily identified there. This also allows us to fully exploit existing knowledge of Coulomb wave functions in coordinate space. Indeed, one  major feature of this work is the development of EFT matrix elements in position space. It will be interesting to explore higher-order loop diagrams using this method.

We applied the Lagrangian, power counting, and calculational methods developed in our simple EFT to the $^7\mathrm{Be}(p,\gamma)^8\mathrm{B}$ reaction, including all the complications necessary for a realistic calculation, i.e., spin degrees of freedom and a $\bes$ core excitation at low energy. The latter modifies the corresponding effective range expansion within the EFT: the $s$-wave acquires a small inelasticity, while the $p$-wave becomes an elementary coupled-channel problem. The $\be$ ANCs were also computed in terms of the underlying EFT parameters. This connection was used in our previous LO work to fix the EFT parameters to ANCs computed {\it ab initio} and produce a LO estimate of the $S$-factor~\cite{Zhang:2014zsa}.

We then computed the reaction amplitude using time-ordered perturbation theory and found that loop diagrams are finite in calculations using the spatial coordinate. The final results bear similarities to those of quantum-mechanical models, but with the crucial difference that there is an explicit hierarchy of contributions--external capture, core excitation, and short distance terms--that had never before been studied as systematically. This displays the power of EFT, which provides a systematic way to organize the matrix element. The resulting amplitude is model independent in the sense that it has no regulator dependence and a minimal parameter set for a specified accuracy.  The generality of the formalism places  estimates of theory uncertainties on firmer ground.  

As an explicit demonstration of this, we showed that several published models can be defined as specific points in the EFT parameter space---at least as far as the capture amplitude up to center-of-mass energies of 500 keV is concerned.  The $S$-factors and phase shifts of each model permit extraction of an unambiguous set of EFT parameters:  ANCs from the $d$-wave $S$-factors, $s$-wave couplings from phase shifts, and contact couplings from the $s$-wave $S$-factors.  The coordinates of these models in the space of EFTs (i.e., corresponding values of EFT couplings) in general agree with power-counting expectations.  From the difference between the fitted EFTs and the original models outside the fitted energy range, we estimate that the truncation error of our calculation is actually N$^3$LO and thus $< 1\%$. 

The halo EFT developed here thus covers the space of low-energy theories of $^7\mathrm{Be}(p,\gamma)^8\mathrm{B}$ and has omitted terms that are negligible over the energy range important for extrapolation from laboratory to astrophysical conditions.  This facilitates our purely data-driven extrapolation, wherein we use Bayesian methodology to sample the space of EFT parameters and compute the posterior PDF in that parameter space based on data on $^7\mathrm{Be}(p,\gamma)^8\mathrm{B}$ and the scattering lengths extracted in Ref.~\cite{Angulo2003}. The resulting extrapolant does not include the tacit assumptions of a potential model and produces a smaller (combined theory and experiment) uncertainty than that of previous evaluations.  

Finally, we point out that the EFT and Bayesian methodology used here is applicable to other systems with similar features.  
 
\acknowledgments
We thank Barry Davids, Pierre Descouvemont, and Stefan Typel for sharing details of their calculations with us. We thank Carl Brune for several useful discussions on the physics of this reaction. 
We are grateful to the Institute for Nuclear Theory for support under Program INT-14-1, ``Universality in few-body systems: theoretical challenges and new directions", Workshop INT-15-58W, ``Reactions and structure of exotic nuclei", and Program INT-16-2a, "Bayesian Methods in Nuclear Physics". During all three we made significant progress on this project.  
X.Z.~and D.R.P.~acknowledge support from the US Department of Energy under grant DE-FG02-93ER-40756. X.Z. also acknowledges support from the US Department of Energy under grant DE-FG02-97ER-41014. K.M.N.~acknowledges support from the Institute of Nuclear and Particle Physics at Ohio University, and from U.S.~Department of Energy Awards No.~DE-SC 0010 300 and No.~DE-FG02-09ER41621 at the University of South Carolina.

\appendix

\section{Lippmann-Schwinger expansion} \label{app:Hformalism}

Following the canonical quantization procedure (e.g.~\cite{Weinberg:1995mt}), we can derive a Hamiltonian from the toy-model Lagrangian in expression~(\ref{eqn:toylag}). Without showing the details, we list the free Hamiltonian density and the strong-interaction potential density:
\begin{subequations}
\begin{alignat}{2}
\mathcal{H}_0 (\vec{x}) &=  c^{\dagger} \left(- \frac{\overset{\rightarrow}{\nabla}^2}{2 M_c} \right) c 
+n^{\dagger} \left( - \frac{\overset{\rightarrow}{\nabla}^2}{2 M_n}  \right) n -\phi^{\dagger} \left( -\frac{\overset{\rightarrow}{\nabla}^2}{2 M_{nc}} - \Delta_\phi \right)  \phi \notag \\
  & \quad + \pi^{\dagger\, i}  \left( - \frac{\overset{\rightarrow}{\nabla}^2}{2 M_{nc}} -\Delta_\pi \right)  \pi_i   \label{eqn:toyham0} \\
\mathcal{W}_s(\vec{x})  & = -  h_s \phi^{\dagger} n c  + H.C.\, ,   \label{eqn:toyhams}  \\
\mathcal{W}_p (\vec{x}) & = - h_p \pi^{\dagger\, i}\, n \,\widetilde{\vec{V}}_{Ri}\,  c + H.C. \ .  \label{eqn:toyhamp}
\end{alignat}
\end{subequations}

The free Hamiltonian and potential $H_0 = \int d {\vec{x}} \mathcal{H}_0(\vec{x})$, $W_s = \int d {\vec{x}} \mathcal{W}_s (\vec{x})$, and $W_p = \int d {\vec{x}} \mathcal{W}_p (\vec{x})$. In principle, the daggered fields should be represented by the corresponding conjugate momentum fields, e.g., $\Pi_\phi \equiv \partial \mathcal{L}/\partial \dot{\phi} $, but for simplicity we just use the daggered fields themselves. After quantization, the fields are operator-valued functions depending on the space coordinate $\vec{x}$; when acting on a state in Fock space the fields annihilate particles while the daggered fields create particles, for example $\phi(\vec{x}) = \int \frac{d \vec{P}_{\phi}}{\left(2 \pi \right)^3} e^{i \vec{P}_{\phi}\cdot \vec{x}} \phi_{\vec{P}_{\phi}}$  and $\phi^{\dagger}(\vec{x}) = \int \frac{d \vec{P}_{\phi}}{\left(2 \pi \right)^3} e^{-i \vec{P}_{\phi}\cdot \vec{x}} \phi^\dagger_{\vec{P}_{\phi}}$. We stress that no anti-particle degrees of freedom exist in this theory.   

The subtlety due to the extra minus sign for the $s$-wave dimer ($\phi$ field) free Hamiltonian in expression~(\ref{eqn:toyham0}) should be properly dealt with. For this field we impose the canonical commutation relation $[\phi(\vec{x})\, , \, \phi^{\dagger}(\vec{x}')]= -\delta\left(\vec{x}-\vec{x}'\right)$. Then, if $|0\rangle$ is the vacuum state and $|\vec{P}_\phi\rangle \equiv \phi^{\dagger}_{\vec{P}_\phi} |0\rangle$, it follows that $\langle\vec{P}_\phi'|\vec{P}_\phi\rangle = - (2 \pi)^3 \delta \left(\vec{P}_\phi- \vec{P}_\phi' \right) $. This won't cause any problems in our EFT, because the particle number is conserved and finite, and therefore the energy is bounded from below. However, the completeness relation for the $\phi$-mode subspace is: 
\begin{equation}
I_{\phi} = - \int \frac{d\vec{P}_{\phi}}{\left(2 \pi \right)^3} |\vec{P}_\phi\rangle\langle\vec{P}_\phi|.
\end{equation}
An alternative approach to deal with the ``$-$'' sign is to introduce an extra sign for defining the ket state, $\langle\vec{P}_\phi| \equiv \langle0| (-) \phi_{\vec{P}_\phi}$, so that the state's norm and the expectation value of Hamiltonian are positive. Physical results, e.g., the scattering $T$-matrix, computed in this way are the same as using the approach we have given here. All the other fields have conventional norms/completeness relations, e.g., $[c(\vec{x})\, , \, c^{\dagger}(\vec{x}')]= \delta\left(\vec{x} -\vec{x}'\right)$, $\langle\vec{P}_c'|\vec{P}_c\rangle = (2 \pi)^3 \delta \left(\vec{P}_c- \vec{P}_c' \right) $, and  $I_{c} = \int \frac{d\vec{p}}{\left(2 \pi \right)^3} |\vec{P}_c\rangle\langle\vec{P}_c|$. (Of course, for a fermion, the anti-commutator should be used.)

Quantization of the electromagnetic sector of the theory needs more care and won't be discussed here (see e.g.~\cite{Weinberg:1995mt}). After quantization the Hamiltonian can be written as:
\begin{eqnarray}
W_\gamma  = \sum_{\lambda =1,2} \int \frac{d \vec{P}_{\gamma}}{(2 \pi )^3} \omega_{\vec{P}_\gamma,\lambda} a^{\dagger}_{\vec{P}_\gamma,\lambda} a_{\vec{P}_\gamma,\lambda} - e \int d\vec{x}\, \vec{J}_N(\vec{x}) \cdot \vec{A} (\vec{x}) + \frac{e^2}{8\pi} \int d\vec{x} d\vec{x}' \frac{\rho_N(\vec{x}) \rho_N(\vec{x}') }{\vert \vec{x}-\vec{x}' \vert}  \label{eqn:toyhamphoton}
\end{eqnarray}
The first term is the free Hamiltonian for a transverse photon with two polarizations.  The last term is the pure Coulomb potential $W_C$, while the second term is what remains of the canonical $J^\mu A_\mu$ interaction, which elsewhere in the text we refer to as $-L_{EM}$. The current and charge densities associated with the matter fields, $\vec{J}_N$ and $\rho_N$, can be derived from the EFT Lagrangian. 

Since the two-particle state $|n,c\rangle$ appears frequently in the calculation, we now list several important relations between the individual-particle coordinates $\vec{R}_n$ and $\vec{R}_c$ (momenta $\vec{P}_n$ and $\vec{P}_c$) on the one hand, and the center-of-mass coordinate $\vec{R}_{nc}$ (momentum $\vec{P}_{nc}$) and relative coordinate $\vec{r}_{nc}$ (momentum $\vec{p}_{nc}$) on the other:
\begin{eqnarray}
\vec{R}_{n}&=&\vec{R}_{nc}+ f \vec{r}_{nc} \ , \\ 
\vec{R}_{c}&=&\vec{R}_{nc}-(1-f) \vec{r}_{nc} \ , \\
\vec{P}_{n}&=& (1-f)\vec{P}_{nc}+\vec{p}_{nc} \ , \\
\vec{P}_{c}&=& f\vec{P}_{nc}-\vec{p}_{nc} \ , 
\end{eqnarray}
As mentioned before, $f=\mc/\mnc$. Based on the single-particle Fock-state definition, the two-particle-state normalizations are 
\begin{eqnarray}
\langle\vec{R'}_{n} \vec{R'}_{c}|\vec{R}_{n} \vec{R}_{c}\rangle&=& \delta (\vec{R'}_{n}-\vec{R}_{n}) \delta (\vec{R'}_{c}-\vec{R}_{c}) \ , \label{eq:norm1}\\
\langle\vec{P'}_{n} \vec{P'}_{c}|\vec{P}_{n}\vec{P}_{c}\rangle &=& (2\pi)^{3} \delta(\vec{P'}_{n}-\vec{P}_{n})(2\pi)^{3} \delta(\vec{P'}_{c}-\vec{P}_{c}) \label{eq:norm2}\ .
\end{eqnarray}
The  Fock states with given space coordinates are defined as  $|\vec{R}_{n} \vec{R}_{c}\rangle \equiv n^{\dagger}(\vec{R}_n) c^{\dagger}(\vec{R}_c) |0\rangle$. Equivalently, these states can be labeled by variables that manifestly separate out CM motion, i.e., $ |\vec{P}_{nc}\, \vec{p}_{nc}\rangle \equiv |\vec{P}_{n}\vec{P}_{c}\rangle$ and $|\vec{R}_{nc}\, \vec{r}_{nc}\rangle \equiv |\vec{R}_{n} \vec{R}_{c}\rangle $. Since 
\begin{eqnarray}
d\vec{R}_{n} d\vec{R}_{c}&=& d\vec{R}_{nc} d\vec{r}_{nc} \ , \\
d\vec{P}_{n} d\vec{P}_{c}&=& d\vec{P}_{nc} d\vec{p}_{nc} \ ,
\end{eqnarray}
the normalizations for the states of Eq.~(\ref{eq:norm1}) and (\ref{eq:norm2}) can be rewritten as 
\begin{eqnarray}
\langle\vec{R'}_{nc} \vec{r}_{nc}'|\vec{R}_{nc} \vec{r}_{nc}\rangle&=&\delta(\vec{R}_{nc}'-\vec{R}_{nc}) \delta(\vec{r}_{nc}'-\vec{r}_{nc}) \ , \\
\langle\vec{P'}_{nc} \vec{p}_{nc}'|\vec{P}_{nc} \vec{p}_{nc}\rangle&=&(2\pi)^{3}\delta(\vec{P}_{nc}'-\vec{P}_{nc}) (2\pi)^{3} \delta(\vec{p}_{nc}'-\vec{p}_{nc}) \ .
\end{eqnarray}
Plane waves are then written as:
\begin{eqnarray}
\langle\vec{R}_{n} \vec{R}_{c}|\vec{P}_{n}\vec{P}_{c}\rangle= e^{i\left(\vec{P}_n\cdot\vec{R}_n +\vec{P}_c\cdot\vec{R}_c \right)}= e^{i\left(\vec{P}_{nc}\cdot\vec{R}_{nc} +\vec{p}_{nc}\cdot\vec{r}_{nc} \right)}=\langle\vec{R}_{nc}\, \vec{r}_{nc} |\vec{P}_{nc}\, \vec{p}_{nc}\rangle.
\end{eqnarray}

We make extensive use of the so-called Coulomb distorted states, defined in terms of the plane waves as:
 \begin{eqnarray}
|\vec{P}_{nc}, \cw{\vec{p}_{nc}}{\pm}\rangle  \equiv  \left(1+  \frac{1}{E-H_C + i 0^{\pm}} W_C \right) |\vec{P}_n,\vec{P}_c\rangle \ ,\label{eqn:CoulombWFdef}
\end{eqnarray}
with $H_C\equiv H_0 + W_C$. These states can be computed analytically, and are:
\begin{eqnarray}
\langle\vec{R}_{nc}\, \vec{r}_{nc}  |\vec{P}_{nc}, \cw{\vec{p}_{nc}}{\pm}\rangle = e^{i \vec{P}_{nc}\cdot \vec{R}_{nc}} \cw{\vec{p}_{nc}}{\pm}(\vec{r}_{nc}) \label{eqn:coulombcoordinatewf} \ .
\end{eqnarray}
with $ \cw{\vec{p}_{nc}}{\pm}(\vec{r}_{nc})$ the so-called Coulomb wave functions in coordinate space~\cite{GoldbergerQM, Kong:1999sf}, details of which can be found in Appendix~\ref{app:coulombwf}. 

Now we are in a position to compute several matrix elements used in the scattering and reaction calculations. For $W_s$, we have 
\begin{eqnarray}
\langle\vec{P}_{\phi} | W_s |\vec{P}_{nc}, \cw{\vec{p}_{nc}}{\pm}\rangle  &=& \langle\vec{P}_{\phi} | \int d \vec{r} (-)h_s \phi^{\dagger}(\vec{r}) n(\vec{r}) c(\vec{r}) |\vec{P}_{nc}, \cw{\vec{p}_{nc}}{\pm}\rangle \notag \\ 
 & = &  h_s (2 \pi)^3
\delta(\vec{P}_{nc} - \vec{P}_{\phi}) \cw{\vec{p}_{nc}}{\pm}(0)  \label{eqn:Wsmatrixelement}
\end{eqnarray}
Note the subscripts of momentum variables in the Fock state indicate the particle type therein. In the 2nd step of the derivation, the sign is flipped because of the $\phi$ state's negative norm. Hermitian conjugation of this vertex's matrix element amounts to complex conjugation of the expression~(\ref{eqn:Wsmatrixelement}). 

If some care is taken with the operation of derivatives, we can compute the matrix element for $W_p$ in a similar fashion:
\begin{eqnarray}
\langle\vec{P}_{\pi}, \pi^j  | W_p |\vec{P}_{nc}, \cw{\vec{p}_{nc}}{\pm}\rangle  &=& \langle\vec{P}_{\pi}, \pi^j| \int d \vec{r} (-) h_p \pi^{\dagger}_i(\vec{r}) n(\vec{r})  \tilde{\vec{V}}_R^i c(\vec{r}) |\vec{P}_{nc}, \cw{\vec{p}_{nc}}{\pm}\rangle \notag \\ 
 & = & i \frac{ h_p}{\mr } (2 \pi)^3 \delta(\vec{P}_{nc} - \vec{P}_{\phi}) \int d\vec{r} \delta(\vec{r}) \partial_j\cw{\vec{p}_{nc}}{\pm}(\vec{r}) \notag \\
&= &  i \frac{ h_p}{\mr } (2 \pi)^3 \delta(\vec{P}_{nc} - \vec{P}_{\phi})  \partial_j\cw{\vec{p}_{nc}}{\pm}(0) \ . \label{eqn:Wpmatrixelement}
\end{eqnarray}
The index $j$ corresponds to the dimer state with spin projection $j$ on the $\hat{z}$ axis. 
\begin{figure}
\centering
\includegraphics[width=10cm, angle=0]{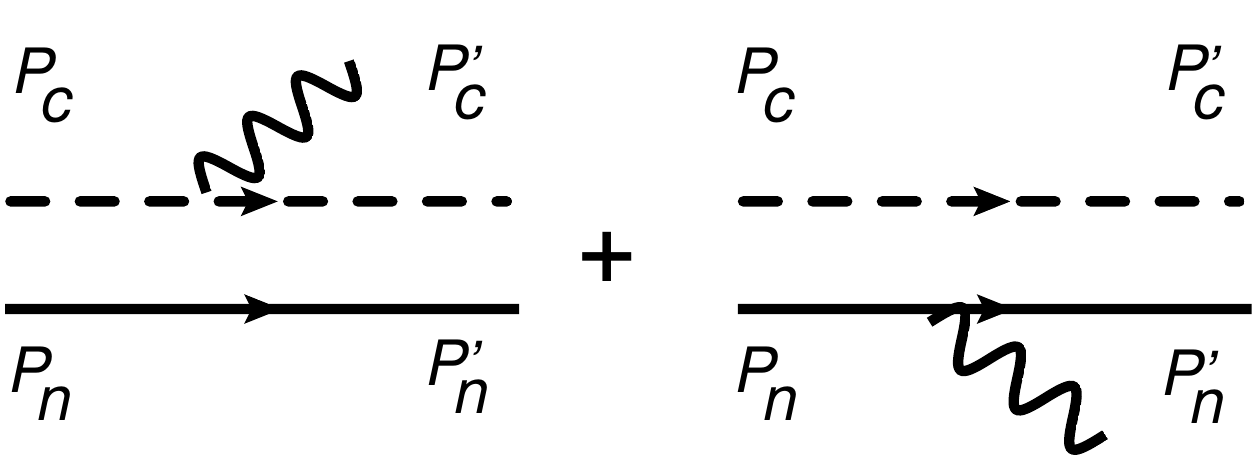}
\caption{The matrix element of the EM vertex $L_{EM}$--defined in expression~(\ref{eqn:toyhamphoton})--between initial free $n-c$ state and final free $n-c$ plus one transverse photon state. } \label{fig:EMvertex}
\end{figure}

We can also calculate the vertex shown in Fig.~\ref{fig:EMvertex} in the plane-wave basis (in Coulomb gauge): 
\begin{eqnarray}
&& \langle\vec{P}_n',\vec{P}_c',\, \vec{P}_{\gamma}, A^{\lambda} |L_{EM}|\vec{P}_n,\vec{P}_c \rangle =\notag \\ 
&& \qquad (2\pi)^{6}\delta\left(\vec{P}_{nc}-\vec{P'}_{nc}-\vec{P}_{\gamma}\right) \delta\left(\vec{p}_{nc}-\vec{p'}_{nc}-f\vec{P}_{\gamma}\right) \frac{Q_{n}}{\mn} \es_{\lambda} \cdot \left[\vec{p}_{nc}+(1-f)\vec{P}_{nc}\right] + \notag \\ 
&& \qquad (2\pi)^{6}\delta\left(\vec{P}_{nc}-\vec{P'}_{nc}-\vec{P}_{\gamma}\right) \delta\left(\vec{p}_{nc}-\vec{p}_{nc}'+(1-f)\vec{P}_{\gamma}\right) \frac{Q_{c}}{\mc} \es_{\lambda} \cdot \left[-\vec{p}_{nc}+f\vec{P}_{nc}\right]. 
\end{eqnarray}
Again $A^\lambda$ means the transverse photon with polarization $\lambda$. $\vec{P}_{\gamma}$ and $\es_{\lambda}$ are the photon's outgoing momentum and its polarization vector. It is then straightforward to convert this to a matrix element between Coulomb-distorted states in the frame where the total initial momentum  of the $nc$ system, $\vec{P}_{nc}=\vec{0}$.
\begin{eqnarray}
&& \langle\vec{P}_{nc}', \cw{\vec{p}'}{-},\, \vec{P}_{\gamma}, A^\lambda|L_{EM}|\vec{P}_{nc}=\vec{0}, \cw{\vec{p}_{nc}}{+} \rangle 
  = (2\pi)^{3}  \delta\left(\vec{P'}_{nc}+\vec{P}_{\gamma}\right) 
 \times  \nonumber \\
&& \quad \int d\vec{r}_{nc}  \cwc{\vec{p'}_{nc}}{-} (\vec{r}_{nc}) \left[ e^{- i f \vec{P}_{\gamma}\cdot\vec{r}_{nc}} \frac{Q_n}{\mn}  - e^{ i (1-f) \vec{P}_{\gamma} \cdot\vec{r}_{nc}} \frac{Q_c}{\mc}  \right] \left(-i\es_{\lambda} \cdot \partial_{\vec{r}_{nc}}\right)  \cw{\vec{p}_{nc}}{+}(\vec{r}_{nc}) \ . \label{eqn:EMvertexCM}
\end{eqnarray}

We can also define the full propagator for all the fields, which is useful in diagrammatic calculations. For example, we define the full propagator $D_{\phi}$
\begin{eqnarray}
 \langle \vec{P}_\phi' \vert \frac{1}{E-H + i 0^{+}}  \vert \vec{P}_{\phi} \rangle \equiv \left( 2 \pi \right)^3 \delta \left(\vec{P}_\phi-\vec{P}_\phi' \right)D_{\phi}\left(\vec{P}_\phi', \vec{P}_\phi;\, E \right). \label{eqn:DphidefHformalism}
\end{eqnarray}
The states are free dimer states here. The other fields' propagators are defined in the same way and hence won't be shown explicitly here. The matrix elements can be expanded using the LSE.    

\section{Asymptotic Coulomb wave function} \label{app:coulombwf}
The Coulomb-distorted incoming and outgoing wave functions in coordinate space are 
\begin{eqnarray}
\cw{\vec{k}}{\pm}(\vec{r})&=&e^{-\frac{\pi}{2}\eta} e^{i\vec{k}\vec{r}}\Gamma(1\pm i\eta) M(\mp i\eta ,1;\pm i k r-i \vec{k}\vec{r}) \ , 
\end{eqnarray}
with $\eta \equiv \frac{Z_c Z_n \alpha_{\rm em} \mr}{k}= \frac{\kc}{k}$, and $\frac{k^2}{2 \mr}$ for the relative energy of the two-particle system. $M(a,b;z)$ is the Kummer function \cite{MathHandBook1}, and at small $z$: 
\begin{eqnarray}
M(a,b;z)=1+\frac{a}{b} z + \frac{a(a+1)z^{2}}{b(b+1)2!}+ \cdots \ . 
\end{eqnarray} 
By using this expansion, we get
\begin{eqnarray}
\cw{\vec{k}}{\pm}(r=0) &=& C_{\eta,0} e^{\pm i \sigma_{0}} \ , \notag \\ 
\dpartial{j}\cw{\vec{k}}{\pm}(0)&=& 3 C_{\eta,1} e^{\pm i \sigma_{1}} ik_{j} \ . \label{eqn:dcw0}
\end{eqnarray}
with  $ C_{\eta,l}\equiv \frac{2^{l}e^{-\frac{\pi}{2}\eta}|\Gamma(l+1+i\eta)|}{\Gamma(2l+2)} $ and $e^{2i\sigma_{l}} \equiv  \frac{\Gamma(l+1+i\eta)}{\Gamma(l+1-i\eta)}$ . 

As with the plane wave, the $\cw{\vec{k}}{\pm}(\vec{r})$ have a partial-wave decomposition:
\begin{eqnarray}
\cw{\vec{k}}{+}(\vec{r}) &=& 4\pi \sum_{l,m} i^{l} \frac{\cwf{l}(k,r)}{kr} e^{i\sigma_{l}} \dY{l}{m}(\vecpt{k}) \uY{l}{m}(\vecpt{r}) \ , \\ 
\cw{\vec{k}}{-}(\vec{r}) &=& 4\pi \sum_{l,m} i^{l} \frac{\cwf{l}(k,r)}{kr} e^{-i\sigma_{l}} \uY{l}{m}(\vecpt{k}) \dY{l}{m}(\vecpt{r}) \ .  \label{eqn:cwldecomp}
\end{eqnarray}
Here $\uY{l}{m}(\vecpt{r})$ is a conventional spherical harmonic, but $\dY{l}{m}(\vecpt{r}) \equiv \uY{l}{m}(\vecpt{r})^{\ast}$. Meanwhile, 
$\cwf{l}(k,r)$ is the regular solution of the Schr\"odinger equation with pure Coulomb interaction at angular momentum $l$:
\begin{eqnarray}
\cwf{l}(k,r)=C_{\eta, l} e^{ikr} (kr)^{l+1} M(l+1+i\eta;2l+2,-2ikr) \ .
\end{eqnarray}
It is a real function for real $k$, $r$, and $\kc$, as can be checked using $M(a,b,z)=e^{z}M(b-a,b,-z)$.  Associated with it is an irregular real solution, known as $\cwg{l}(k,r)$. (More details can be found in Ref.~\cite{MathHandBook1}.) The two can be related to the Whittaker function \cite{MathHandBook1} for real $k$, $r$, and $\kc$: 
\begin{subequations}
\begin{alignat}{2}
\cwg{l}(k,r)+i\cwf{l}(k,r)&=& e^{i\sigma_{l}} e^{\frac{\pi}{2}\eta} (-i)^{l} W_{-i\eta,l+\frac{1}{2}}(-2ikr) \ , \label{eqn:FGvsW1} \\
\cwg{l}(k,r)-i\cwf{l}(k,r)&=& e^{-i\sigma_{l}} e^{\frac{\pi}{2}\eta} (i)^{l} W_{i\eta,l+\frac{1}{2}}(2ikr) \ .  \label{eqn:FGvsW2}
\end{alignat}
\end{subequations}
Here $W_{\kappa,\mu}(z)$ is analytic in $\kappa$, $\mu$, and $z$. An important property about this function needs to be pointed here, i.e. $k^{l}\Gamma(l+1-i\eta) W_{i\eta,l+\frac{1}{2}}(2ikr)$ as a function of $k$ is analytic on the whole lower complex plane, including the real axis. To prove this statement first observe that $\Gamma(l+1-i\eta)$ is analytic when $\mathrm{Im}~k \leq 0$, and, as mentioned before, the Whittaker function is analytic for non-zero $k$. Therefore, to prove the proposed analyticity property, we only need to focus on $k\rightarrow 0 - i 0^+$ from the lower plane. The Whittaker function can be represented in integral form \cite{ConfluentHypergeometricBuchholz} (for $z\neq 0$, $ - \pi < \arg{z} < \pi$, and $\mathrm{Re}(\frac{\mu+1}{2}-\kappa) > 0$):
\begin{eqnarray}
W_{\kappa, \frac{\mu}{2}}(z) =\frac{z^{k}e^{-z}}{\Gamma(\frac{\mu+1}{2}-\kappa)} \int_{0}^{+\infty} e^{-w} w^{\frac{\mu-1}{2}-k} (1+\frac{w}{z})^{\frac{\mu-1}{2}+k} dw  \ , 
\end{eqnarray}
which leads to  
\begin{eqnarray}
W_{i\eta,l+1/2}(z) \Gamma(l+1-i\eta)z^{l} |_{z=2ikr}
&=&e^{-z} \int_{0}^{+\infty} e^{-w} w^{l} (w+z)^{l} \left[1+\frac{2ikr}{w}\right]^{i \frac{\kc}{k}} dw   \notag \\ 
&=& \int_{0}^{+\infty} e^{-w} w^{2l} e^{-\frac{2r \kc }{w}} dw \ (\mathrm{when}\ k\rightarrow 0 - i 0^+) \ ,
\label{eq:finitelimit}
\end{eqnarray}
which is finite. Note that as long as $\mathrm{Im} (k) \leq 0$, the $ - \pi < \arg{z} < \pi$ and $\mathrm{Re}(\frac{\mu+1}{2}-\kappa) > 0$ conditions are satisfied. Eq.~(\ref{eq:finitelimit}) indicates that we can analytically continue $k^l \Gamma(l+1-i\eta) W_{i\eta,l+\frac{1}{2}}(2ikr)$ from $k$'s lower half-plane  ($\mathrm{Im}(k) < 0$) to the real axis ($\mathrm{Im}(k)=0$). Following the same arguments, we can analytically continue $ k^l \Gamma(l+1+i\eta) W_{-i\eta,l+\frac{1}{2}}(-2ikr)$ from $k$'s upper half-plane ($\mathrm{Im}(k) > 0$) down to the real axis ($\mathrm{Im}(k)=0$). This proves the analyticity properties claimed in the main text. 

\bibliography{./nuclear_reaction-7-26-2015}

\end{document}